\documentclass[a4paper,fleqn,usenatbib]{mnras}

\usepackage{newtxtext,newtxmath}

\usepackage[T1]{fontenc}
\usepackage{ae,aecompl}

\usepackage{graphicx}	
\usepackage{amsmath}	

\usepackage[usenames]{color}

\title[Rotation of convective core]{Rotation of the convective core in $\gamma$ Dor stars measured by dips in period spacings of g modes coupled with inertial modes} 

\author[H. Saio et. al]{
Hideyuki Saio,$^{1}$ \thanks{E-mail: saio@astr.tohoku.ac.jp}
Masao Takata,$^{2}$
Umin Lee,$^{1}$
Gang Li,$^{3}$
Timothy Van Reeth$^{4}$
\\
$^{1}$Astronomical Institute, Graduate School of Science, Tohoku University, Sendai 980-8578, Japan\\
$^{2}$Department of Astronomy, School of Science, The University of Tokyo, Bunkyo-ku, Tokyo 113-0033, Japan \\
$^{3}$Sydney Institute for Astronomy, School of Physics, 2006 University of Sydney, Australia\\
$^{4}$Institute of Astronomy, KU Leuven, Celestijnenlaan 200D, B-3001 Leuven, Belgium\\
}

\date{Accepted XXX. Received YYY; in original form ZZZ}

\pubyear{2020}

\begin{document}
\label{firstpage}
\pagerange{\pageref{firstpage}--\pageref{lastpage}}
\maketitle

\begin{abstract}
The relation of period spacing ($\Delta P$) versus period ($P$) of dipole prograde g modes is known to be useful to measure rotation rates in the g-mode cavity of rapidly rotating $\gamma$ Dor and slowly pulsating B (SPB) stars.
In a rapidly rotating star, an inertial mode in the convective core can resonantly couple with g modes propagative in the surrounding radiative region.
The resonant coupling causes a dip in the $P$\,-\,$\Delta P$ relation, distinct from the modulations due to the chemical composition gradient.  
Such a resonance dip in $\Delta P$ of prograde dipole g modes appears around a frequency corresponding to a spin parameter $2f_{\rm rot}{\rm(cc)}/\nu_{\rm co-rot} \sim 8-11$  with $f_{\rm rot}$(cc)  being the rotation frequency of the convective core and $\nu_{\rm co-rot}$ the pulsation frequency in the co-rotating frame.
The spin parameter at the resonance depends somewhat on the extent of core overshooting, central hydrogen abundance, and other stellar parameters.
We can fit the period at the observed dip with the prediction from prograde dipole g modes of a main-sequence model, allowing the 
convective core to rotate differentially from the surrounding g-mode cavity.    
We have performed such fittings for 16 
selected $\gamma$ Dor stars having well defined dips, and found that
the majority of $\gamma$ Dor stars we studied rotate nearly uniformly, while convective cores tend to rotate slightly faster than the g-mode cavity in less evolved stars.
\end{abstract}




\section{Introduction}
Many F to B type main-sequence stars show light variations attributed to many  simultaneously excited nonradial low-degree g mode oscillations with periods ranging from $\sim0.5$ to $\sim2$~days.
These variables are called $\gamma$~Doradus ($\gamma$~Dor) stars and Slowly Pulsating B (SPB) stars. 
It is known that g modes in the SPB stars are excited by the Fe-Ni opacity bump \citep{gau93,Dzi93}, while the excitation mechanism for g modes in the $\gamma$~Dor stars is not settled yet \citep[see, e.g.,][]{kah20}.

The densely and regularly distributed periods of g modes, whose propagation zone is located in the deep interior including chemically inhomogeneous layers surrounding the convective core, are ideal for asteroseismology to probe the deep interior structure of a star.
The long pulsation periods, which are often comparable to rotation periods, are significantly affected by the Coriolis force.
This property, in turn, can be used to probe the rotation in the deep interior, although ground-based photometric observations are not suitable for such long and multiple periodic pulsations.
The obstacle has been resolved by the recent advents of space photometry from satellites. 
In particular, up to four-year long ultra-accurate space photometry by the Kepler satellite \citep{kepler}
brought revolutionary developments in asteroseismology for $\gamma$~Dor and SPB stars \citep[e.g.,][]{kur14,kee15,tri15,mur16,sch16,vanr16,LiG20,pap17}. 

Period spacings ($\Delta P$) of intermediate to high order g modes in a slowly rotating chemically homogeneous (ZAMS) star  are roughly constant (in the co-rotating frame) with respect to period or frequency. In evolved main-sequence stars, however, period spacings show modulations caused by a steep gradient of the hydrogen abundance exterior to the convective core \citep{mig08}. The modulation amplitude is sensitive to diffusive mixing of chemical composition in radiative layers \citep{bou13}. 
In addition, the mean value of the period spacing (in the co-rotating frame) decreases with evolution (due to an increase in the Brunt-V\"ais\"al\"a frequency in the g-mode cavity).
These properties are useful to infer the strength of diffusion in the deep interior and  the evolutionary stages of $\gamma$ Dor stars, 
although dependences on initial chemical composition, core-overshooting, stellar mass, and etc. should be taken into account \citep{mom19}.
We will discuss, in this paper, another type of $\Delta P$ modulations (dips) that occur in rapidly rotating stars due to the resonant coupling between g modes in the near-core region and an {\it inertial} mode in the convective core. 

If the rotation period is comparable to the pulsation periods in the co-rotating frame, g-mode period spacings in the inertial frame vary as a function of period \citep{bou13,vanr15,oua17,chr18,LiG19a}. The property has been used to estimate rotation frequencies in the g mode cavity of $\gamma$ Dor stars and SPB (slowly pulsating B) stars
\citep[e.g.,][]{vanr16,oua17,zwi17,pap17,chr18,LiG19,LiG20,tak20}.
To calculate g-mode frequencies of a rotating stars, traditional approximation of rotation \citep[e.g.,][TAR]{lee97}, where the horizontal component of angular velocity of rotation, $\Omega\sin\theta$ is neglected, has been employed in many investigations. It is known that the approximation gives sufficiently accurate frequencies and hence accurate period spacings ($\Delta P$) of g modes 
and their modulations due to a steep chemical-composition gradient.
Recently, however, \citet{sai18b} found that $P$\,-\,$\Delta P$ relation  of dipole prograde g modes calculated without using the TAR has a narrow deviation from that obtained by using the TAR. 
\citet{oua20} found the cause of the deviation (or dip) to be resonance couplings with an {\it inertial} mode in the convective core. 

In this paper, we further investigate the property of the dips in the $P$\,-\,$\Delta P$ relation of g modes caused by the resonance with an inertial mode in the convective core. We first discuss the theoretical property of the period spacing of g-modes and couplings with an inertial mode in \S2 and \S3. Then, we fit theoretical $P$\,-\,$\Delta P$ relations and resonance dips with observational data of some $\gamma$ Dor stars to estimate rotation rates of convective cores.
In \S\ref{sec:rmodes} we show no resonance coupling to occur between an inertial mode and r modes. In Appendix\ref{sec:bigdip} we show an example (KIC~1431379) of large dips caused by chemical composition gradient. 

\section{Period spacings of g modes in a rotating star}
\label{sec:pspace}
The traditional approximation of rotation (TAR) is useful for studying the property of low-frequency nonradial pulsations in rotating stars. 
The TAR neglects the horizontal component of the angular velocity of rotation $\Omega\sin\theta$ (uniform rotation is assumed), centrifugal force of rotation, and the Eulerian perturbations of gravitational potential (i.e., Cowling approximation). Then, the governing equations for low-frequency nonradial adiabatic  pulsations of a rotating star are reduced to those for a non-rotating star except that $\ell(\ell+1)$ is replaced with $\lambda$, the eigenvalue of the Laplace's tidal equation \citep[e.g.,][]{lee97}, where $\ell$ means the latitudinal degree of a nonradial pulsation in a non-rotating star.  While $\ell(\ell+1)$ is constant, $\lambda$ varies as a function of spin parameter,~$s$, defined as
\begin{equation}
s \equiv {2\Omega\over\omega} = {2f_{\rm rot}\over \nu_{\rm co-rot}},
\label{eq:spin}
\end{equation}
 where $\omega$ and $\nu_{\rm co-rot}$ are, respectively, angular and cyclic frequency of pulsation in the co-rotating frame, while $f_{\rm rot}$ is cyclic frequency of rotation.
The eigenvalue $\lambda$ varies significantly if $s > 1$ \citep[see e.g.][]{lee97,tow03a,sai18b}.  Because of the variation of $\lambda$, low-frequency oscillations in moderately to rapidly rotating stars (i.e., $s>1$) have properties significantly different from those in a non-rotating star.

According to the recent analyses of Kepler light curves for low-frequency pulsations in $\gamma$ Dor stars \citep{vanr16,oua17,LiG19,LiG20} and SPB stars \citep{pap15,pap17}, the majority of pulsations in moderately to rapidly rotating stars are prograde sectoral g modes \citep[sometimes called Kelvin modes; see e.g.,][]{tow03a,tak20} and r modes \citep[normal modes of Rossby waves; e.g.][]{sai18}.\footnote{However, no r modes have been found in SPB stars despite that the excitation by the $\kappa$ mechanism is predicted 
\citep[e.g.,][]{tow05,sav05,lee06}. 
}
In this paper, we discuss prograde sectoral g modes in rapidly rotating stars. We adopt, as in \citet{oua20,sai18b} and in \citet{unno}, the convention that negative azimuthal order $m < 0$ corresponds to prograde modes. 
We also note that all the theoretical pulsation frequencies (or periods) in this paper have been obtained using the adiabatic approximation.

Under the TAR, frequency of a prograde sectoral g mode in the co-rotating frame is given as
\begin{equation}
\nu_{\rm co-rot} \approx {|m|\over n_{\rm g}}\nu_0  \quad \mbox{for} \quad s > 1 ~\mbox{and}~ n_{\rm g} \gg 1, 
\label{eq:nuco}
\end{equation}
where 
\begin{equation}
\nu_0 \equiv {1\over 2\upi^2}\int_{r_1}^{r_2}{N\over r}dr
\label{eq:nu0}
\end{equation}
with Brunt-V\"ais\"al\"a frequency $N$ and $n_{\rm g} (> 0)$ being the number of radial nodes \citep{sai18b}. 
In equation~(\ref{eq:nuco}), the property of $\lambda \approx m^2$ (if $s > 1$) for prograde sectoral modes \citep[e.g.][]{lee97,tow03a} is used.
Since the range of g-mode cavity  $r_1 < r < r_2$ is not very sensitive to g-mode frequencies, $\nu_0$ is nearly constant for a star, so that the period spacing in the co-rotating frame, 
\begin{equation}
\Delta P_{\rm co-rot} \approx {1\over |m|\nu_0}
\label{eq:dP}
\end{equation}
is nearly constant.

\begin{figure}
  \includegraphics[width=\columnwidth]{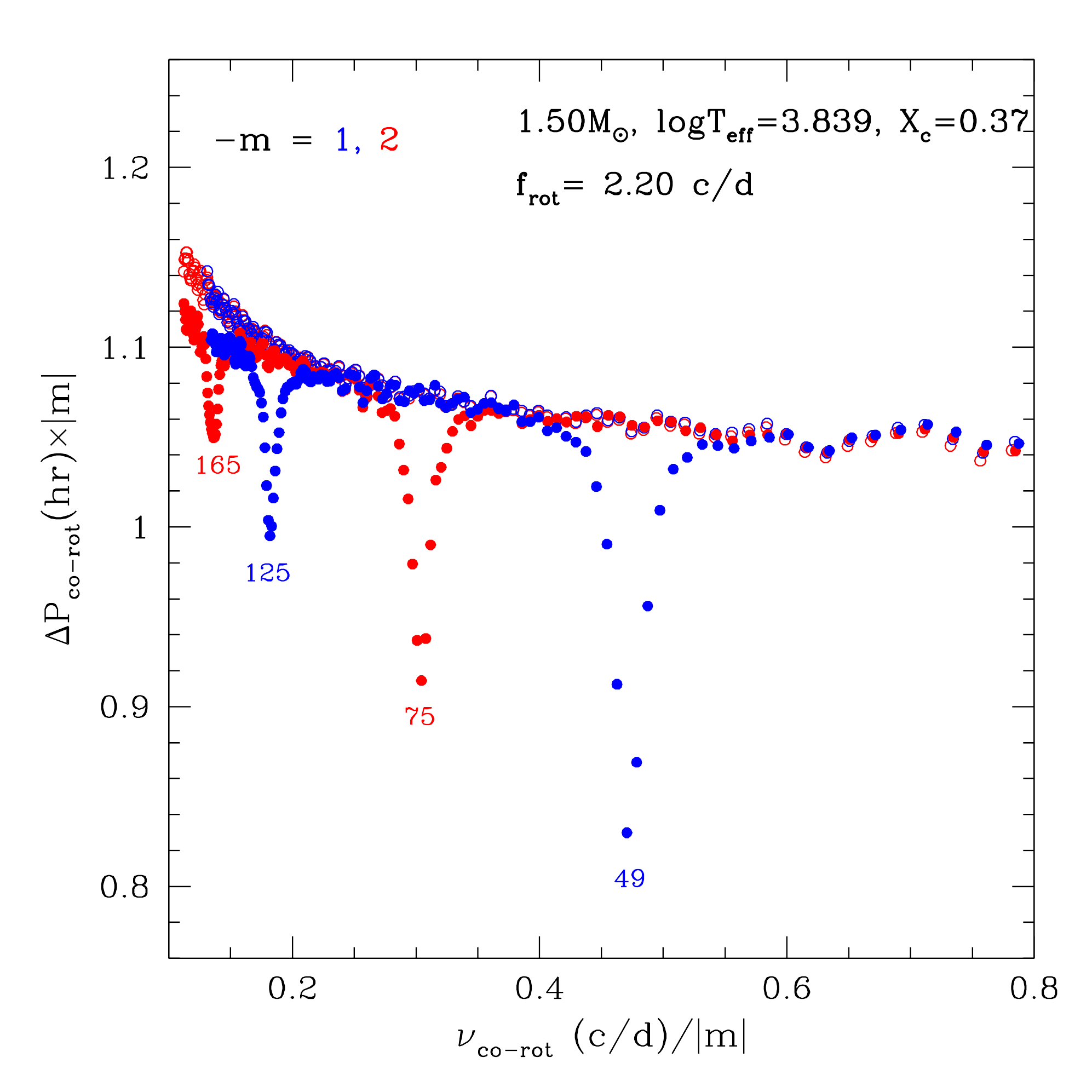} 
  \caption{Period spacing in the co-rotating frame, $\Delta P_{\rm co-rot}$, versus frequency in the co-rotating frame, $\nu_{\rm co-rot}$ for prograde sectoral g modes of $m=-1$ (blue symbols) and $-2$ (red symbols) in a main-sequence model of $1.5~M_\odot$. To compensate for the effect of different azimuthal order $m$, vertical and horizontal axes are multiplied and divided by $|m|$, respectively [see equations~(\ref{eq:nuco}),(\ref{eq:dP})]. Open symbols show results obtained using the traditional approximation of rotation (TAR). Numbers shown at minima in $\Delta P_{\rm co-rot}$ indicate the number of radial nodes in the eigenfunction for the mode at each minimum. The minima occur at spin parameters of $9.3$ and $24$ for $m=-1$ modes, and at $7.2$ and $16$ for $m=-2$ modes. }
  \label{fig:dPco_m1m2}
\end{figure}

Open circles in Fig.~\ref{fig:dPco_m1m2} show period-spacing versus frequency in the co-rotating frame obtained with the TAR for prograde sectoral g modes of $m = -1$ (blue) and $-2$ (red) assuming a rotation frequency of $2.20$~d$^{-1}$ in a $1.5$-$M_\odot$ main sequence model. (An initial composition of $(X,Z)=(0.72,0.014)$ \citep{Geneva2012} is adopted unless stated otherwise 
in the stellar evolution models in this paper.)\footnote{Stellar structure models were obtained by the MESA code \citep[v.7184;][]{pax11,pax13,pax15}, in which convective core boundary was determined by the Schwarzschild criterion, elemental diffusion was activated to have smooth Brunt-V\"ais\"al\"a frequency, and radiation turbulence was also activated to prevent too much helium settling. Rotational deformation was neglected.} 
As predicted by the approximate relations in equations~(\ref{eq:nuco}),(\ref{eq:dP}), the effect of different $m$ is nearly compensated by taking horizontal and vertical axes as $\Delta P_{\rm co-rot}|m|$ and $\nu_{\rm co-rot}/|m|$, respectively. Period spacings modulate due to the presence of a steep hydrogen abundance gradient around the convective core \citep{mig08,bou13}.  
In the inertial frame, period spacing decreases with period for prograde sectoral g modes, which is employed to measure the rotation speed in the g-mode cavity in the envelope \citep{vanr15,oua17,chr18,LiG19a}.

Filled circles in Fig.~\ref{fig:dPco_m1m2} show results obtained without the TAR, by using the expansion method of \citet{lee95}, in which the eigenfunction of a pulsation mode is expressed by a sum of a truncated series of terms proportional to different degrees of spherical harmonics $Y_\ell^m$. We include four to eight spherical harmonics depending on the degree of amplitude spreading among the terms. Thus obtained results (filled circles in Fig.~\ref{fig:dPco_m1m2}) generally agree with those obtained using the TAR (open circles) except for notable deviations at dips of filled circles. Such a deviation around a spin parameter of $9.3$ was first recognized by \citet{sai18b} for the $m=-1$ sequence in the study on the $\gamma$ Dor star KIC 5608334. (Other dips are out of the observed frequency range.)
Although \citet{sai18b} guessed wrongly that the deviation might be caused by a coupling with a tesseral g mode,  \citet{oua20} found from two-dimensional calculations that the dip should be caused by a resonance coupling between g modes in the radiative region and an {\it inertial} mode in the convective core, where waves are propagative due to the effect of the Coriolis force. In the next section, we discuss the resonance couplings between prograde dipole g modes with inertial modes.

\section{Coupling between g modes and inertial modes in the convective core}
\begin{figure}
  \includegraphics[width=\columnwidth]{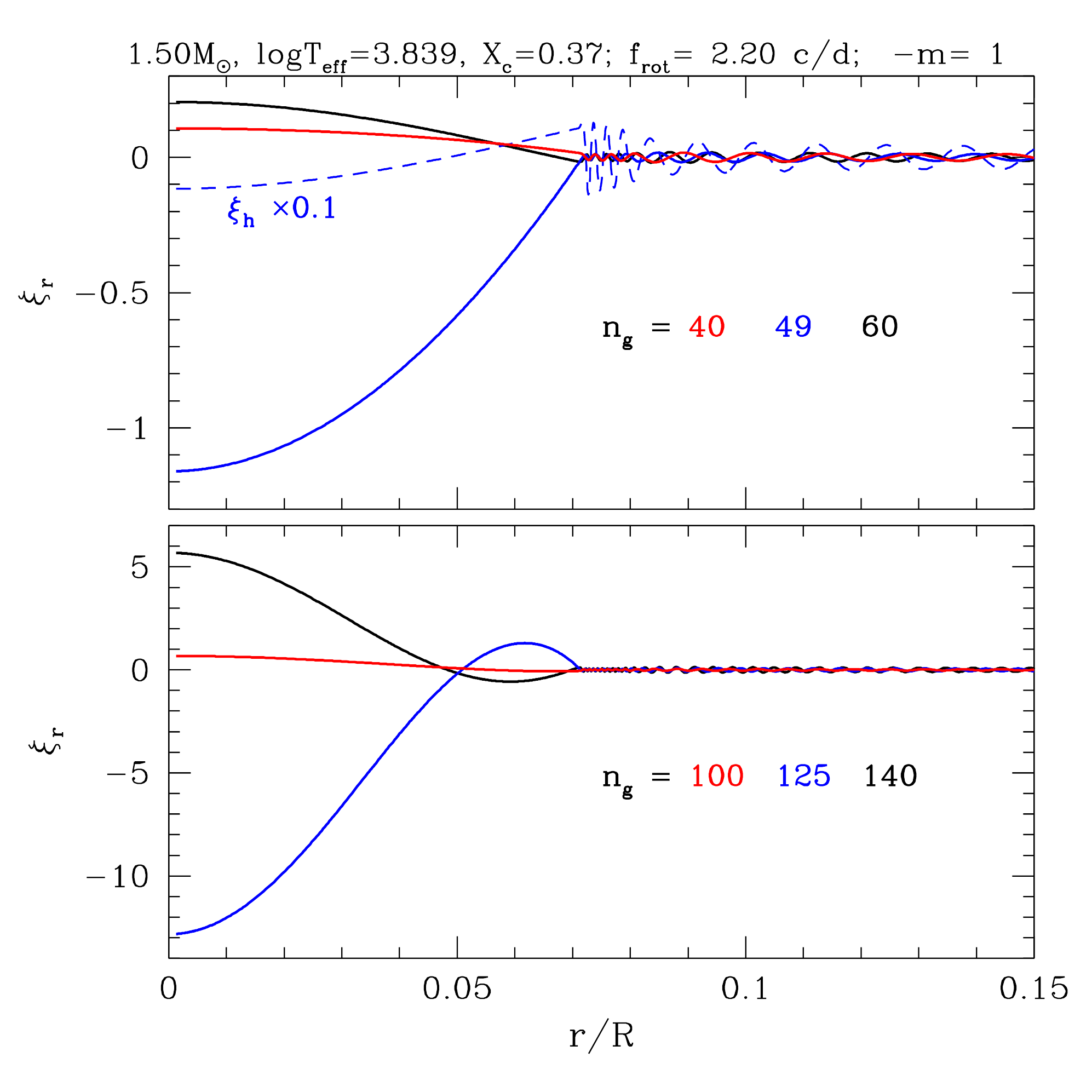}
\caption{Radial displacements of g modes as a function of fractional radius in the central part of the $1.5$-$M_\odot$ model. Plotted are for g modes around period-spacing minima at spin parameters of about $9.3$ (upper panel) and about  $24$ (lower panel). The radial displacement is normalized to unity at the stellar surface. The blue dashed line shows the horizontal displacement of the $n_{\rm g}=49$ mode for comparison. The convective core boundary is located at $r/R=0.071$.}
  \label{fig:eigf}
\end{figure}

In order to confirm that the dips of $\Delta P_{\rm co-rot}$ seen in Fig.~\ref{fig:dPco_m1m2} are caused by resonance couplings with inertial modes in the convective core, we plot eigenfunctions of dipole ($m=-1$) g modes around these  dips in Fig.~\ref{fig:eigf}.
Solid lines are radial displacements $\xi_r$ (the first component of expansion; i.e., with $\ell=1$)\footnote{The other components with $\ell=3,5,\ldots$ are very small compared to the $\ell=1$ component within the convective core except close to the core boundary.} as a function of fractional radius ($0 < r/R \le 0.15$) for some prograde dipole modes around $s=9.3$ ($n_{\rm g}=49$; upper panel) and at $s=24$ ($n_{\rm g}=125$; lower panel). For all cases, $\xi_r$ is normalized to unity at the stellar surface ($r=R$).  
Clearly, the amplitude in the convective core is maximum for the mode at the center of each dip of $\Delta P_{\rm co-rot}$, indicating the resonance coupling between an inertial mode in the convective core and g modes in the surrounding radiative g-mode cavity.
The dashed blue line in Fig.~\ref{fig:eigf} shows the horizontal displacement $\xi_{\rm h}$ (multiplied by $0.1$) of the $n_{\rm g}=49$ mode for comparison with the radial displacement (blue solid line). As is well known, the horizontal displacement of a high-order g mode is much larger than the radial displacement in the g-mode cavity. However, the horizontal and radial displacements are comparable with each other for the inertial mode in the convective core. Therefore, for a g mode coupled with an inertial mode, the contrast between the amplitude in the convective core and in the radiative region is much larger for the radial displacement. 

The mode with $n_{\rm g} = 49$ (upper panel) has no node in the convective core, while the mode with $n_{\rm g} = 125$ (lower panel) has one node in the convective core.  
This indicates that the 'fundamental' inertial mode in the convective core couples with g modes in the upper panel, while the 'first-overtone' inertial mode couples with g modes in the lower panel.  
When the frequency of an inertial mode in the convective core becomes very close to the frequency of a g mode in the radiative region, the amplitude in the core becomes comparable to or larger than the surface value of the g mode by coupling, then the period spacing attains a minimum. The frequency of the mode with $n_{\rm g}=49$ (no node in the convective core) is larger (by a factor of $2.6$) than the frequency of the mode with $n_{\rm g} = 125$ (one node in the convective core). This indicates that a larger radial-wavelength in the convective core
corresponds to a higher frequency of the inertial mode;
the character is the same as that of g modes. 

In a rotating convective core, where the Brunt-V\"ais\"al\"a frequency is almost zero ($N^2\approx 0$),\footnote{In this paper we have assumed $N^2=0$ in the convective core, while overstable convective modes in the core possibly couple with g modes in the envelope if we assume a non-zero super-adiabatic temperature gradient (i.e. $N^2<0$) in the convective core as discussed in \citet{lee20}. } low frequency inertial waves approximately obey the local dispersion relation given (in the co-rotating frame) as
\begin{equation}
\omega^2 = {(2\mathbf{\Omega}\cdot\mathbf{k})^2\over k^2} \quad {\rm or} \quad
\left({\omega\over 2\Omega}\right)^2 = {k_z^2\over k^2}
\label{eq:disp}
\end{equation}
\citep[e.g.,][]{unno,lee97}, where $\mathbf{k}$ means wave number.
This indicates that the property of the inertial modes in the convective core is governed by the spin parameter, $s=2\Omega/\omega$.

\begin{figure}
	\includegraphics[width=\columnwidth]{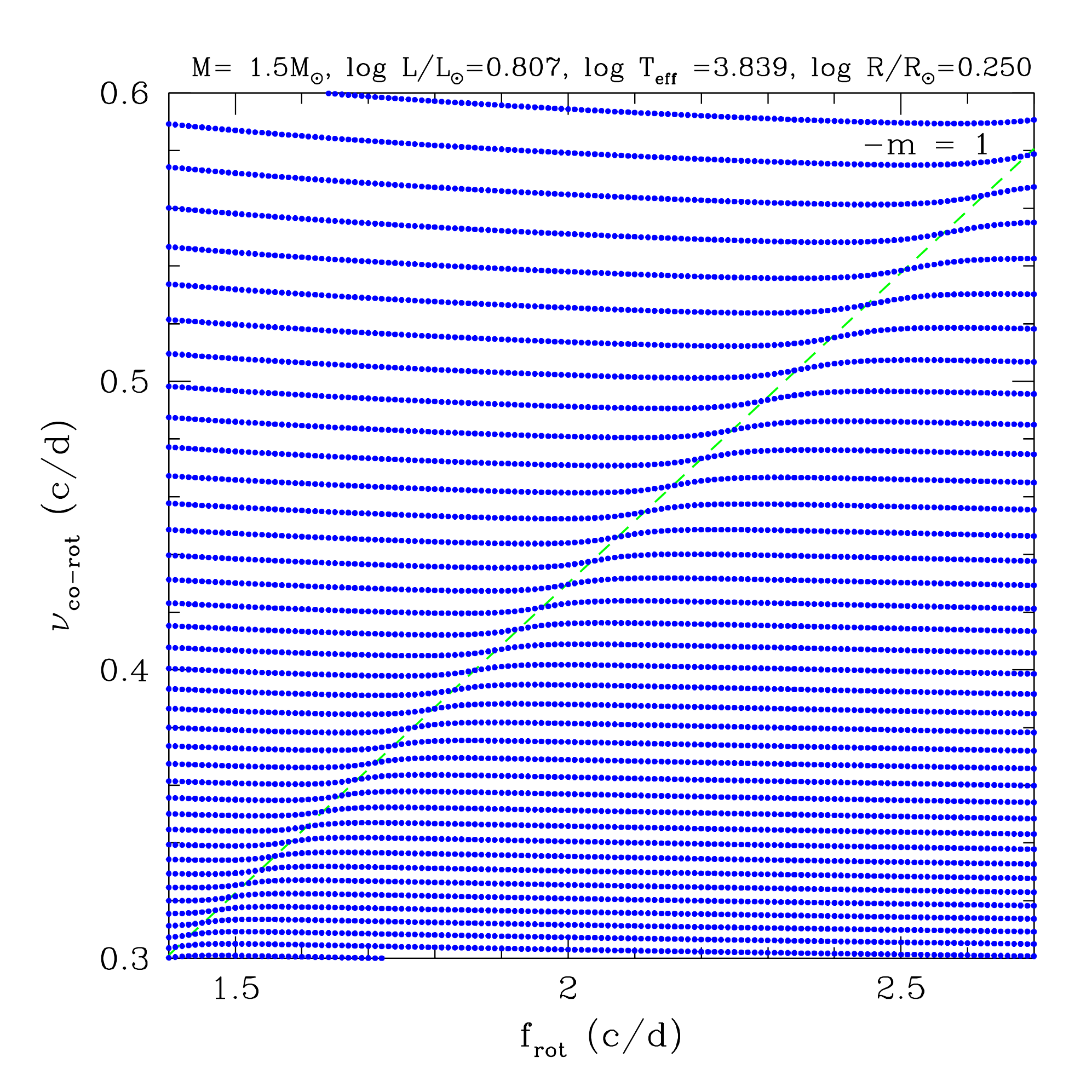}
    \caption{G-mode frequencies in the co-rotating frame versus rotation frequency of a $1.5\,M_\odot$ main-sequence model. A sequence of avoided crossings passes diagonally; i.e., with a constant spin parameter, $s\,(=2f_{\rm rot}/\nu_{\rm co-rot})$, of about 9.3. (Dashed line indicates the locus of $s=9.3$.)  The feature  is caused by the interaction with an inertial mode in the convective core.}
    \label{fig:cross}
\end{figure}

Fig.~\ref{fig:cross} shows frequencies (in the co-rotating frame) of prograde dipole g modes as a function of rotation frequency for the same $1.5$-$M_\odot$ model used in Fig.~\ref{fig:dPco_m1m2}, but with different rotation frequencies.
The diagonal feature is a sequence of avoided crossings between g modes in the radiative g-mode cavity and the fundamental inertial mode in the convective core that occur at a spin parameter of $\approx9.3$ (dashed line in Fig.\,\ref{fig:cross}). 
This is consistent with the dispersion relation [equation~(\ref{eq:disp})] indicating the spin parameter governs the property of the inertial mode  irrespective of the rotation frequency. 

The intrusion of the inertial mode frequency into the g-mode frequency spectrum causes a slight decrease in the spacings of g mode frequencies
around the avoided crossing at a fixed rotation frequency. 
The maximum mode density corresponds to a dip of $\Delta P_{\rm co-rot}$ at $\nu_{\rm co-rot}= 0.47$\,d$^{-1}$ seen in Fig.~\ref{fig:dPco_m1m2}.
This phenomenon is similar to the period-spacing dips of mixed modes in the red giants, in which coupling occurs between g modes in the core and p modes in the envelope \citep[e.g.,][]{mos12,chr12}.

\begin{figure}
\includegraphics[width=\columnwidth]{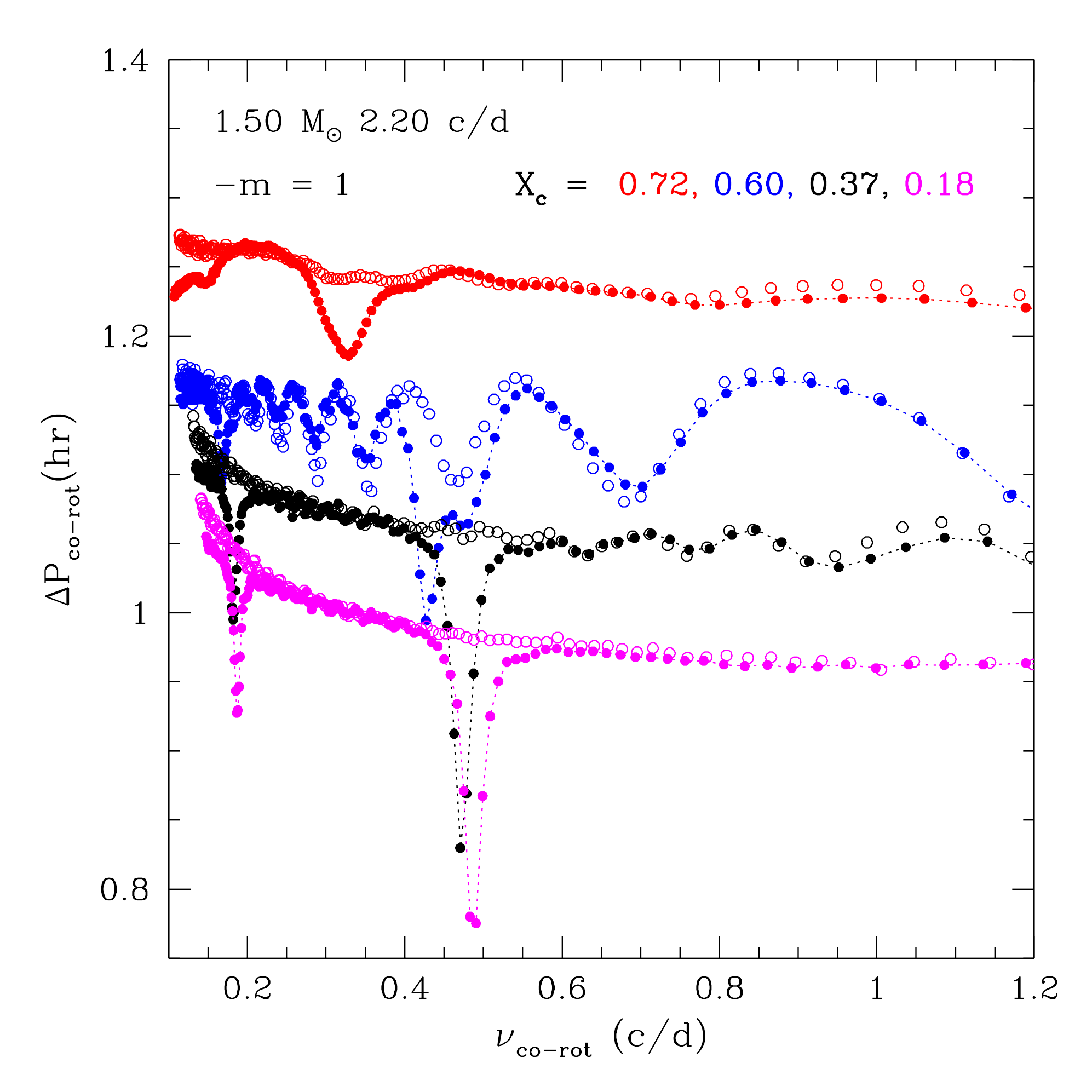}
\caption{Period spacings of dipole prograde g modes (in the co-rotating frame) at various evolutionary stages of a $1.50$-$M_\odot$ model with a rotation frequency of $2.20$~d$^{-1}$. Each evolutionary stage is designated by the central hydrogen abundance $X_{\rm c}$ and color coded as indicated. Open circles are results obtained using the TAR. The gradual increase in period spacings toward low-frequency limit (in evolved models in particular) is probably due to a poor resolution in our numerical calculations. }
\label{fig:evol}
\end{figure}
Fig.~\ref{fig:evol} shows period spacings of dipole prograde g modes in the co-rotating frame for a rotation frequency of $2.20$~d$^{-1}$ at selected evolutionary stages of models, where evolutionary stages are designated by the hydrogen mass fraction at the center, $X_{\rm c}$.
As in the non-rotating case, the mean value of $\Delta P_{\rm co-rot}$ decreases with evolution, and the modulation due to the steep gradient in hydrogen abundance is largest when $X_{\rm c}\sim0.6$.\footnote{The convective core mass grows until $X_{\rm c}$ decreases to $\sim0.5$. Although determining the physical convective core boundary is not  
very simple \citep{Gabriel14}, the numerical model stability and precision suffice for our qualitative evaluation of $\Delta P$ modulations during the stellar evolution. }
The modulation gets weaker  in later evolutionary stages, because diffusion rounds off a sharp edge of chemical composition distributions \citep{mig08,bou13}. 
Comparing variations in $\Delta P_{\rm co-rot}$ with the results obtained with TAR (open circles) makes the resonance dips obvious even for the $X_{\rm c}=0.6$ case with strong modulations due to the chemical composition gradient. 

The frequency at a resonance dip is smaller (i.e., the spin parameter is larger) in the ZAMS model than those in evolved models.
This is consistent with the result of \citet{oua20}, who obtained that the spin parameter of the dipole prograde inertial mode in a homogeneous density core is considerably higher than for a ZAMS model and evolved models.
They argued that a density gradient in the convective core makes the frequency of the inertial mode larger.
In evolved models the resonance frequencies only slightly increase as the evolution proceeds (i.e., as $X_{\rm c}$ decreases).

\begin{figure}
\includegraphics[width=\columnwidth]{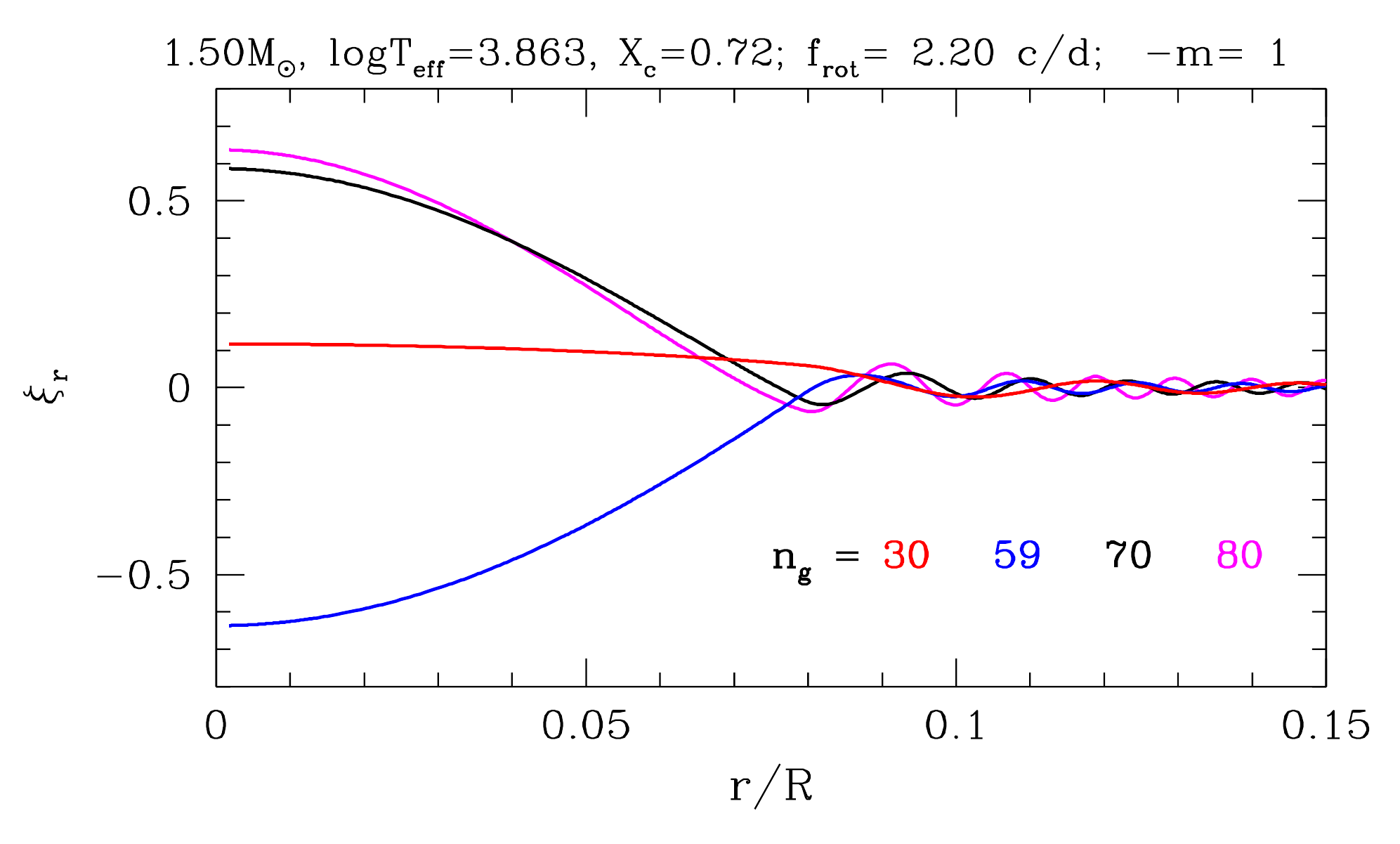}
\caption{Eigengunction $\xi_r$ versus fractional radius, $r/R$ in the ZAMS model of $1.5~M_\odot$ around a minimum of $\Delta P_{\rm co-rot}$, which corresponds to $n_{\rm g}=59$. Different line color indicates the number of radial nodes ($n_{\rm g}$) for each mode. The convective core boundary is located at $r/R=0.077$.}
\label{fig:eig_zams}
\end{figure}

In the ZAMS model, the resonance dip is broader and shallower compared with evolved models.
To understand the cause of the difference, we plot, in Fig.~\ref{fig:eig_zams}, $\xi_r$ as a function of $r/R$ for some modes in and around the dip at $\nu_{\rm co-rot}= 0.33$~d$^{-1}$. 
An apparent difference from the functions plotted in Fig~\ref{fig:eigf} is the longer  wavelength of a g mode in the radiative zone surrounding the convective core in the ZAMS model.
The long wavelength is caused by the smaller Brunt V\"ais\"al\"a frequency without a gradient in hydrogen abundance.  
A smaller phase difference at the core boundary between adjacent g modes due to the longer wavelength might be the reason for the broad dip in the ZAMS model. 

\begin{figure}
\includegraphics[width=\columnwidth]{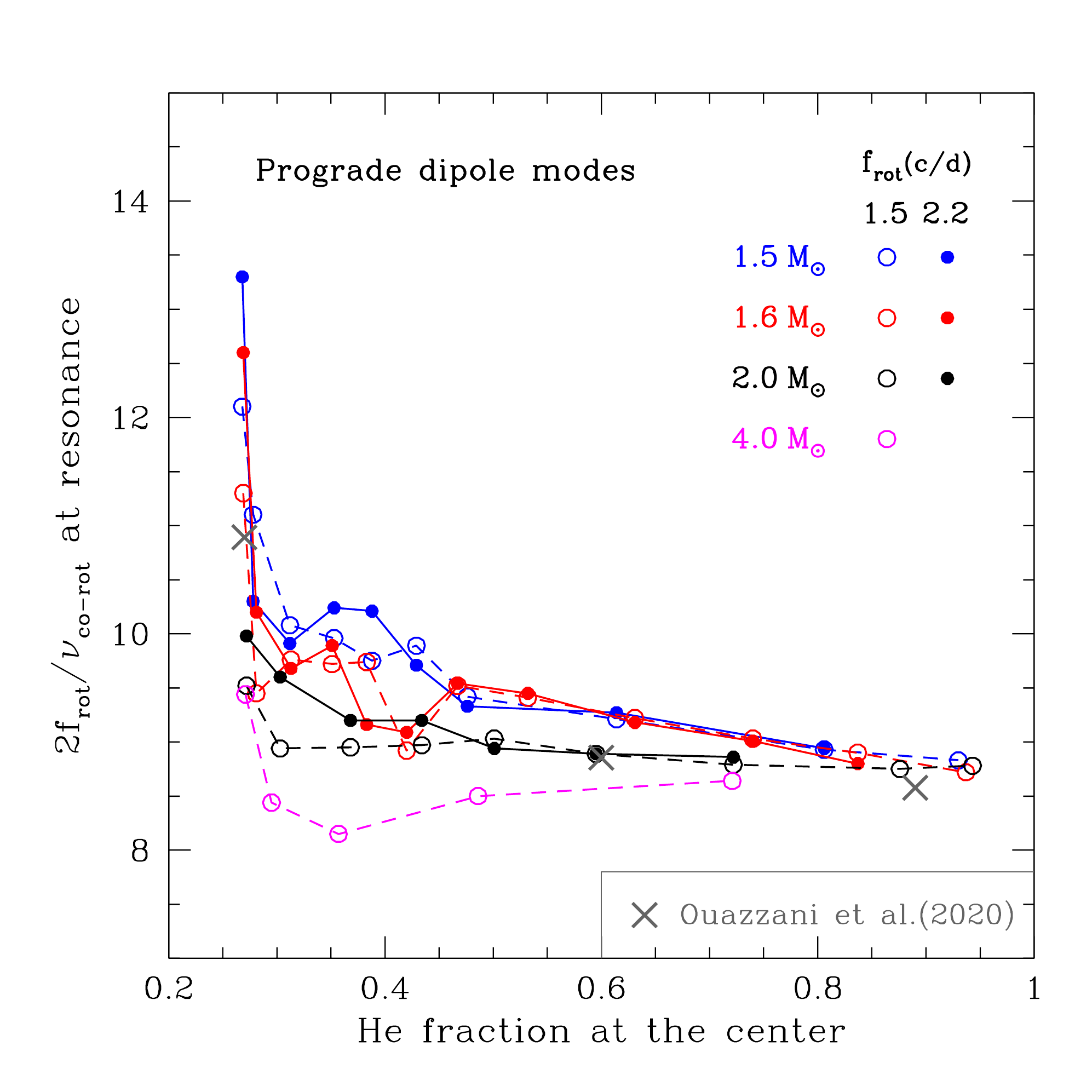}
\caption{Spin parameter at resonance with a 'fundamental' prograde dipole inertial mode in the convective core versus central helium mass fraction.  Stellar masses are color coded as indicated. Filled and open circles are for models with rotation frequencies of $1.5$~d$^{-1}$ and $2.2$~d$^{-1}$, respectively. Crosses show the results obtained by \citet{oua20} for prograde dipole modes in main-sequence models.
}
\label{fig:spinYc}
\end{figure}

Fig.~\ref{fig:spinYc} shows the spin parameter ($2f_{\rm rot}/\nu_{\rm co-rot}$) at the center of the dip caused by the resonance with the fundamental inertial mode of the convective core for various model masses at various stages of main-sequence evolution (denoted by the central helium abundance $Y_{\rm c}$).
For each case, the spin parameter at the resonance is highest (i.e.,$\nu_{\rm co-rot}$ is smallest) in the ZAMS model. It rapidly decreases with evolution and attains to nearly a constant value ($\sim\!\!8.8$) in the late stage of main-sequence evolution.
(Some wiggles at $Y_{\rm c}\sim 0.4$ are related to large modulations of $\Delta P_{\rm co-rot}$ due to chemical composition gradient; cf. Fig.~\ref{fig:evol}.) 
Gray crosses in Fig.~\ref{fig:spinYc} show  
results of \citet{oua20}, who  obtained, for dipole prograde modes, the spin parameters $s=10.9, 8.9$, and 8.6 for $X_{\rm c}=0.68$ (1.40\,$M_\odot$), 0.35 ($1.60~M_\odot$), and 0.06 ($1.86~M_\odot$), respectively, which are perfectly consistent with our results.

Generally, the spin parameter at the resonance is insensitive to the assumed rotation frequency in the late stage of evolution (see filled and open circles in Fig.\ref{fig:spinYc}), which was also found by \citet{oua20}.
The spin parameter at the resonance tends to be smaller for more massive models.
This can be understood as follows; a larger mass model has a larger convective core so that the radial wavelength of the 'fundamental' inertial mode and hence its oscillation frequency should be larger;\,i.e., the spin parameter at the resonance is lower for a higher mass.

\section{Effects of core overshooting}
Asteroseismic analyses for $\gamma$ Dor stars often favour models with convection overshooting \citep[e.g.,][]{mur16,sch16}.
To see the effects of overshooting in g-mode period spacings, we have included, in some models, diffusive overshooting from the convective core boundary, whose mixing at distance $z$ from the boundary is proportional to 
\begin{equation}
\exp[-2z/(h_{\rm os}H_{\rm p})] 
\label{eq:h_os}
\end{equation}
\citep{her00} as implemented in MESA, adopting $h_{\rm os} = 0.01$ (OS01) and $0.02$ (OS02).

The overshooting produces a radiative zone around the core boundary with little gradient of chemical composition, in which the Brunt-V\"ais\"al\"a frequency is much lower than in the surrounding layers with a steep gradient of chemical composition.
The presence of such a zone affects the propagation of g modes which couple with an inertial mode in the convective core.

\begin{figure*}
\centering 
\includegraphics[width=0.33\textwidth]{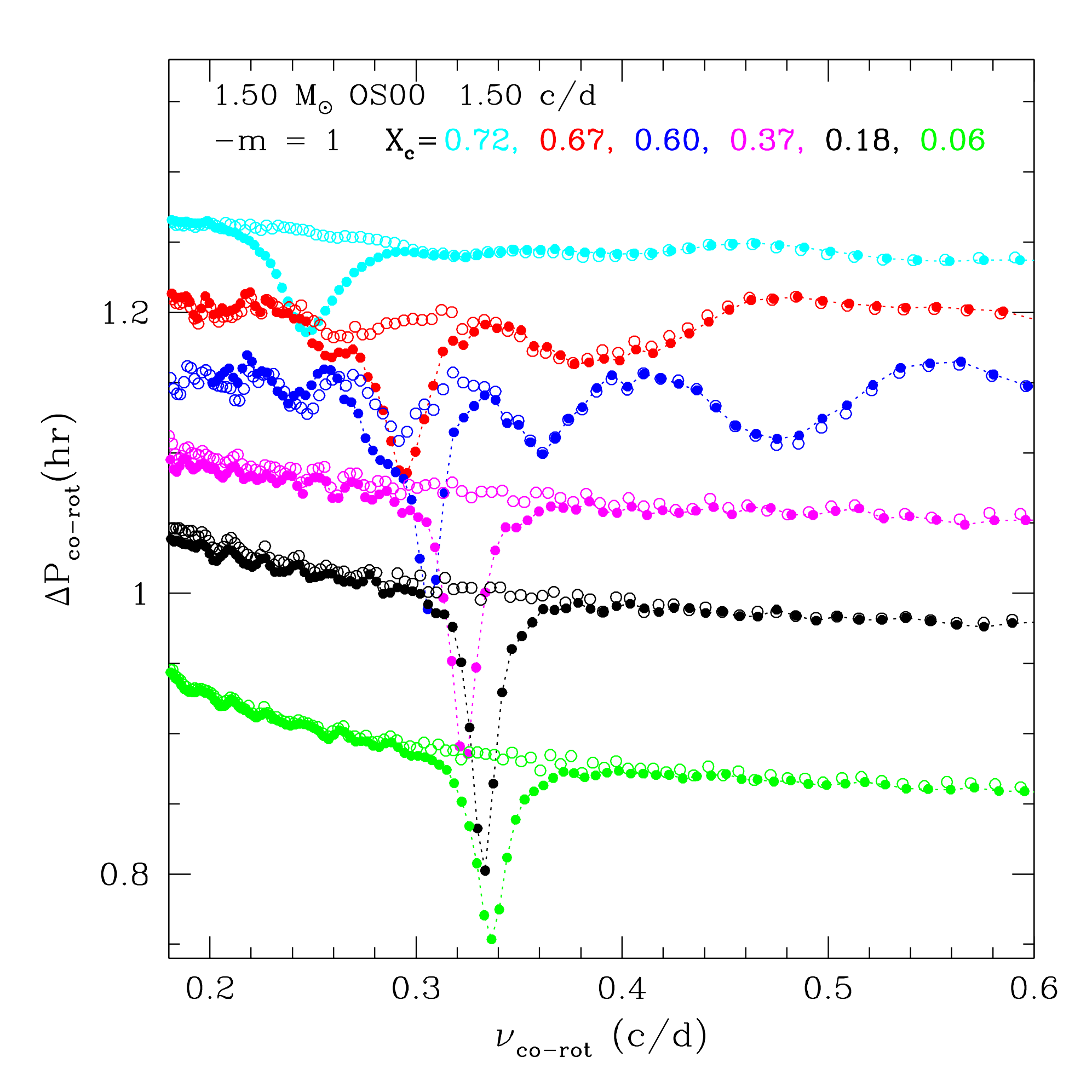}
\includegraphics[width=0.33\textwidth]{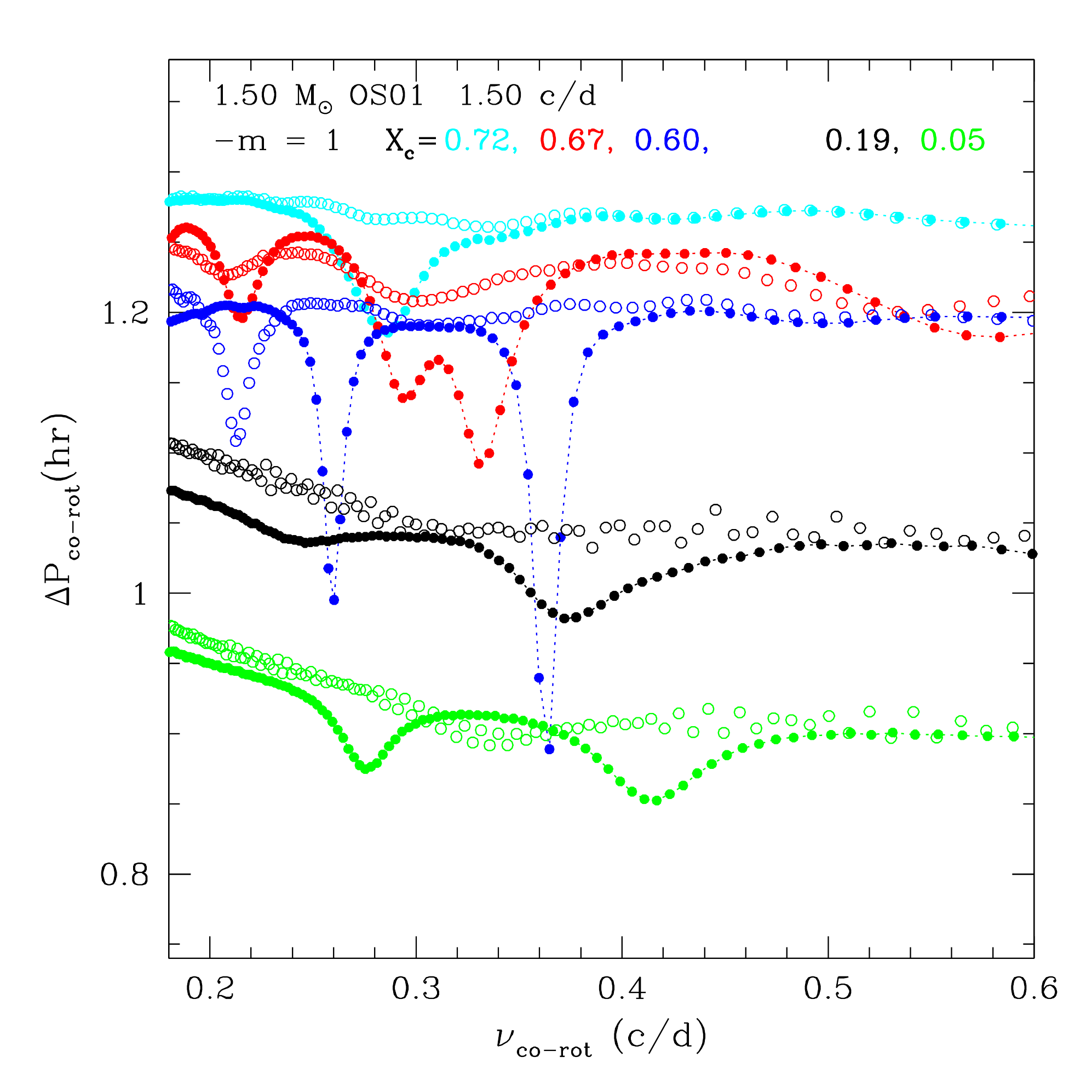}
\includegraphics[width=0.33\textwidth]{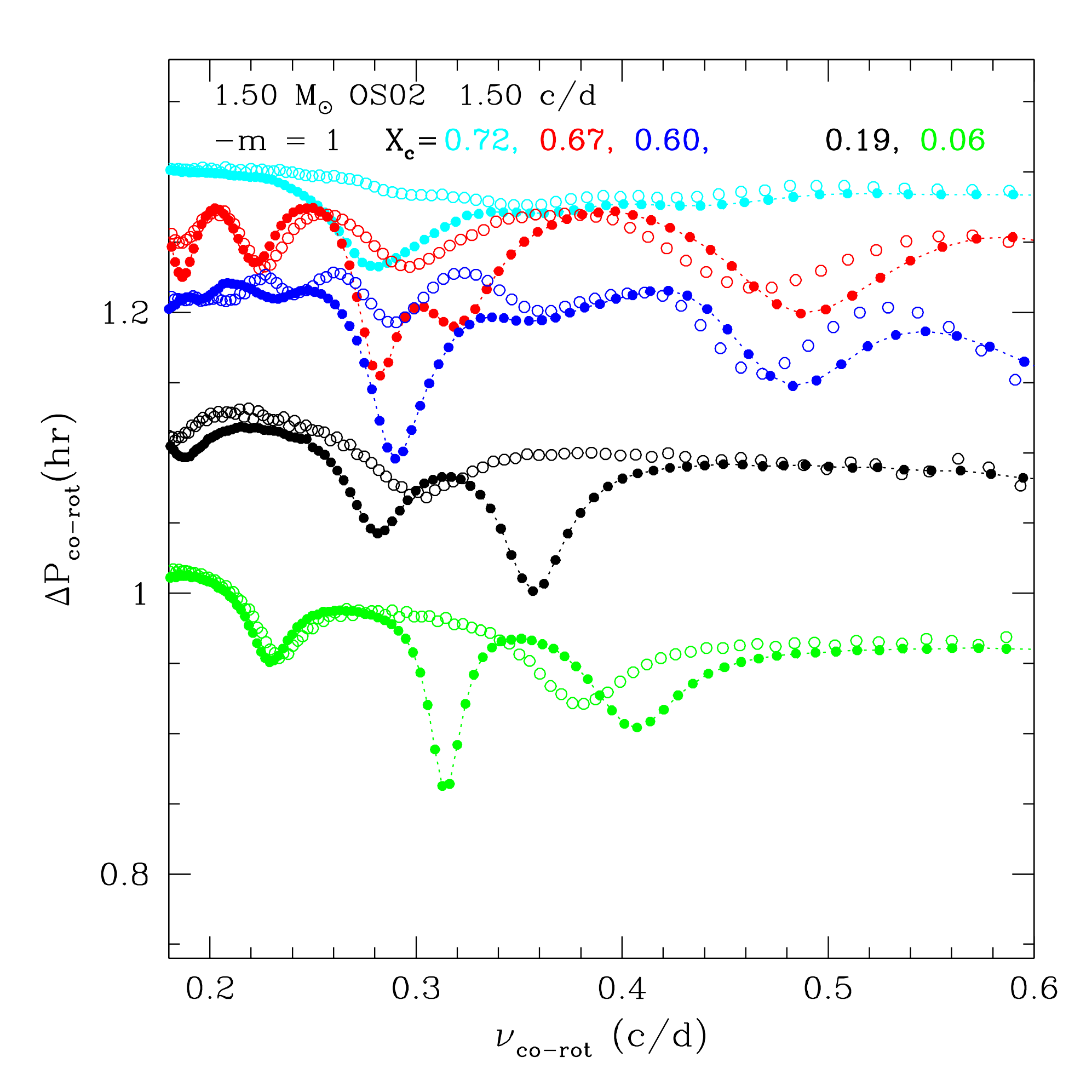}
\includegraphics[width=0.33\textwidth]{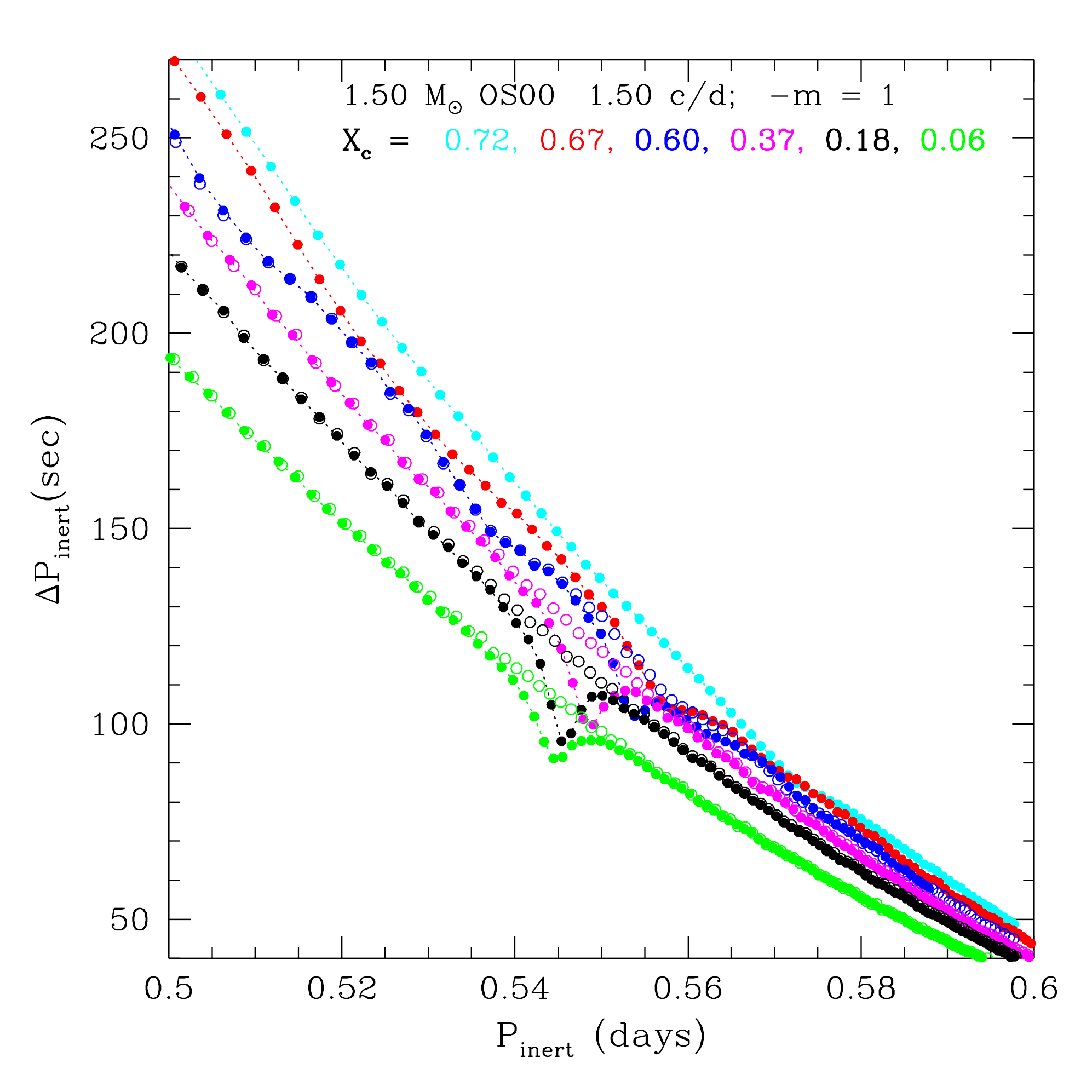}
\includegraphics[width=0.33\textwidth]{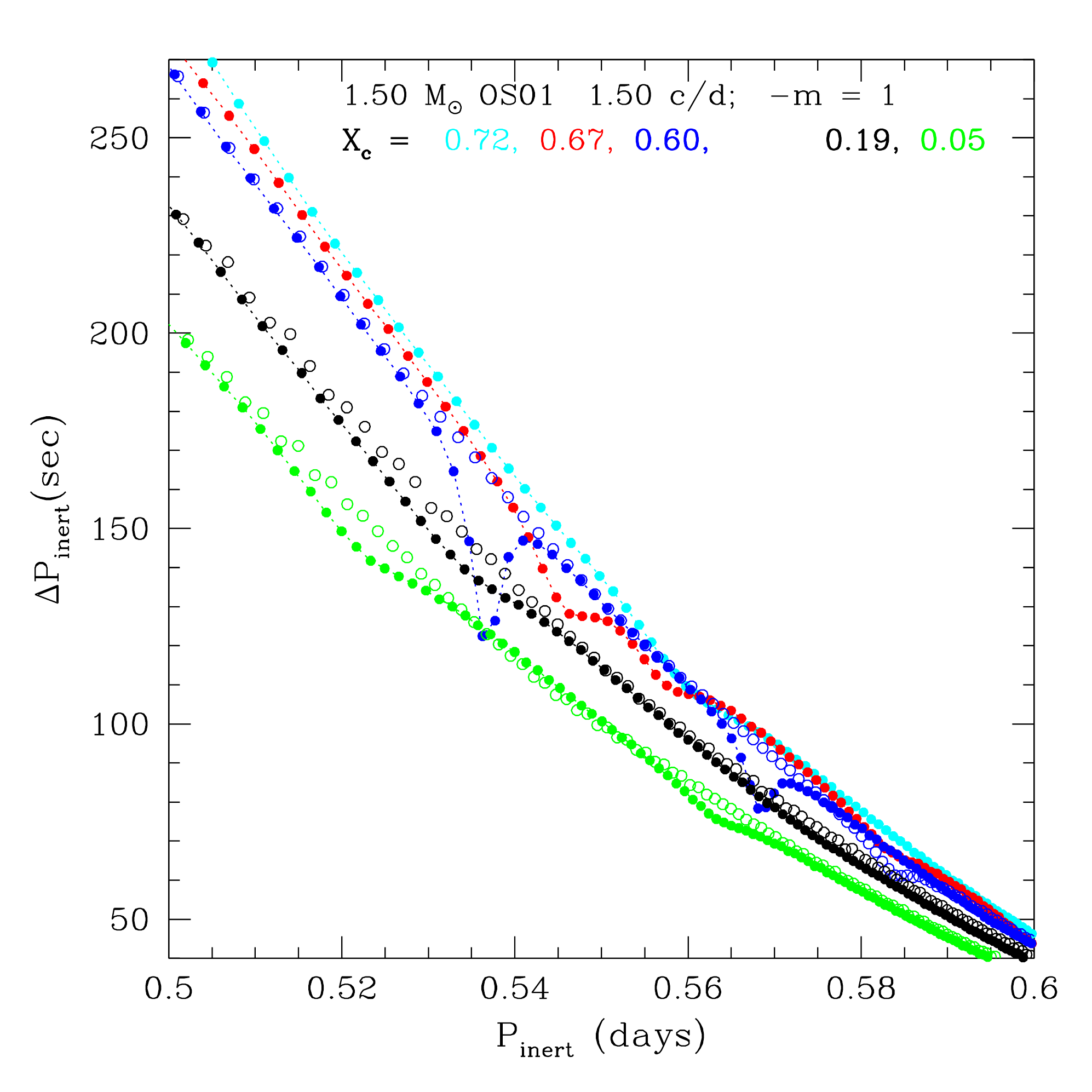}
\includegraphics[width=0.33\textwidth]{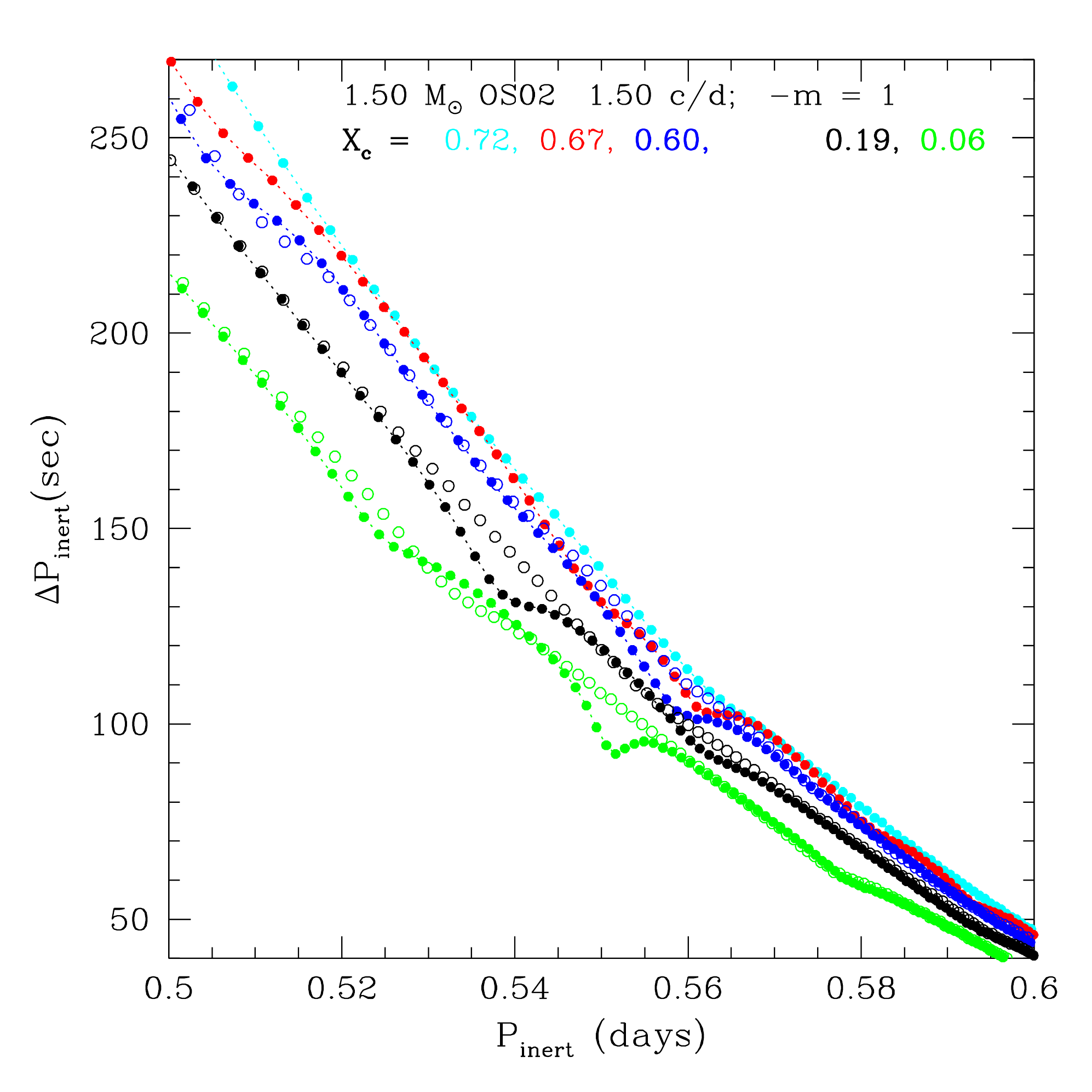}
\caption{Upper panels show period-spacing ($\Delta P_{\rm co-rot}$) of dipole prograde g modes versus frequency in the co-rotating frame ($\nu_{\rm co-rot}$), while lower panels show the same information as above but in the relation between period-spacing ($\Delta P_{\rm inert}$) and period ($P_{\rm inert}$) in the inertial frame. 
Plotted are prograde dipole g modes in $1.5~M_\odot$ models for a rotation frequency of 1.5~d$^{-1}$ at selected evolutionary stages (designated  with central hydrogen mass fraction $X_{\rm c}$). The leftmost panels are for the models without  overshooting, while the middle and rightmost panels are for models including overshooting of $h_{\rm os}=0.01$ (OS01) and $0.02$ (OS02), respectively. 
Open circles show results obtained using the TAR. 
For models with overshooting, evolution stages at $X_{\rm c}\approx 0.37$ are not shown for better visibility.
}
\label{fig:pdp_overshoot}
\end{figure*}

Fig.~\ref{fig:pdp_overshoot} shows period-spacing versus frequency or period of dipole prograde modes in $1.5~M_\odot$ models with and without overshooting at selected evolution stages (color coded) at a rotation frequency of 1.5\,d$^{-1}$. 
Upper panels show $\Delta P_{\rm co-rot}$ versus $\nu_{\rm co-rot}$ in the co-rotating frame, while lower panels show $\Delta P_{\rm inert}$ versus $P_{\rm inert}$ in the inertial reference frame. Open circles denote results obtained using the TAR. 
Because of the outstanding general trend of $\Delta P_{\rm inert}$ versus period $P_{\rm inert}$ in the inertial frame, dips are less prominent in the inertial frame (lower panels) compared with the appearances in the co-rotating frame (upper panels).

The upper-left panel for models without overshooting is very similar to Fig.~\ref{fig:evol} for a faster rotation frequency of 2.20~d$^{-1}$. Since a resonance coupling with an inertial mode of the convective core occurs at a spin parameter ($2f_{\rm rot}/\nu_{\rm co-rot}$), dips of $\Delta P_{\rm co-rot}$ occur at larger $\nu_{\rm co-rot}$ by about 50\% in Fig.~\ref{fig:evol}. (Frequencies of dips associated with the first-overtone inertial mode are too small in the models with $f_{\rm rot}=1.50$~d$^{-1}$  in Fig.~\ref{fig:pdp_overshoot}.) 

The $\nu_{\rm co-rot}$\,-\,$\Delta P_{\rm co-rot}$ relations for models with overshooting (middle and right panels of Fig.~\ref{fig:pdp_overshoot})  are more complex; additional dips tend to appear, while most dips are broader but sometimes sharp dips appear (e.g., $X_c = 0.60$ model with OS01; middle panels of Fig.~\ref{fig:pdp_overshoot}).
In addition, period spacings calculated using the TAR often do not agree with the results obtained by the expansion method even outside dips. Furthermore, for some models with overshooting,  dips appear even in the period spacings obtained with
the TAR (open circles), which never happens in models without overshooting.  

\begin{figure*}
\centering
\includegraphics[width=0.33\textwidth]{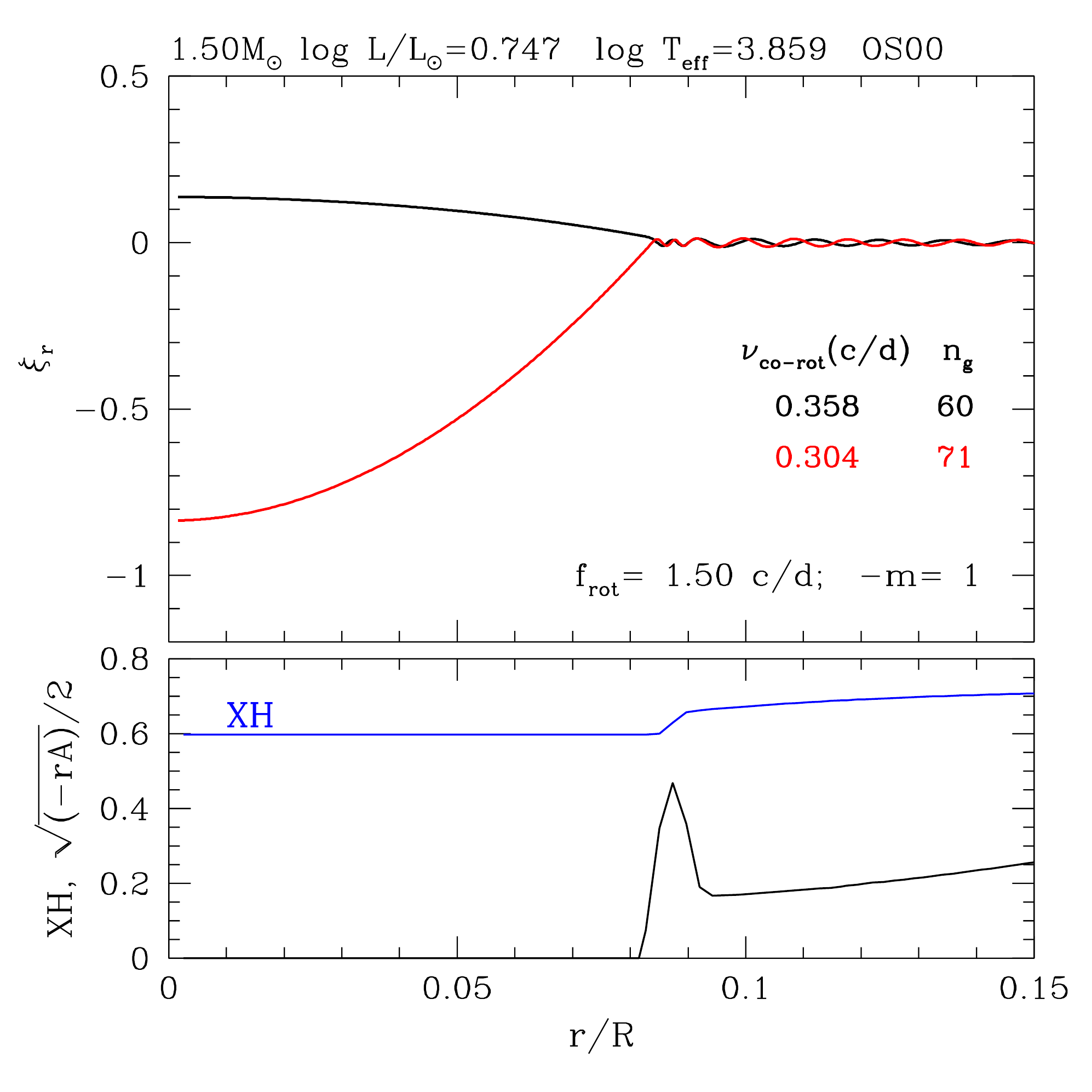}
\includegraphics[width=0.33\textwidth]{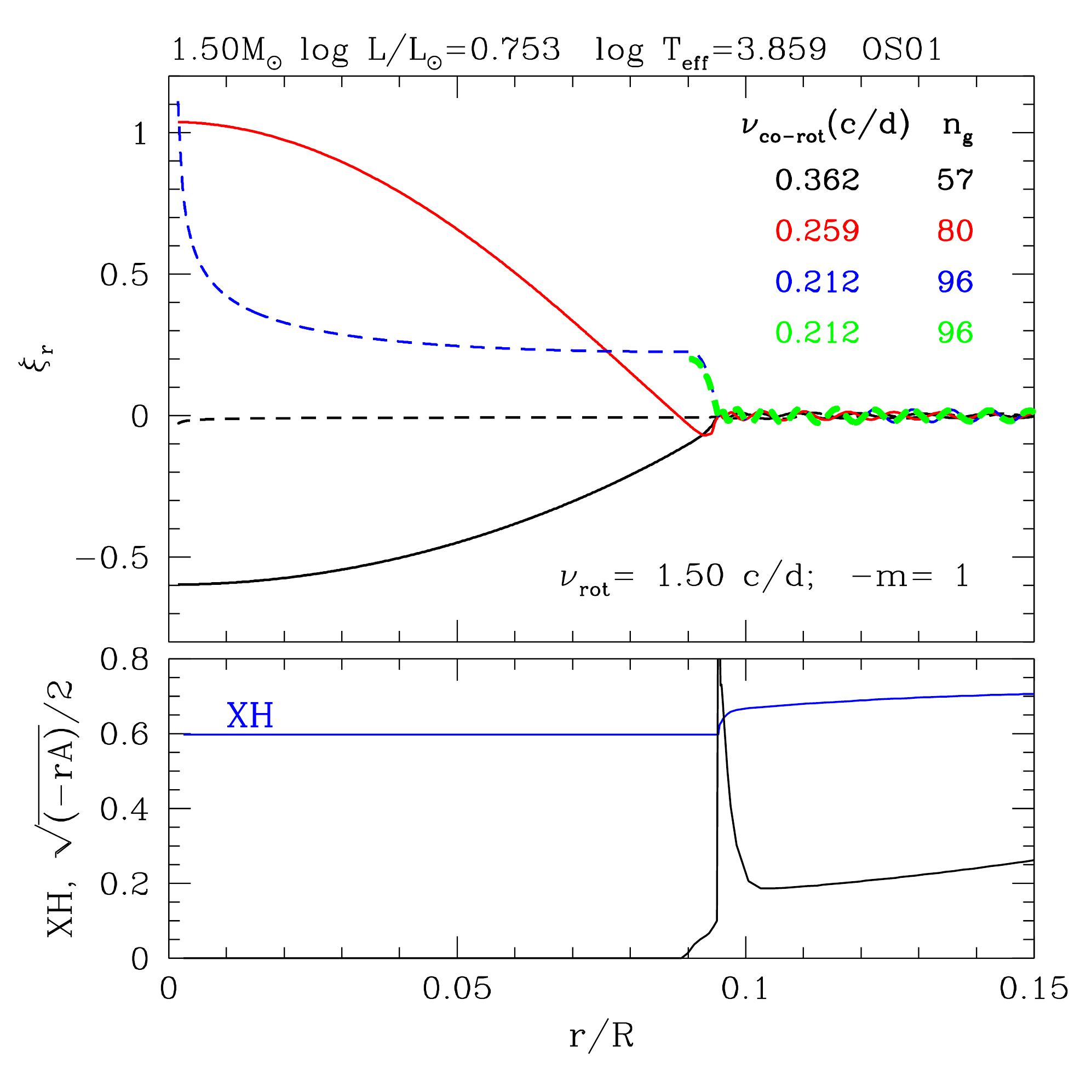}
\includegraphics[width=0.33\textwidth]{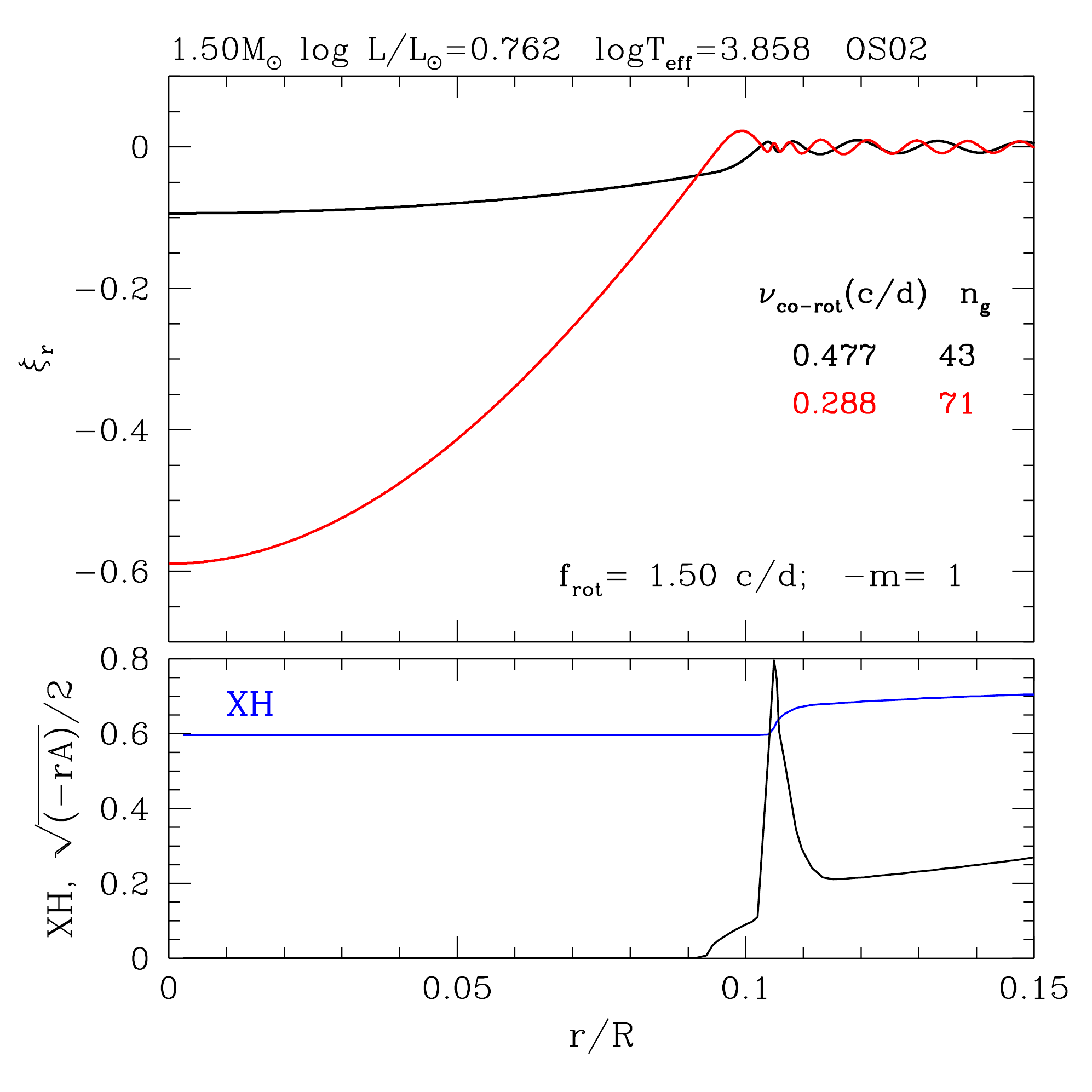}
\caption{{\bf Upper panels:} Radial displacements of selected prograde dipole modes as a function of fractional radius in $1.50$-$M_\odot$ main-sequence models without overshooting (left), with overshooting of OS01 (middle) and OS02 (right) at $X_{\rm c}=0.60$. The color of each line refers to the frequency $\nu_{\rm co-rot}$ and the number of radial nodes $n_{\rm g}$ written in the same color in the upper-right corner of each panel.
Dashed lines show radial displacements obtained using the TAR for the modes corresponding to colors. The TAR is not a good approximation near the center, which causes $|\xi_r|$ to steeply increase towards the center. Green dashed line shows the radial displacement of the same TAR mode shown by the blue dashed line but the former is obtained by imposing the inner boundary condition at the convective core boundary;i.e., removing the effect of the convective core.  {\bf Lower panels:} Hydrogen mass fraction (blue lines denoted as XH) and normalized Brunt-V\"ais\"al\"a frequency $-rA= N^2r/g$ with local gravity $g$. We note that $(-rA)^{1/2}/2 < 0.005$ for the frequencies of the modes shown in the upper panels.}
\label{fig:eigf_1p50cd_xc60}
\end{figure*}

\begin{figure*}
\centering
\includegraphics[width=0.33\textwidth]{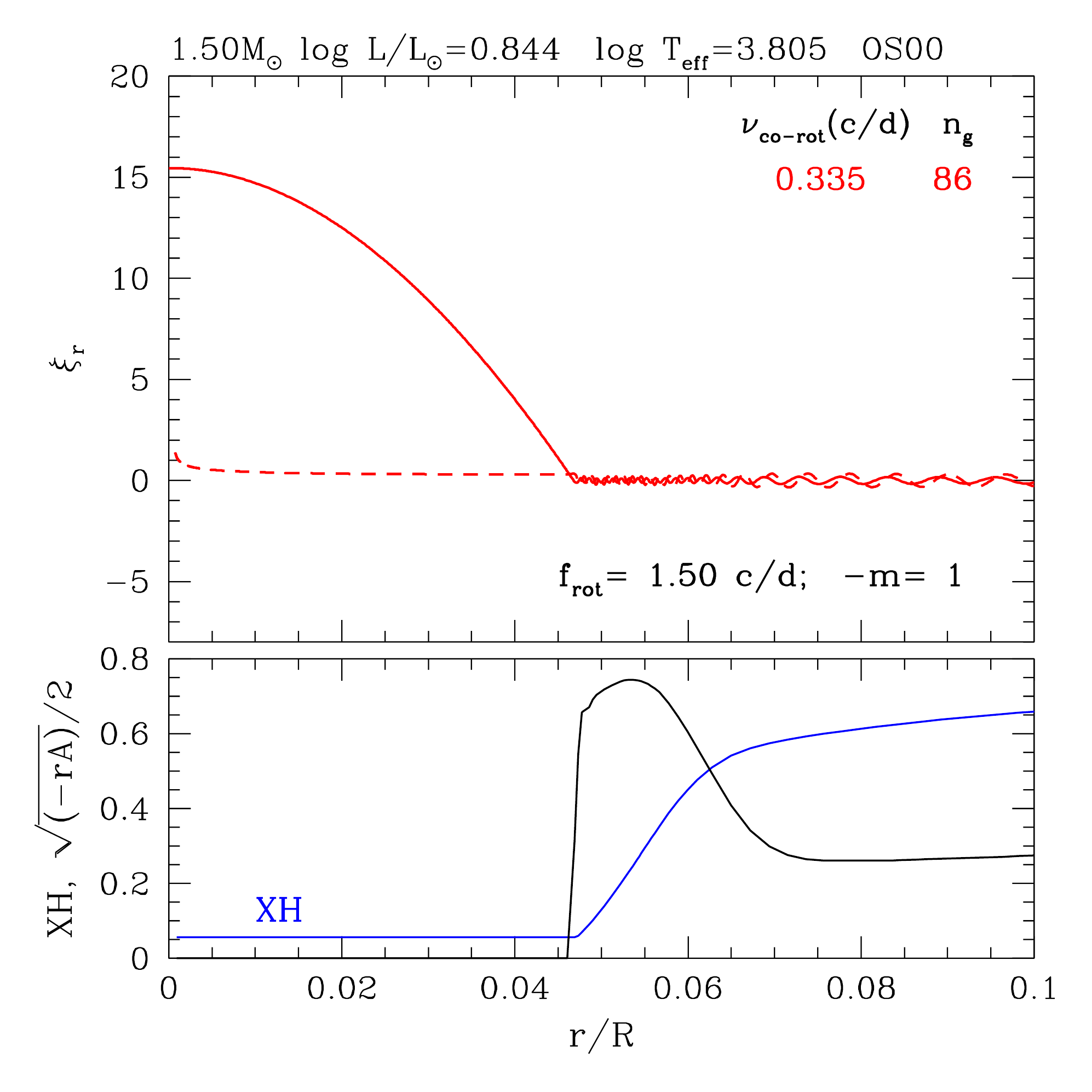}
\includegraphics[width=0.33\textwidth]{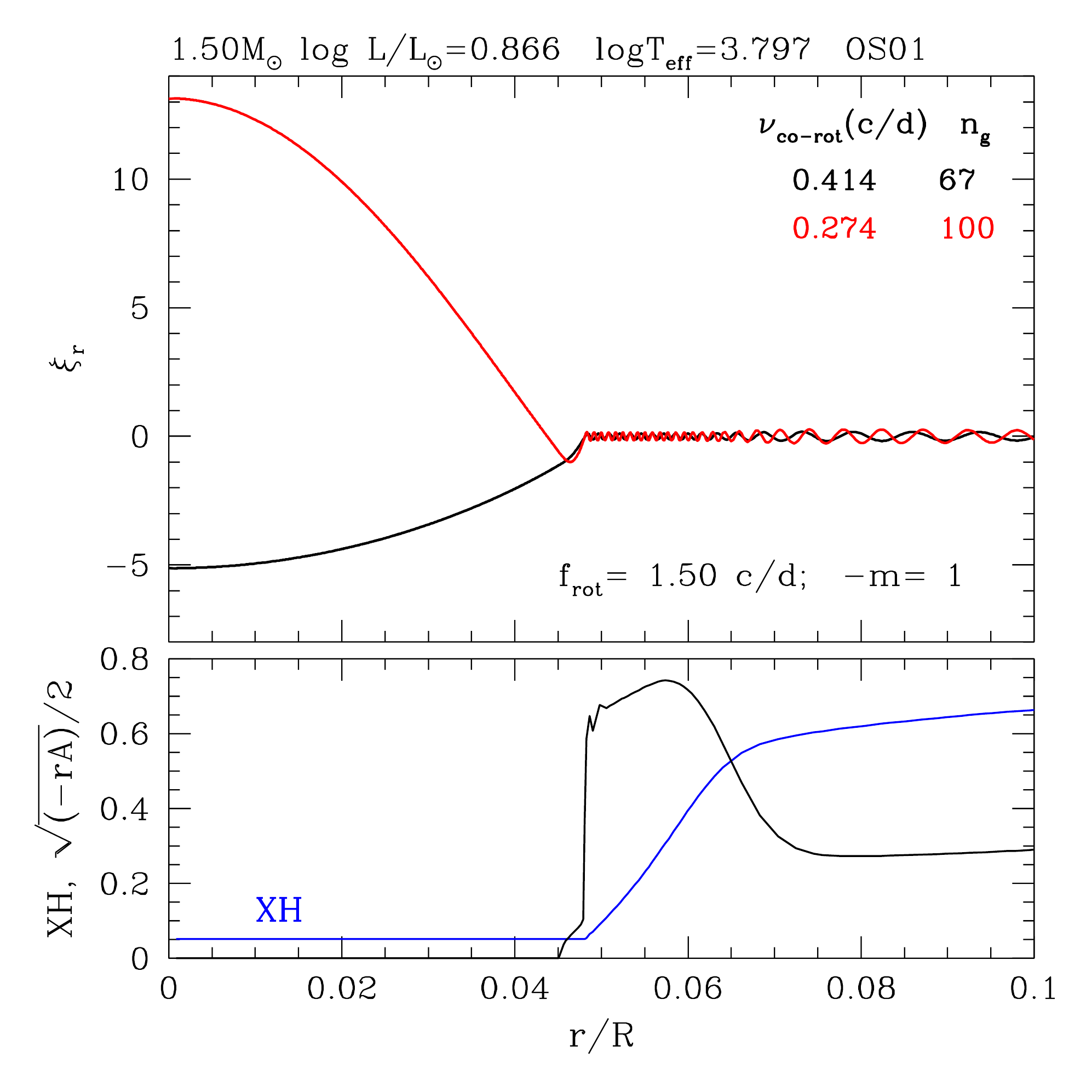}
\includegraphics[width=0.33\textwidth]{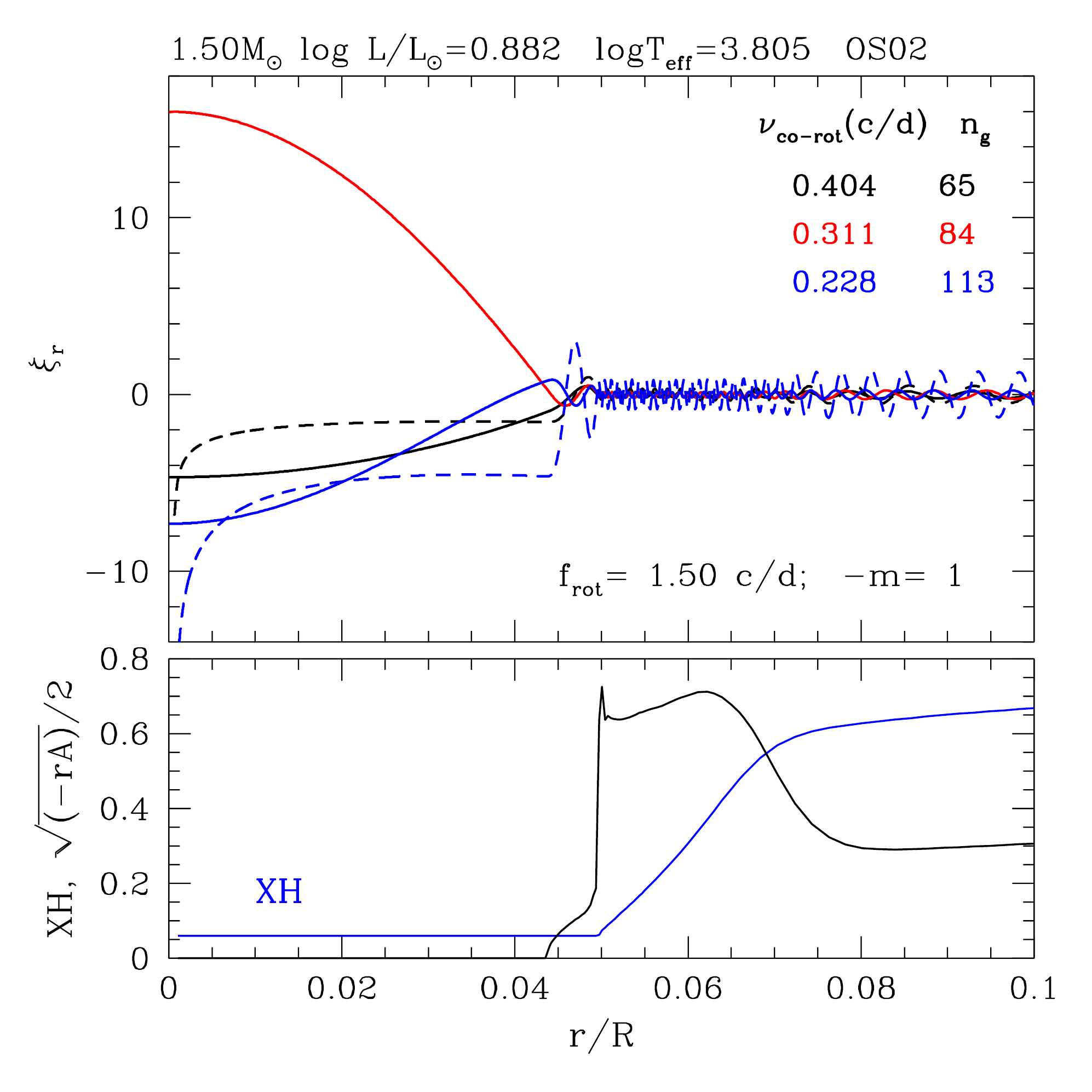}
\caption{Same as Fig.~\ref{fig:eigf_1p50cd_xc60} but for models at $X_{\rm c}\approx 0.05$.}
\label{fig:eigf_1p50cd_xc05}
\end{figure*}

In order to understand the cause of these effects of overshooting, we plot in Fig.~\ref{fig:eigf_1p50cd_xc60} (models at $X_{\rm c}\approx 0.60$) and in Fig.~\ref{fig:eigf_1p50cd_xc05} (models at $X_{\rm c}\approx 0.05$) the radial displacements of modes at dips in the $\nu_{\rm co-rot}$\,-\,$\Delta P_{\rm co-rot}$ relation as a function of fractional radius (upper panels).
The lower panels show the normalized Brunt-V\"ais\"al\"a frequency and mass-fraction of hydrogen profile.

The left upper panel of Fig.~\ref{fig:eigf_1p50cd_xc60} shows $\xi_r$ as a function of fractional radius for the two modes at $\nu_{\rm co-rot}= 0.304$, and  0.358~d$^{-1}$ in the model  at $X_{\rm c}\approx 0.60$ without overshooting. While these modes are located at dips of $\Delta P_{\rm co-rot}$ (see blue dots in Fig.~\ref{fig:pdp_overshoot} upper left panel), only the 0.304~d$^{-1}$ (spin parameter $s= 9.87$) mode is in resonance with an inertial mode having large amplitude in the convective core. The dip around the other mode, which have small amplitude in the convective core, should be caused by the steep gradient of the Brunt-V\"ais\"al\"a frequency \citep{mig08}. 

A similar phenomenon occurs also in the OS02 model at $X_{\rm c}=0.60$ (right upper panel of Fig.~\ref{fig:eigf_1p50cd_xc60}); i.e., among the two dips of $\Delta P_{\rm co-rot}$ at $\nu_{\rm co-rot}=0.288$ and $0.477$~d$^{-1}$, only the $0.288~{\rm d}^{-1}$ ($s=10.4$) mode resonantly couples with an inertial mode. The resonance dip in the OS02 model, however, is broader than the resonance dip in the OS00 model (cf. left and right panels of Fig.~\ref{fig:pdp_overshoot}). 
The broader resonance dip is related with the presence of a chemically homogeneous radiative zone produced by overshooting (OS02).
The Brunt-V\"ais\"al\"a frequency is small there so that radial wavelength of a g mode is much larger than that in the surrounding layers with inhomogeneous chemical composition ($\mu$-gradient zone with $\mu$ being the mean molecular weight).  
The center of a resonance dip of $\Delta P_{\rm co-rot}$ corresponds to a maximum coupling with an inertial mode, which occurs at an optimal spatial wave phase at the convective core boundary. 
If the wavelength in the radiative zone surrounding the convective core is much longer than the wavelength in the $\mu$-gradient zone, increasing or decreasing the number of nodes by one in the latter zone would cause only a slight change in the wave phase at the core boundary so that the coupling strength would change only slightly.     
For this reason, the large wavelength in the overshooting zone would cause a broader resonance dip than in the models without overshooting.

The effect of overshooting is more complex in the OS01 model at $X_{\rm}= 0.60$ (blue symbols in the upper middle panel of Fig.~\ref{fig:pdp_overshoot}), in which two deep dips of $\Delta P_{\rm co-rot}$ appear at $\nu_{\rm co-rot}= 0.362$ and 0.259~d$^{-1}$, and a dip appears at 0.212~d$^{-1}$ even in the period spacing calculated using the TAR, which never occurs in OS00 models.  
Radial displacements of these modes are shown in the middle panel of Fig.~\ref{fig:eigf_1p50cd_xc60}. 
The inertial mode bounded by the steep increase of the Brunt-V\"ais\"al\"a frequency at $r/R=0.095$ (black solid line) is responsible for the 0.362-d$^{-1}$ dip of $\Delta P_{\rm co-rot}$, while the inertial mode bounded by the convective core boundary at $r/R=0.090$ (red solid line) is responsible for the dip at 0.259\,d$^{-1}$.
The frequency of the former mode is slightly larger because the wavelength of the inertial mode is slightly larger than the latter mode for which the thickness of the narrow overshooting zone coincides with the half of the radial wavelength. 

Dashed lines stand for $\xi_r$ of the modes obtained using the TAR.
Under the TAR, no wave propagation in the convective core should occur, so that the amplitude should be constant there (except near the center where the amplitude increases steeply toward the center, indicating the TAR to be inappropriate there). 
For this reason, the radial displacement should be anti-node at the convective core boundary. 
Since the wavelength of a g mode is generally very small in the near-core region, the amplitude at an anti-node and hence the amplitude in the convective core tend to be very small as illustrated by the black dashed line in the middle panel of Fig.~\ref{fig:eigf_1p50cd_xc60}. 
However, the wavelength of a g mode in the zone produced by overshooting, is large and hence the amplitude at the anti-node can be appreciably large as in the case of $\nu_{\rm co-rot}= 0.212$~d$^{-1}$ (blue dashed line). The frequency is at the center of the dip in the $\Delta P_{\rm co-rot}$-$\nu_{\rm co-rot}$ relation obtained with the TAR (blue open circles in the upper middle panel of Fig.~\ref{fig:pdp_overshoot}). 
Thus, even the period spacing calculated using the TAR can have a dip if the stellar interior has a chemically homogeneous radiative zone surrounding the convective core, although the dip is nothing to do with the inertial mode.
As we mentioned above, the steep increase of the displacement under the TAR (dashed lines) near the center is caused by the breakdown of the TAR. However, the amplitude in the convective core does not affect at all the frequency of a mode under the TAR, which we have confirmed by re-calculating the 0.212-d$^{-1}$ mode, imposing the inner (regularity) boundary condition at the convective-core boundary; i.e., removing the effect of the convective core.
The result is shown by the green dashed line and the frequency written in green in the middle panel of Fig.\,\ref{fig:eigf_1p50cd_xc60}. 
This indicates that the dip under the TAR at 0.212~d$^{-1}$ should be caused by the peculiar behaviour of the eigenfunction in the narrow overshooting zone.

Fig.~\ref{fig:eigf_1p50cd_xc05} shows the same information as Fig.~\ref{fig:eigf_1p50cd_xc60} but for models at $X_{\rm c}\approx 0.05$ in the late stage of main-sequence evolution. In these models the convective core is geometrically more compact and surrounded by a thicker $\mu$-gradient region, compared to the models with $X_{\rm c} = 0.60$. In spite of the considerable differences between models at $X_{\rm c} \approx 0.05$ and 0.60, prograde dipole pulsations show similar phenomena associated with the chemically homogeneous zone produced by overshooting. Eigenfunctions at two dips at $\nu_{\rm co-rot}= 0.414$ and 0.274\,d$^{-1}$ in the OS01  model at $X_{\rm c}=0.05$ (Fig.~\ref{fig:eigf_1p50cd_xc05}) are similar to those for $\nu_{\rm co-rot}= 0.362$ and 0.259~d$^{-1}$ modes, respectively, at $X_{\rm c}=0.60$ (Fig.\,\ref{fig:eigf_1p50cd_xc60}).

We have found above that the core overshooting affects significantly (and complicatedly) the $\nu_{\rm co-rot}$\,-\,$\Delta P_{\rm co-rot}$ and $P_{\rm inert}$\,-\,$\Delta P_{\rm inert}$ relations. This comes from the fact that the core overshooting produces a chemically (nearly) homogeneous radiative zone where g modes propagate with a wavelength much longer than in the surrounding $\mu$-gradient zone.
While we assume in this paper that convective overshoot at the core boundary leaves a radiative zone, such a zone may be convective (at least partially) if the thermal time there is sufficiently long, as reviewed by \citet{zahn02}, who called it  `convective penetration'. 
Although matter mixing and hence stellar evolution would be little affected by whether such a narrow zone is convective or radiative, it affects significantly  the period spacings of g modes and resonance couplings with inertial modes. If the convective-core overshoot occurs only as {\it penetration}, it would slightly increase the size of convective core without significantly affecting the structure of g-mode cavity surrounding the adiabatic region.
In this case, period spacings of g modes might increase slightly, caused by a slight increase in the buoyancy radius, $P_0$ ($=1/\nu_0$ in eq.\ref{eq:nu0}). 

Probably, we should consider layers surrounding a convective core to consist of a inner 'convective penetration' zone and a radiative zone of overshooting, although the fraction of each zone is not known.
In this paper we specify the extent of overshooting by $h_{\rm os}$  (eq.~\ref{eq:h_os}), which  should be regarded to specify the radiative overshooting zone above the possible penetration zone.  For this reason, even if a comparison of $P_{\rm inert}$\,-\,$\Delta P_{\rm inert}$ relation with observations happen to prefer a very small $h_{\rm os}$, it does not necessarily mean that overshooting is negligible, because a substantial penetration zone may be present below the thin radiative zone.

\section{Comparison with observation}
In agreement with the results of \citet{oua20}, our calculations using the expansion method predict a dip in period spacings of prograde dipole g modes at a period corresponding to a spin parameter $(2f_{\rm rot}/\nu_{\rm co-rot})\sim\!\!9$ (Fig.~\ref{fig:spinYc}). The dip is caused by the resonance coupling with the fundamental inertial mode of the convective core.
The exact spin parameter at a dip depends on the stellar parameters, the evolution stage, and the assumption of overshooting.
If we identify such a dip in observed g mode period spacings, we can determine the rotation frequency of the convective core
by finding a model which reproduces the period and depth of the dip. 
However, no clear resonance dips in g-mode pulsators (i.e., $\gamma$ Dor, SPB stars) had been found previously, except for an indication in the $\gamma$ Dor star KIC~5608334 \citep{sai18b}.
A spin parameter of about 9 corresponds to a period of $0.82P_{\rm rot}$ in the inertial frame, at which we expect to find a resonance dip if the observed g-mode period range extends beyond the period. $\gamma$ Dor stars in the sample analysed by \citet{LiG20} indicate that the observed maximum spin parameters of (prograde dipole) g modes can be larger than 9 if the rotation periods are shorter than one day. In other words, it is possible to find a resonance dip in some of the stars rotating faster than $\sim 1$~d$^{-1}$.

We have searched $P$\,-\,$\Delta P$ (period versus period-spacing) relations obtained from Kepler data by \citet{vanr16,LiG19,LiG20} for possible resonance dips, and found many $\gamma$~Dor stars having possible resonance dips. 
From these $\gamma$ Dor stars, we have selected 16 stars (Table~\ref{tab:sum}) which have relatively clean $\Delta P$ dips likely caused by the resonance without much $\Delta P$ modulations due to chemical composition gradients.\footnote{We have also searched   $P$\,-\,$\Delta P$ relations of SPB stars obtained by \citet{pap17}. However, we found no convincing cases.}

For each of the selected 16 $\gamma$ Dor stars, we try to find a model (with an assumed extent of overshooting) consistent with the $P$\,-\,$\Delta P$ sequence with a resonance dip.
We adopt the rotation frequency, $f_{\rm rot}$, obtained by \citet{vanr16,LiG19,LiG20} from the observational $P$\,-\,$\Delta P$ relations, as the rotation frequency throughout the layers exterior to the boundary of the convective core.
In other words, we assume that no strong differential rotations are present in the radiative g-mode cavity. This may be justified by the previous studies on $\gamma$ Dor stars which found the differential rotation from the near-core to the surface to be weak \citep{kur14,sai15,sch15,mur16,vanr18,LiG20}.

The theoretical $P_{\rm inert}$\,-\,$\Delta P_{\rm inert}$ relation for a given value of $f_{\rm rot}$ shifts downward  (i.e., $\Delta P_{\rm inert}$ decreases) with decreasing mass and/or advancing evolution stage.
For each $\gamma$ Dor star, we guess the mass of the star from the global parameters ($T_{\rm eff}$,$L/L_\odot$) given in \citet{mur19} (or in \citet{vanr15apjs}), and calculate $P_{\rm inert}$\,-\,$\Delta P_{\rm inert}$ relations of prograde dipole g modes for the $f_{\rm rot}$ obtained by \citet{LiG19,LiG20}, using the TAR at various evolutionary stages, to find the relation closest to the observed one by eye.

Then, using the expansion method, we calculate the $P_{\rm inert}$\,-\,$\Delta P_{\rm inert}$ relation of the model with the same value of $f_{\rm rot}$. 
Thus obtained $P_{\rm inert}$\,-\,$\Delta P_{\rm inert}$ relation has a resonance dip, but its position does not necessarily agree with the observational position (i.e., period).
Then, we assume a differential rotation between the convective core and the surrounding g-mode cavity.

To calculate g-mode periods by the expansion method for a differentially rotating star, we have adopted the method of \citet{lee88}, in which the rotation profile is expressed as a function of fractional radius $x$, 
\begin{equation}
\Omega(x) = \Omega(1)\left(1+ {b-1 \over 1+\exp[200(x-x_{\rm cc})]}\right),
\label{eq:difrot}
\end{equation}
where $x_{\rm cc}$ is the fractional radius at the boundary of the convective core, and $b$ is the parameter which determines the rotation rate in the convective core relative to the surface. 
The factor $200$ in the denominator is arbitrarily chosen to have a rapid transition around the core boundary.
The rotation rate changes steeply at the boundary of the convective core from $\approx b\Omega(1) = 2\pi f_{\rm rot}({\rm cc})$ to $\approx\Omega(1)= 2\pi f_{\rm rot}$, where $f_{\rm rot}{\rm (cc)}$ and $f_{\rm rot}$ are cyclic rotation frequencies in the convective core and in the radiative envelope including g-mode cavity, respectively. 
The value of parameter $b$ is chosen to fit the resonance $\Delta P_{\rm inert}$ dip with the observed one.

For each star, adopting a standard initial chemical composition of $(X,Z)=(0.72,0.014)$, we tried to fit models with three assumptions of core overshooting; i.e., $h_{\rm os} = 0.0$\,(OS00), 0.01\,(OS01), and 0.02\,(OS02).
In addition, in order to see the effects of different initial chemical composition, we have also performed the same analysis employing OS00 models with $(X,Z)=(0.724,0.010)$.
The results of the fittings for the 16 $\gamma$ Dor stars are summarized in Table\,\ref{tab:sum}. For some stars lines for models with $h_{\rm os} >0$ are missing, for which we could not find good models.

\subsection{Examples of model fittings}\label{sec:fit}
In this subsection we discuss four examples of fitting theoretical predictions (for the standard initial composition) with observational $P$\,-\,$\Delta P$ relations and dips.
Fittings with other 12 $\gamma$ Dor stars are shown in Appendix\,\ref{sec:otherfits}.

\begin{figure*}
\includegraphics[width=0.33\textwidth]{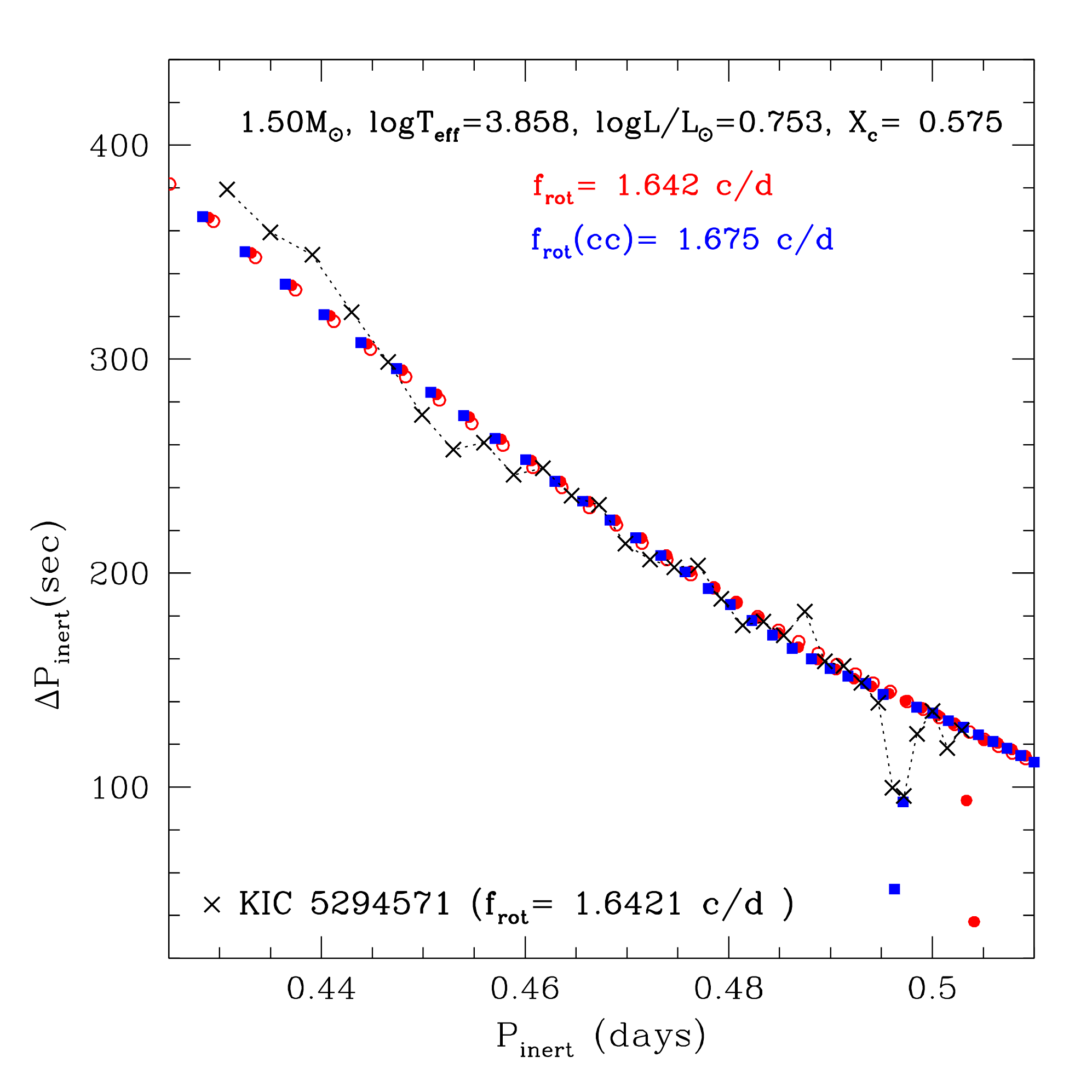}
\includegraphics[width=0.33\textwidth]{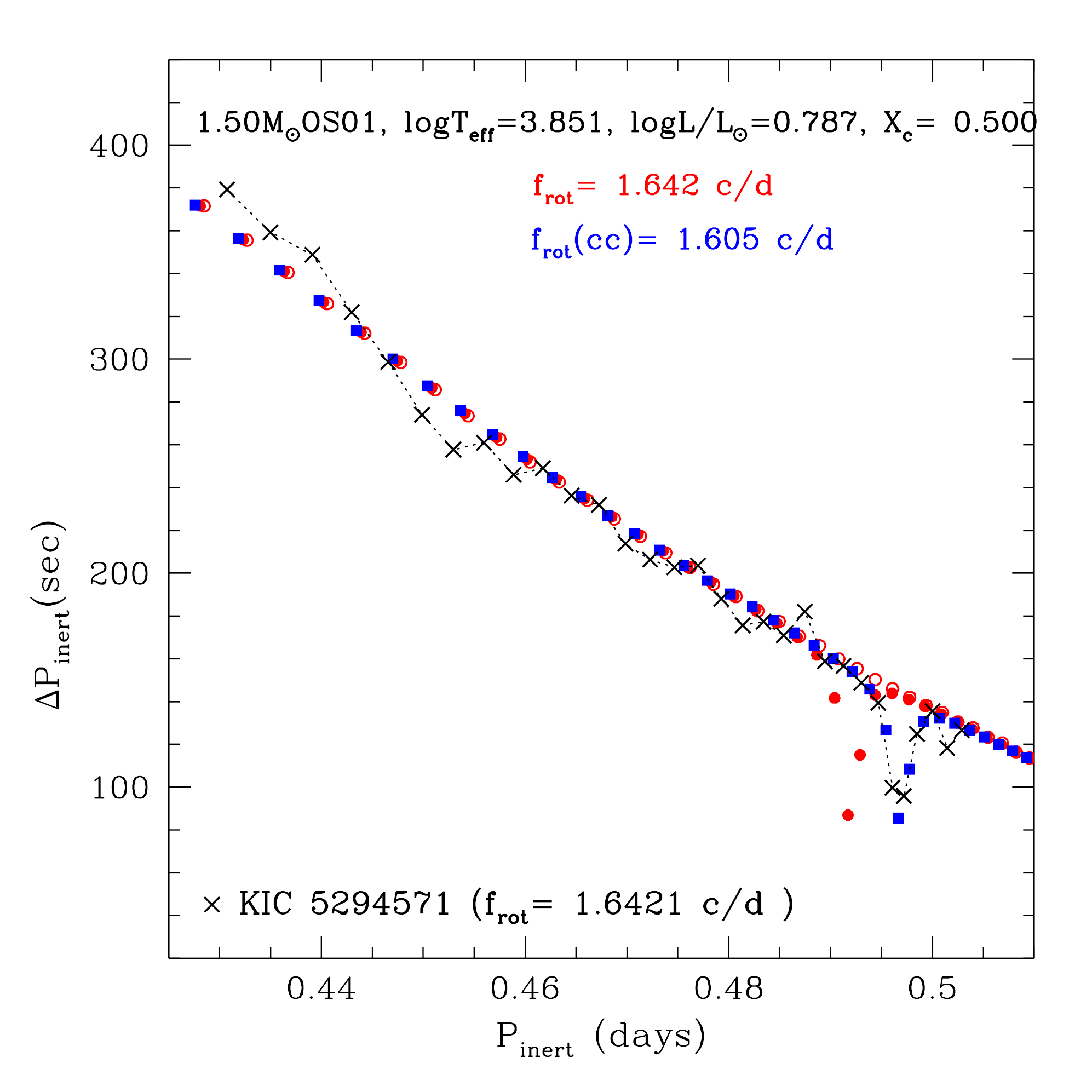}
\includegraphics[width=0.33\textwidth]{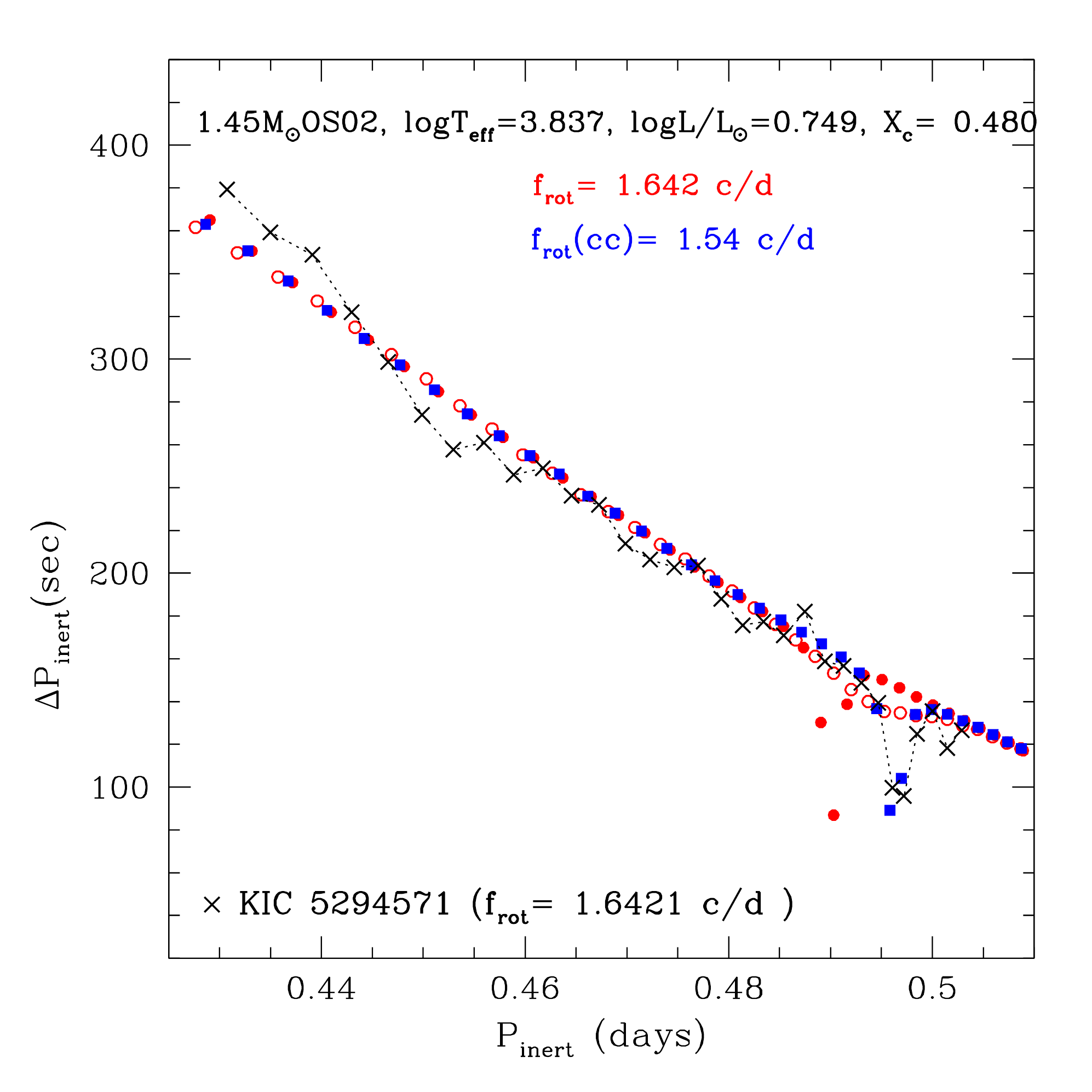}
\caption{Period-spacing ($\Delta P$) versus period ($P$) of KIC~5294571 in the inertial frame \citep[crosses][]{LiG19} compared with models with no overshooting ($h_{\rm os}=0$; left panel), with $h_{\rm os}=0.01$ (OS01; middle panel), and 0.02 (OS02; right panel). Circles and squares are theoretical $(P,\Delta P)$ obtained for the model whose parameters are shown in the upper part of each panel. Red symbols are results obtained assuming uniform rotation with a rotation frequency of 1.642\,d$^{-1}$ derived by \citet{LiG19}. Open circles are $(P,\Delta P)$ obtained using the TAR, while filled ones are those calculated by the expansion method. Blue squares show $(P,\Delta P)$ calculated by the expansion method for the same models but  with a convective core rotating at a different rate of $f_{\rm rot}{\rm (cc)} = 1.675$ (left panel), 1.605 (middle panel) and 1.54~d$^{-1}$ (right panel). }
\label{fig:k529}
\end{figure*}

\subsubsection{KIC 5294571 (Fig.\,\ref{fig:k529})}
The dipole prograde g-mode $P$\,-\,$\Delta P$ relation of KIC~5294571 obtained by \citet{LiG19} from the Kepler light curve is shown by crosses in Fig.~\ref{fig:k529}.  \citet{LiG19} obtained a rotation frequency of $1.6421\pm0.0009$\,d$^{-1}$ from the g- and r-mode $P-\Delta P$ relations. 
The three panels show theoretical $P_{\rm inert}$\,-\,$\Delta P_{\rm inert}$ relations of dipole prograde g modes for models with $h_{\rm os}=0.0$ (no overshooting;left panel), $h_{\rm os}=0.01$ (OS01;middle panel) and $=0.02$ (OS02; right panel).
KIC~5294571 has a clear dip of $\Delta P$  at a period (in the inertial frame) of 0.495~days. The dip is likely caused by the resonance coupling with an inertial mode in the convective core. However, model predictions for the uniform rotation at 1.642~d$^{-1}$ (red filled circles) disagree with the dip of KIC~5294571; the degree of the discrepancies depends on the assumptions of overshooting.  

In order to fit the period at the resonance dip of KIC~5294571 with the theoretical prediction of each model with $h_{\rm os}$, we have searched for a best value of  $f_{\rm rot}{\rm (cc)}$ (rotation frequency in the convective core), which is  different from the $f_{\rm rot}$ of KIC~5294571 attributed to the g-mode cavity.
The best fit result is shown by blue squares in each panel and the adopted $f_{\rm rot}{\rm (cc)}$ is written in blue in Fig.~\ref{fig:k529}.
By changing $f_{\rm rot}{\rm (cc)}$ the period at the dip shifts, while $\Delta P$s in other period range change little because they are mostly determined in the g mode cavity. 
The best fit $f_{\rm rot}{\rm (cc)}$s are  1.675, 1.605, and 1.54~d$^{-1}$ for models with $h_{\rm os} = 0.0$ (left panel), 0.01 (middle panel) and 0.02 (right panel), respectively.
The depth of the dip observed in KIC~5294571 agrees better with the  models including overshooting, while the convective core rotates slightly slower than the radiative layers.

\begin{figure*}
\includegraphics[width=0.33\textwidth]{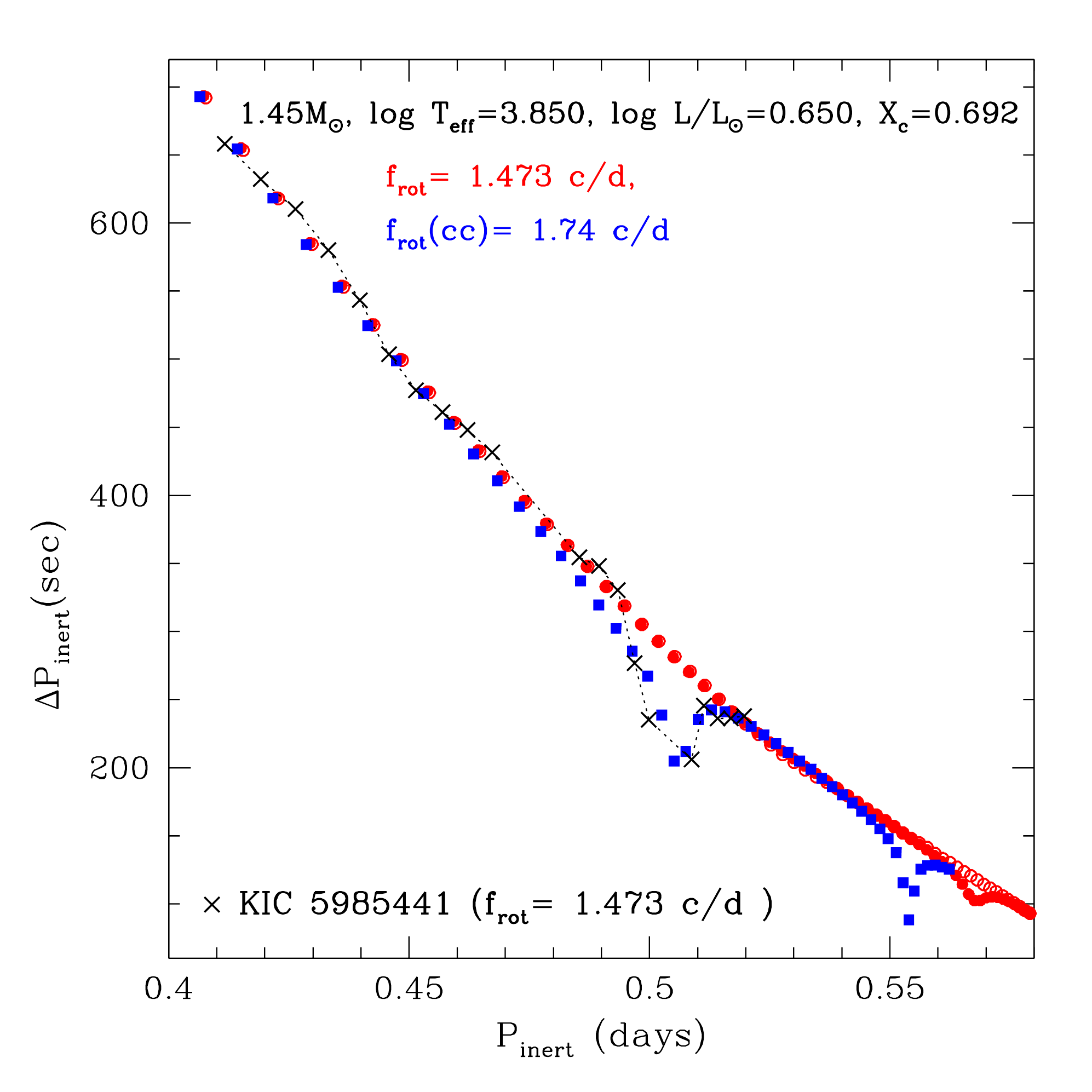}
\includegraphics[width=0.33\textwidth]{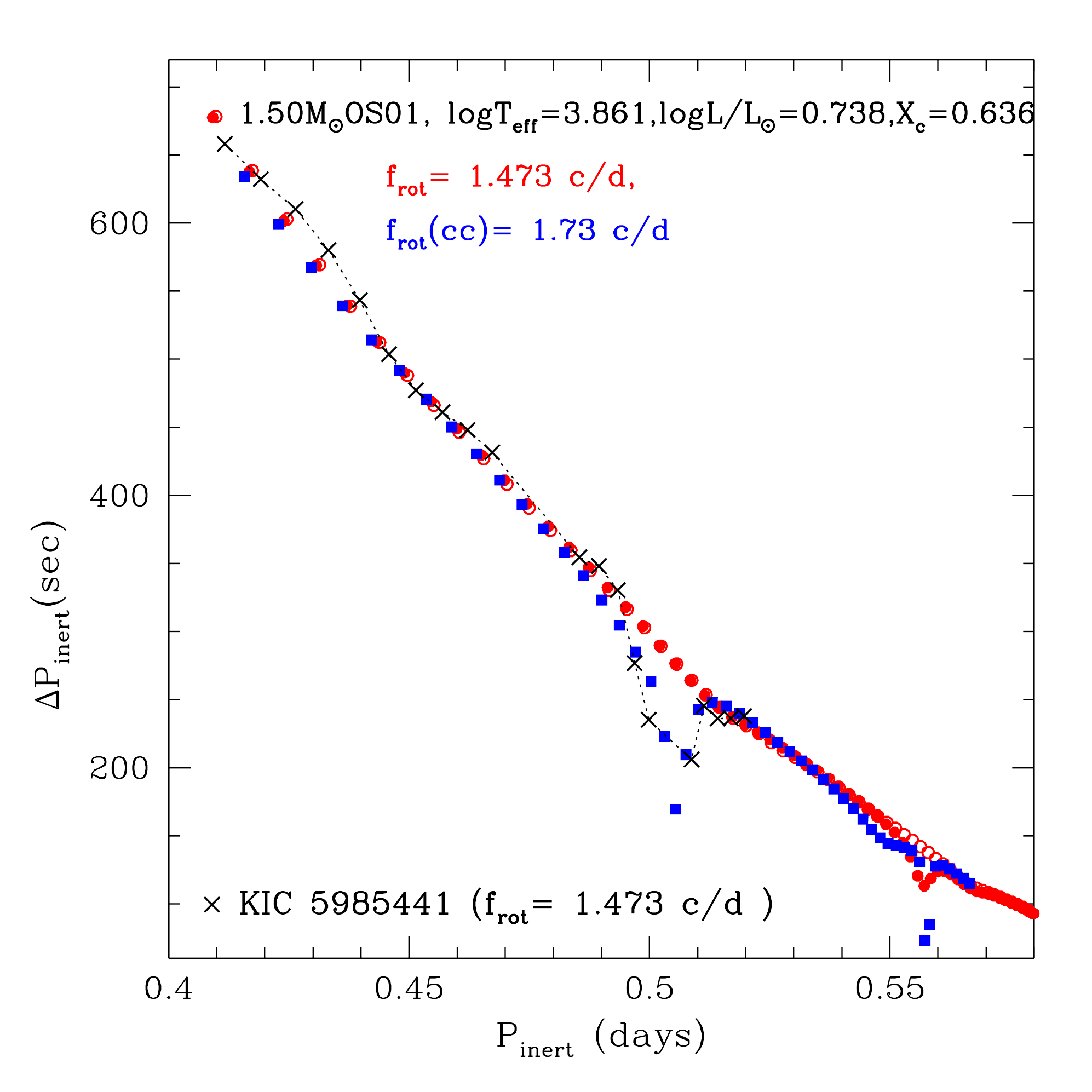}
\includegraphics[width=0.33\textwidth]{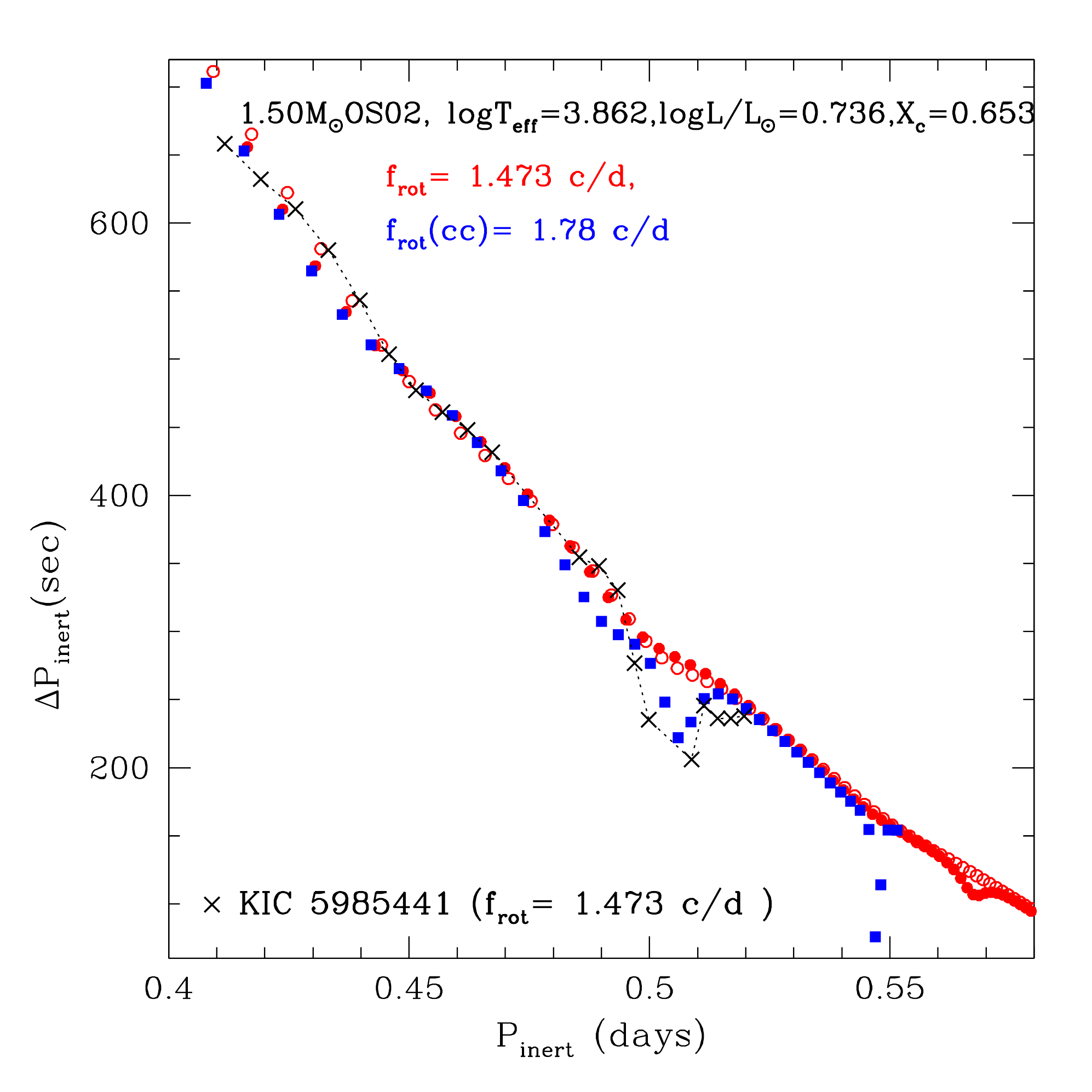}
\caption{The same as Fig.~\ref{fig:k529} but for KIC~5985441. Models with  uniform rotation at 1.473\,d$^{-1}$ (filled red circles) predict resonance dips at periods of 0.56--0.57\,days, which are longer than the observed dip at about 0.5\,days. To fit the dip, the convective core has to rotate about 20\,\% faster (exact values depend on the assumed extent of overshooting) as shown by filled blue squares.
The additional dip at $\sim0.55$ predicted by the differential rotation corresponds to the resonance with the first-overtone inertial mode of the convective core, although observed periods do not extend there.}
\label{fig:k5985}
\end{figure*}

\subsubsection{KIC 5985441 (Fig.\,\ref{fig:k5985})}
From the $P$\,-\,$\Delta P$ relations (for prograde dipole and quadrupole g modes) of KIC~5985441, \citet{LiG20} obtained $f_{\rm rot}=1.473\pm0.008$\,d$^{-1}$, which corresponds to the rotation rate of the g-mode cavity in this star. The $P$\,-\,$\Delta P$ relation for the  prograde dipole g modes  has a large dip at a period of about 0.5\,days (crosses connected by dotted line; Fig.\,\ref{fig:k5985}), which is attributable to the resonance with an inertial mode in the convective core. 
To fit the pronounced dip with a model it is necessary to assume a differential rotation  of  $\sim20$\%  between the convective core and the radiative g-mode cavity as shown by blue squares in Fig.\,\ref{fig:k5985}. While models with or without overshooting can fit reasonably well the $P$\,-\,$\Delta P$ sequence of KIC~5985441, the model of $1.45\,M_\odot$ without overshooting (left panel) reproduces best the $P$\,-\,$\Delta P$ pattern as well as the resonance dip at $\sim0.502$\,days. The differential rotation required in KIC~5985441 is largest among the selected $\gamma$~Dor stars in this paper (see \S\ref{sec:sum} below), while   the star is least studied probably due to its faintness with a Kepler magnitude of 15.8 mag.

\begin{figure*}
\includegraphics[width=0.33\textwidth]{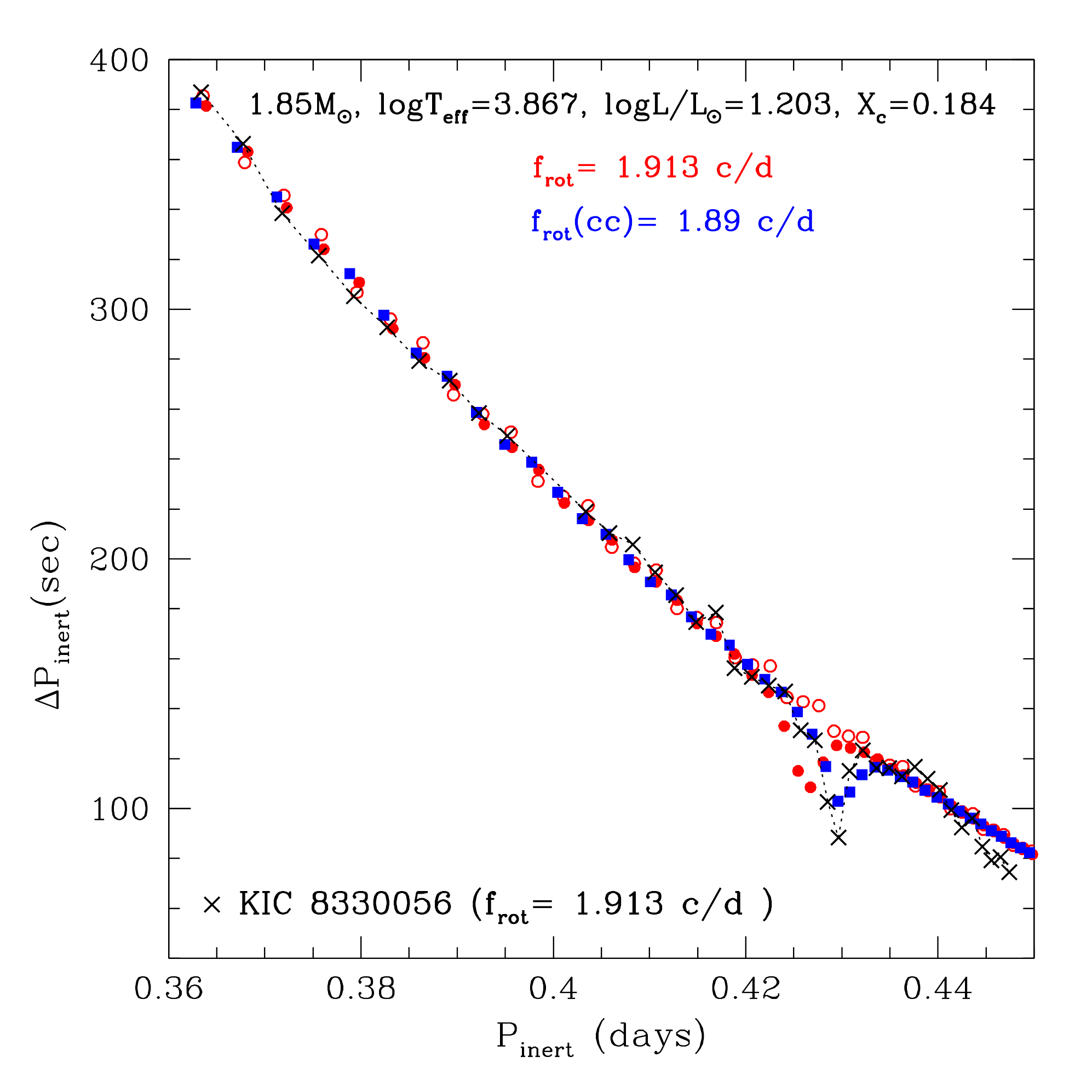}
\includegraphics[width=0.33\textwidth]{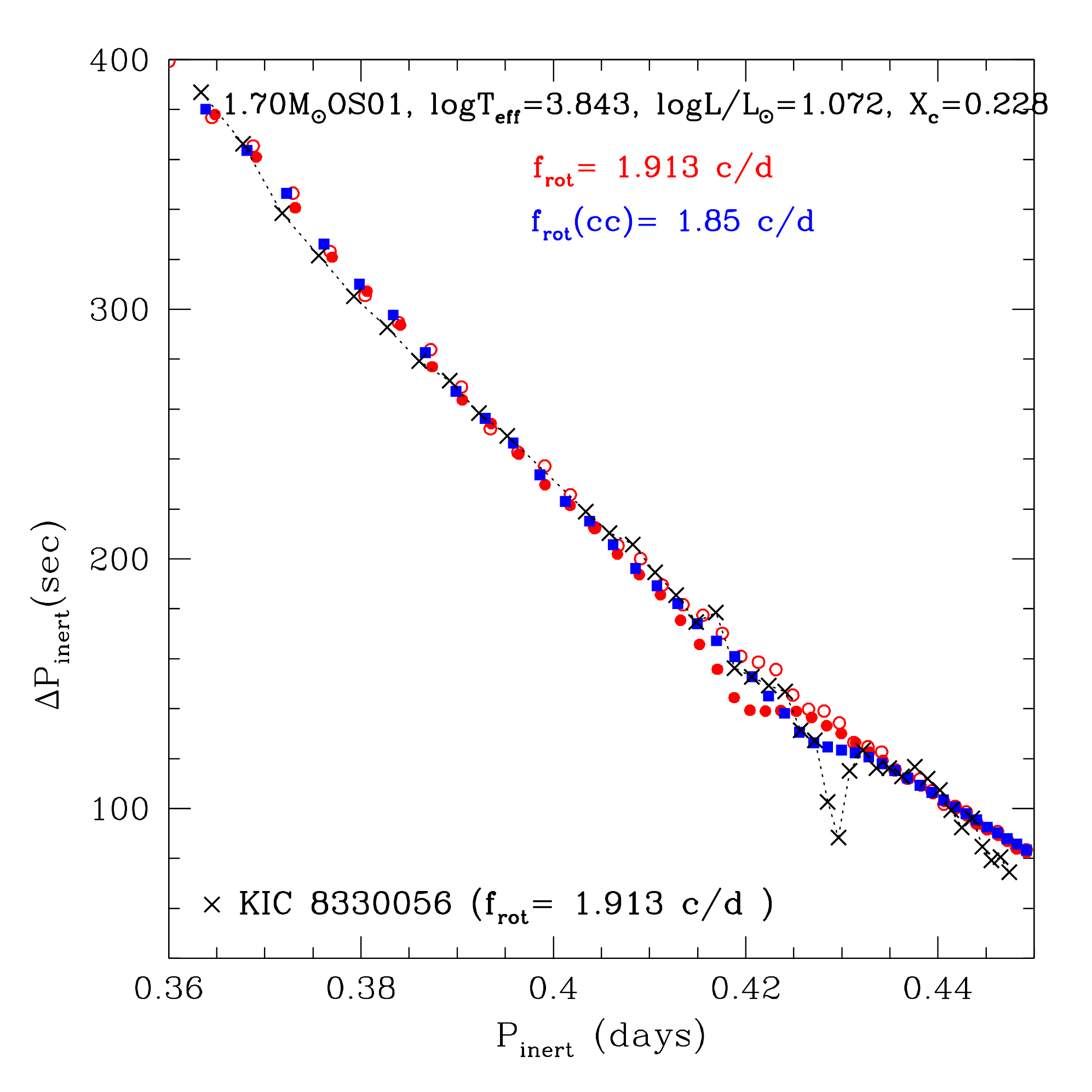}
\includegraphics[width=0.33\textwidth]{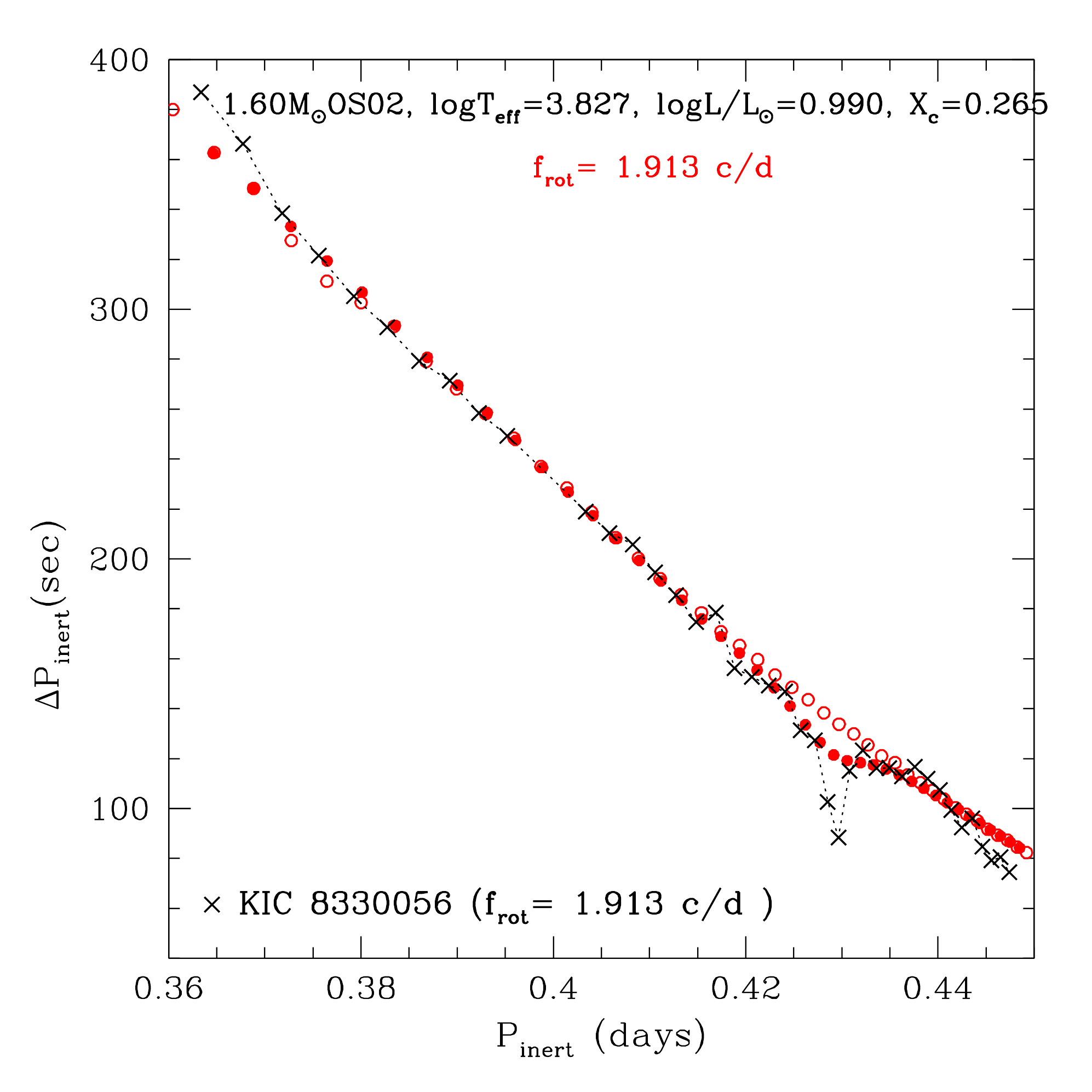}
\caption{The same as Fig.~\ref{fig:k529} but for KIC~8330056, for which \citet{LiG19} obtained a rotation frequency of 1.913~d$^{-1}$.
Crosses connected by dotted line show the observational $P$\,-\,$\Delta P$ relation of prograde dipole g modes, which has a dip at a period of 0.43~days attributable to the resonance coupling with the fundamental inertial mode of the convective core. 
The dip can be fitted well with a 1.85-$M_\odot$ model having a convective core (with no overshooting) rotating at 1.89~d$^{-1}$ (blue squares in the left panel). The resonance dips predicted by the models with core overshooting are too broad and shallow compared with the dip of KIC~8330056.  }
\label{fig:k833}
\end{figure*}

\subsubsection{KIC 8330056 (Fig.\,\ref{fig:k833})}\label{sec:k833}
\citet{LiG19} obtained $f_{\rm rot}=1.913\pm0.001$\,d$^{-1}$ for KIC~8330056 from the  $P$\,-\,$\Delta P$ relations of g and r modes.
Crosses plotted in Fig.~\ref{fig:k833} present the g-mode relation from \citet{LiG19}, where a resonance dip appears at a period of 0.43~days.   
Model fittings for KIC~8330056 are shown in Fig.~\ref{fig:k833} in the same format as in the previous cases.
The 1.85~$M_\odot$ model without overshooting (left panel) reasonably fit the period and the depth of the resonance dip if $f_{\rm rot}{\rm (cc)}=1.89\,{\rm d}^{-1}$ is assumed, which is very close to $f_{\rm rot}=1.913\,{\rm d}^{-1}$ for the g-mode cavity, indicating KIC~8330056 to rotate nearly uniformly.  
The central hydrogen mass fraction of this model, $X_{\rm c}= 0.167$, indicates a late stage of main-sequence evolution.
We note that the rotation rate of this model corresponds to about 83\,\% of the critical Roche model, while we expect little effects of rotational deformation on the g modes and inertial modes which reside in the deep interior.    

Models with core-overshooting are less successful for KIC~8330056 (middle and right panels of Fig.~\ref{fig:k833}).
The position of the $\Delta P$ dip can be approximately fitted assuming a convective-core rotation rate of 1.85~d$^{-1}$ for the OS01 model and uniform rotation for the OS02 model.
However, the predicted dips of these models are too shallow and broad compared with the observed one.
(We have already seen such broad and shallow dips for evolved models with overshooting in Fig.\,\ref{fig:pdp_overshoot}.)  
For this reason, parameters of OS01 and OS02 models are not listed in Table~\ref{tab:sum}.

\begin{figure*}
\includegraphics[width=0.33\textwidth]{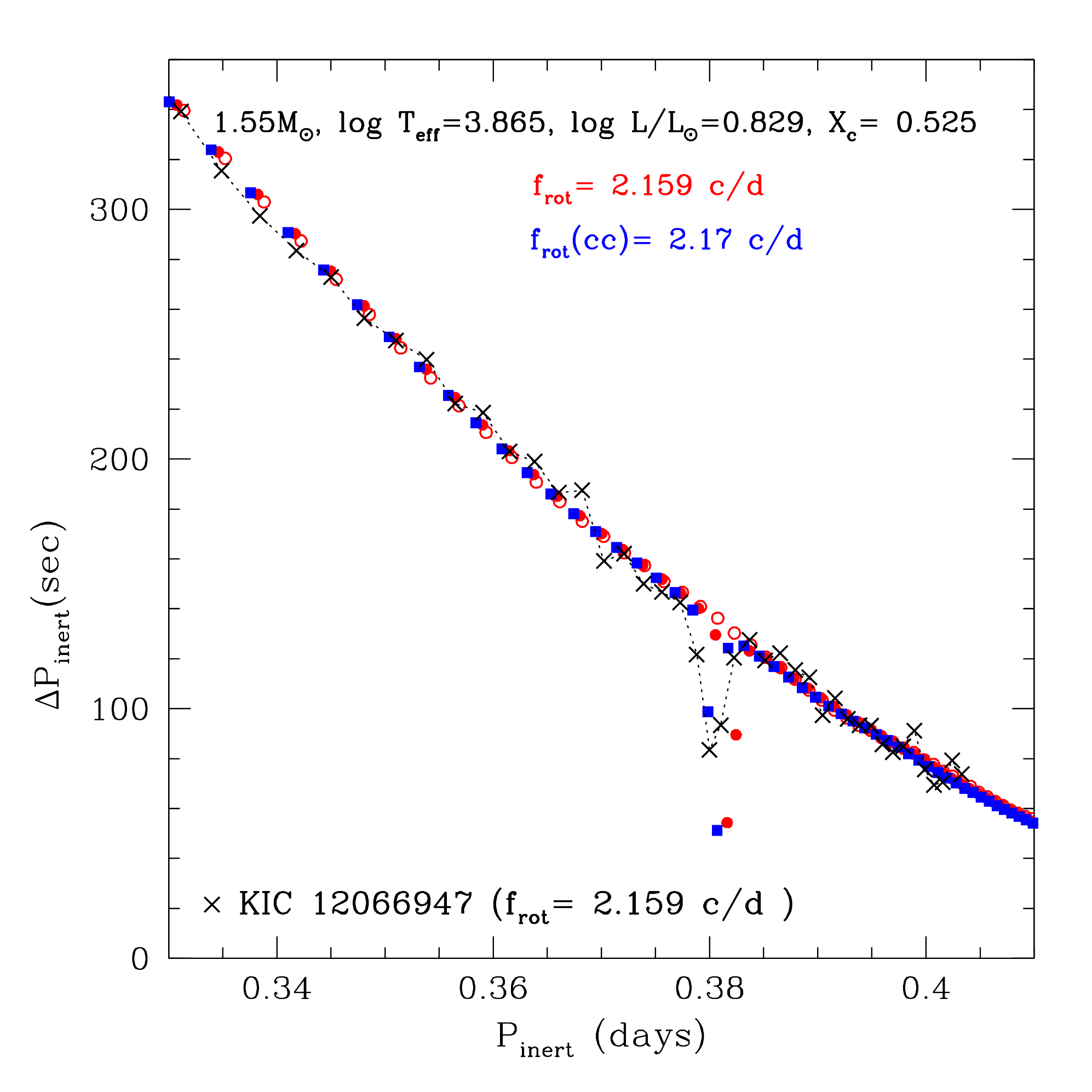}
\includegraphics[width=0.33\textwidth]{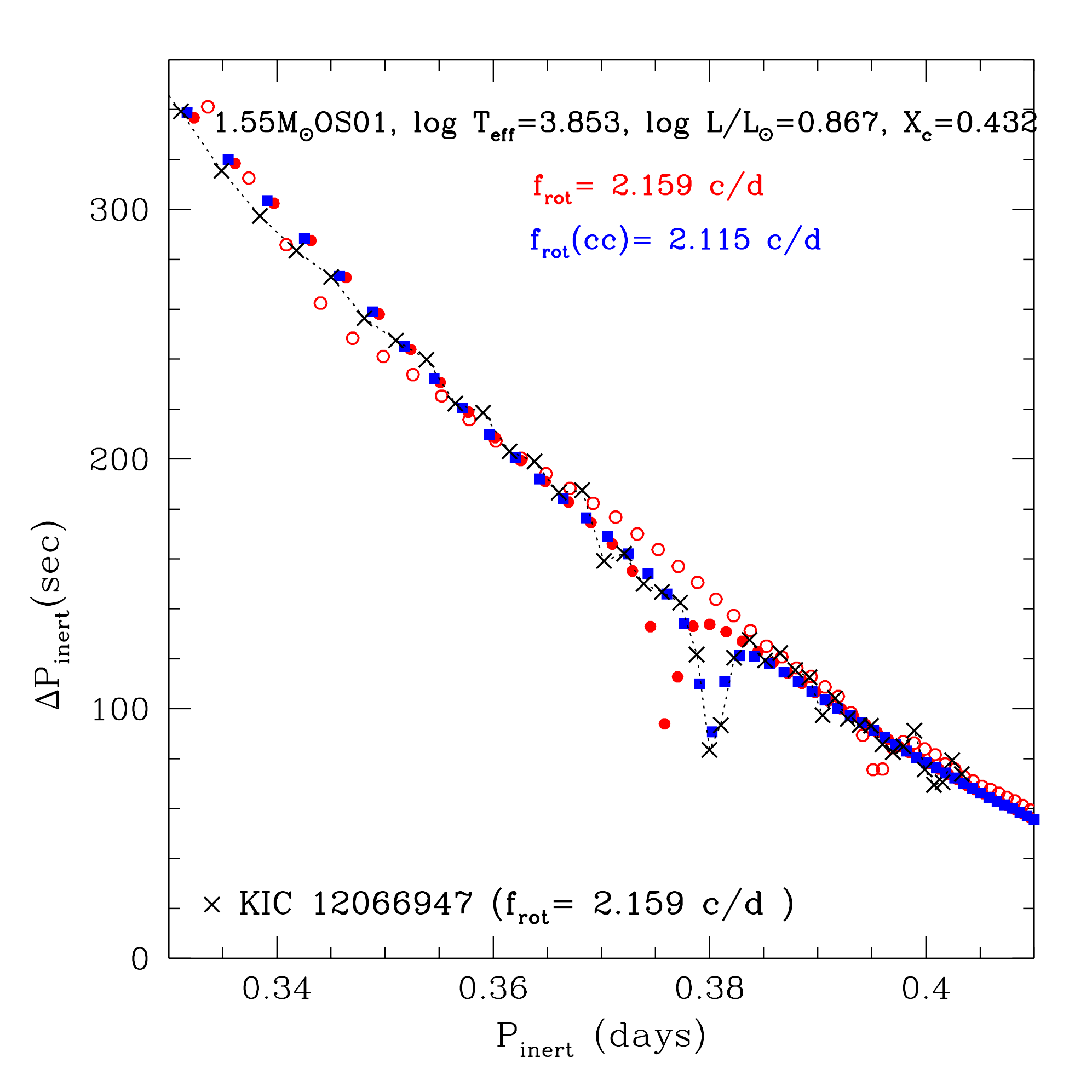}
\includegraphics[width=0.33\textwidth]{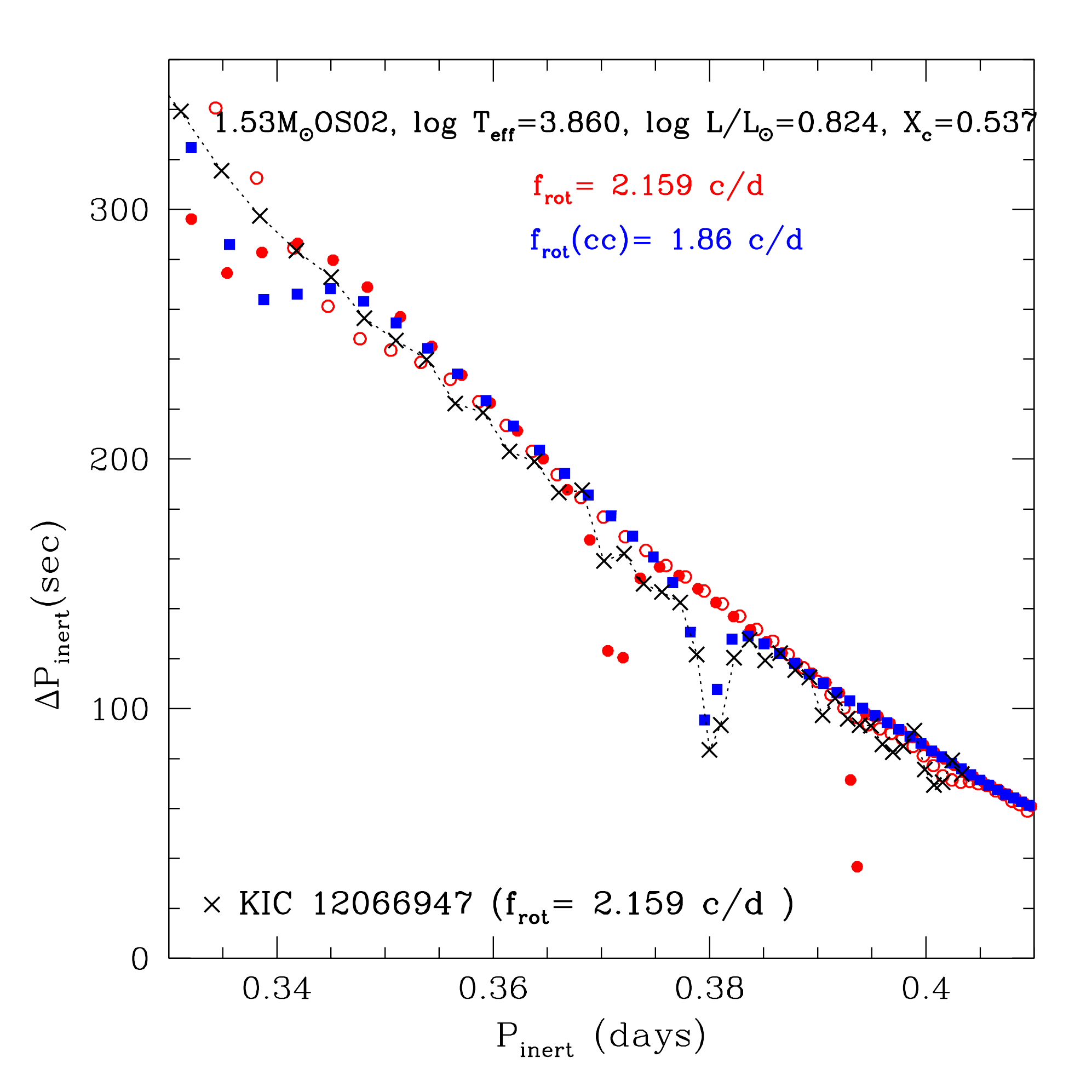}
\caption{The same as Fig.~\ref{fig:k529} but for KIC~12066947. Crosses connected with dotted lines are observational $P-\Delta P$ relation having a resonance dip at 0.38~days \citep{vanr16}. Rotation rate in the g-mode cavity of KIC~12066947 is 2.159~d$^{-1}$.
The resonance dip is best reproduced by the 1.55-$M_\odot$ model with $h_{\rm os}=0.01$ (middle panel) with a convective-core rotation of $f_{\rm rot}{\rm (cc)}=2.115$~d$^{-1}$, which is slightly slower than the rate in the g-mode cavity.
 }
\label{fig:k120}
\end{figure*}

\subsubsection{KIC 12066947 (Fig.\,\ref{fig:k120})}
KIC 12066947 is a rapidly rotating $\gamma$ Dor star
\citep[$V\sin i= 133.3\pm 5.6$~km\,s$^{-1}$][]{vanr15apjs}.
The g-mode $P$\,-\,$\Delta P$ relation of KIC~12066947 has been studied by many authors.
\citet{vanr16} obtained a rotation frequency of $2.160 \pm 0.008$~d$^{-1}$, while \citet{chr18} obtained $2.156\pm0.004$~d$^{-1}$, \citet{LiG19} obtained $2.159\pm 0.002$~d$^{-1}$, and \citet{tak20} obtained $2.15 \pm 0.09$~d$^{-1}$.
These results are all consistent with each other. We have adopted 2.159~d$^{-1}$ as the rotation frequency in the g-mode cavity of KIC~12066947.
The $P$\,-\,$\Delta P$ relation of prograde dipole g modes (crosses connected with dotted line in Fig.~\ref{fig:k120}) has a dip at a period of $0.38$~days \citep{vanr16} which is attributable to the resonance coupling with an inertial mode in the convective core.\footnote{Since period data around the dip are missing in the analysis by \citet{LiG19}, we have adopted the data set obtained by \citet{vanr16} for KIC~12066947.}

The observed $P$\,-\,$\Delta P$ relation and the position of the resonance dip of KIC~12066947 can be fitted with prograde dipole g modes of a 1.55-$M_\odot$ model with no overshooting (left panel) if we adopt a rotation rate of $f_{\rm rot}{({\rm cc})}=2.17~{\rm d}^{-1}$ in the convective core, and $f_{\rm rot}=2.159~{\rm d}^{-1}$ exterior to it  (blue squares in Fig.~\ref{fig:k120}). 
Similarly good fits are obtained for models with overshooting, if we assume slightly different rotation rates in the convective core.
The depth of the dip agrees best with the OS01 model, while $P_{\rm inert}$\,-\,$\Delta P_{\rm inert}$ relation of the OS02 model deviates in a period range of $0.33 - 0.34$~days. 
  
We note that although $f_{\rm rot}= 2.159$\,d$^{-1}$ of KIC~12066947 is larger than the case of KIC~8330056 discussed in \S\ref{sec:k833}, the rotation corresponds to about 55--60\,\% of the critical rotation, less influential than the case of KIC~8330056.

\subsubsection{Summary of model fittings} \label{sec:sum}
   
\begin{table*}
\centering
\caption{Summary of observed parameters and models}
\begin{tabular}{lcccclllll} 
\hline
  & \multicolumn{2}{c}{\citet{mur19}} &\citet{LiG19} & \multicolumn{6}{c}{Models$^{\rm a)}$}\\
  &&&\citet{LiG20} & \multicolumn{6}{c}{\hrulefill}\\  
   KIC   &   $\log T_{\rm eff}$(K) & $\log L/L_\odot$ & $f_{\rm rot}$(obs)~(d$^{-1}$)  & $h_{\rm os}$ & $f_{\rm rot}$(cc)~(d$^{-1}$) & $M/M_\odot$ & $\log T_{\rm eff}$(K) & $\log L/L_\odot$ & $X_{\rm c}$ \\ 
\hline 
 03341457 & $3.841\pm 0.008$ & $0.707\pm 0.03$ & $1.859\pm 0.001$ 
 &~~0.00& 1.87\,(1.88) & 1.45\,(1.45) & 3.846\,(3.872) & 0.696\,(0.740) & 0.56\,(0.62)
 \\ \,(Fig.\,\ref{fig:k334}) &&&
 &~~0.01& 1.83 & 1.40 & 3.834 & 0.645 & 0.54
 \\ &&&
 &~~0.02& 1.84 & 1.40 & 3.814 & 0.730 & 0.34
 \\      
\hline
 04390625 & $3.845\pm 0.016$ & $1.149\pm 0.046$ & $1.12 \pm 0.01$
 &~~0.00& 1.22\,(1.225) & 1.80\,(1.70) & 3.843\,(3.865) & 1.159\,(1.124) & 0.16\,(0.16)  
 \\ \,(Fig.\,\ref{fig:k439}) &&&
 &~~0.02& 1.24 & 1.70 & 3.801 & 1.121 & 0.08 
 \\ 
\hline
 04774208 & $3.847\pm 0.014$ & $0.801\pm 0.033$  & $1.834\pm 0.001$ 
 &~~0.00& 1.915\,(1.98) &  1.55\,(1.50) & 3.872\,(3.887) & 0.803\,(0.769) & 0.61\,(0.71)
 \\ \,(Fig.\,\ref{fig:k477}) &&&
 &~~0.01& 1.90  &  1.53 & 3.863 & 0.804 & 0.56 
 \\ &&&
 &~~0.02& 1.95 &  1.50 & 3.843 &  0.821 & 0.45 
 \\ 
\hline
 05294571 & $3.837\pm 0.013$ & $0.768\pm 0.032$  & $1.6421\pm 0.0009$ 
 &~~0.00& 1.675\,(1.70) &  1.50\,(1.50) & 3.858\,(3.886) & 0.753\,(0.785) & 0.57\,(0.67) 
 \\ \,(Fig.\,\ref{fig:k529}) &&&
 &~~0.01& 1.605 &  1.50 & 3.851 & 0.787 & 0.50
 \\ &&&
 &~~0.02& 1.54 &  1.45 & 3.837 &  0.749 & 0.48  
 \\ 
\hline
 05391059 & $3.831\pm 0.013$ & $0.767\pm 0.026$  & $1.796\pm 0.001$  
 &~~0.00& 1.825\,(1.90) & 1.50\,(1.45) & 3.858\,(3.873) & 0.753\,(0.704) & 0.57\,(0.71)
 \\ \,(Fig.\,\ref{fig:k539}) &&&
 &~~0.01& 1.785 & 1.50 & 3.856 & 0.768 & 0.56 
 \\ 
\hline
 05985441 & $3.848\pm 0.016$ & $0.741\pm 0.10$ & $1.473\pm 0.008$
 &~~0.00& 1.74\,(1.78) & 1.45\,(1.40) & 3.850\,(3.859) & 0.650\,(0.636) & 0.69\,(0.72)
 \\ \,(Fig.\,\ref{fig:k5985})&&&
 &~~0.01& 1.73 & 1.50 & 3.861 & 0.738 & 0.64 
 \\ &&&
 &~~0.02& 1.78 & 1.50 & 3.862 & 0.736 & 0.65 \\
\hline
07968803 & $3.867\pm 0.015$ & $0.823\pm 0.023$ & $1.94\pm 0.01$
 &~~0.00& 1.98\,(2.02) & 1.55\,(1.50) & 3.864\,(3.884) & 0.831\,(0.804) & 0.52\,(0.61) 
\\ \,(Fig.\,\ref{fig:k796})&&&
 &~~0.01& 1.93 & 1.50 & 3.842 & 0.811 & 0.42 
\\ &&&    
 &~~0.02& 2.00 & 1.50 & 3.827 & 0.855 & 0.34 \\
\hline
08326356 & $3.861\pm 0.015$ & $0.991\pm 0.047$ & $2.38\pm 0.02$
 &~~0.00& 2.40\,(2.40) & 1.65\,(1.65) & 3.878\,(3.915) & 0.957\,(0.986) & 0.46\,(0.58) 
 \\ \,(Fig.\,\ref{fig:k832})&&&
 &~~0.01& 2.33 & 1.65 & 3.864 & 0.991 & 0.38
 \\&&&
 &~~0.02& 2.48 & 1.65 & 3.869 & 1.001 & 0.43 \\        
\hline
 08330056 & $3.873 \pm 0.015$ & $1.245\pm 0.035$ &  $1.913 \pm 0.001$  
 &~~0.00& 1.89\,(1.90) &  1.85\,(1.80) & 3.867\,(3.893) & 1.203\,(1.217)  & 0.18\,(0.22) 
 \\ \,(Fig.\,\ref{fig:k833})&&&
\\
\hline
 09962653 & $3.861\pm 0.014$ & $0.786\pm 0.021$ &  $1.763\pm 0.001$ 
 &~~0.00&  1.795\,(1.815) & 1.53\,(1.50) & 3.867\,(3.886) & 0.781\,(0.790) & 0.60\,(0.65) 
 \\ \,(Fig.\,\ref{fig:k996})&&&
 &~~0.01&  1.75   &  1.50 & 3.852 & 0.782 & 0.51
\\
\hline
 11017637 & $3.863\pm 0.014$ & $0.771\pm 0.019$  & $1.6153\pm 0.0008$ 
 &~~0.00& 1.70\,(1.70) &  1.53\,(1.50) & 3.871\,(3.887) & 0.753\,(0.779) & 0.69\,(0.68) 
 \\ \,(Fig.\,\ref{fig:k110})&&&
 &~~0.01& 1.67 &  1.53 & 3.869 & 0.773 & 0.64
 \\ &&&
 &~~0.02& 1.725 & 1.53 & 3.869 &  0.771 & 0.66  \\
\hline
 11550154 & $3.874\pm 0.015$ & $0.774\pm 0.019$  & $2.017\pm 0.001$ 
 &~~0.00& 2.075\,(2.12) &  1.53\,(1.50) & 3.867\,(3.886) & 0.781\,(0.790) & 0.60\,(0.65)
 \\ \,(Fig.\,\ref{fig:k1155})&&&
 &~~0.01& 2.04  &  1.50 & 3.858 & 0.758 & 0.64
\\
\hline
 11649699 & $3.856\pm 0.012$ & $0.742\pm 0.018$  & $1.729\pm 0.002$ 
 &~~0.00& 1.855\,(1.89) & 1.50\,(1.45) & 3.861\,(3.873) & 0.737\,(0.716) & 0.63\,(0.68) 
 \\ \,(Fig.\,\ref{fig:k116})&&&
 &~~0.01& 1.82  & 1.50 & 3.858 & 0.758 & 0.58
\\
\hline
 11907454 & $3.857\pm 0.013$ & $0.732\pm 0.018$  & $1.3387\pm 0.0006$ 
 &~~0.00& 1.435\,(1.48) & 1.50\,(1.45) & 3.860\,(3.872) & 0.746\,(0.731) & 0.60\,(0.64) 
 \\ \,(Fig.\,\ref{fig:k119})&&&
 &~~0.01& 1.435 & 1.50 & 3.860 & 0.743 & 0.62 
 \\ &&&
 &~~0.02& 1.48  & 1.50 & 3.859 &  0.757 & 0.61\\
\hline
 12066947 & $3.865\pm 0.004$ & $0.851\pm 0.025$  & $2.159\pm 0.002$ 
 &~~0.00& 2.17\,(2.185) & 1.55\,(1.55) & 3.866\,(3.894) & 0.828\,(0.875) & 0.53\,(0.58) 
 \\ \,(Fig.\,\ref{fig:k120})&&&
 &~~0.01& 2.115 & 1.55 & 3.853 & 0.867 & 0.43
\\
\hline
 12303838 & $3.864 \pm 0.015$ & $0.868 \pm 0.03$ &  $1.3301 \pm 0.0007$ 
 &~~0.00& 1.415\,(1.42) & 1.60\,(1.55) & 3.871\,(3.895) & 0.898\,(0.872) & 0.48\,(0.59)
 \\ \,(Fig.\,\ref{fig:k123})&&&
 &~~0.01& 1.36  & 1.60 & 3.875 & 0.895 & 0.53
\\   
\hline
\end{tabular}
$^{\rm a)}$ The initial chemical composition $(X,Z)=(0.72,0.014)$ is adopted for standard models, while numbers in parentheses are from models with $(X,Z)=(0.724,0.01)$ 
\label{tab:sum}
\end{table*}

Table~\ref{tab:sum} summarizes results of model fittings shown in this section and in Appendix B.  The table lists observational parameters \citep{mur19}, rotation frequency in the g-mode cavity for each star obtained by \citet{LiG19,LiG20}, and model parameters and the rotation frequency in the convective core to fit observed $\Delta P$ dip for each model.
Results of metal-poor models with $Z=0.010$, are shown as parenthesised numbers in the first row (OS00) of each star in Table~\ref{tab:sum}. 

For some stars models with core-overshooting of $h_{\rm os}=0.02$ and/or $0.01$ are not listed, because these models have resonance dips too broad and shallow, while in many cases dips tend to be reproduced well by models with $h_{\rm os}\le 0.01$.
This could indicate the overshooting from the convective core to be largely 'penetrative' type \citep{zahn02} producing mostly adiabatic layers with little radiative zone even if matter in a substantial range is mixed.  
We note, however, that the preference of a smaller $h_{\rm os}$ could be due to a bias in the selection of stars, because we have selected, in this paper, stars having a strong dip. Further detailed studies would be needed.

\begin{figure}
\includegraphics[width=\columnwidth]{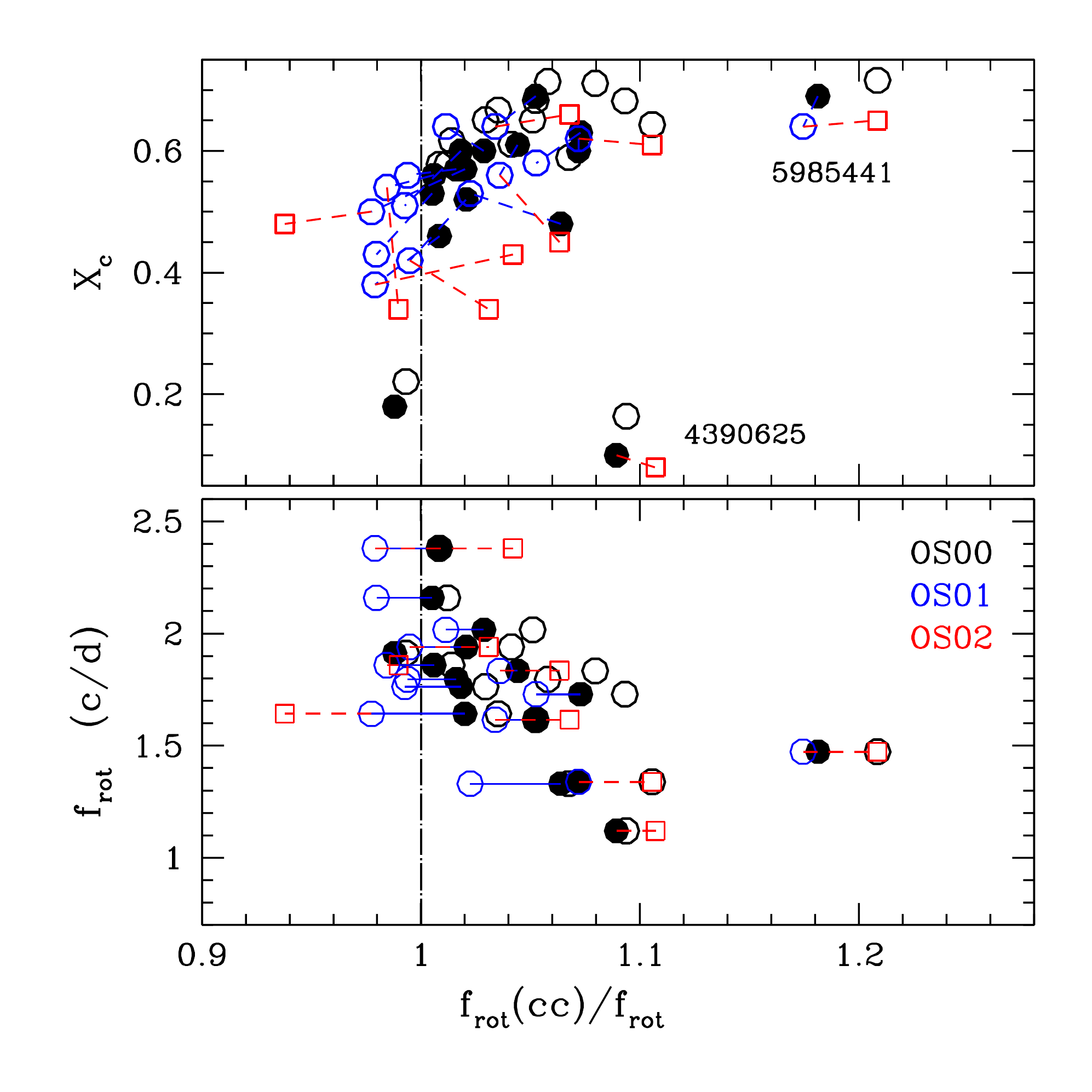}
\caption{The ratios of the rotation frequency in the convective core $f_{\rm rot}$(cc) to that of radiative g-mode cavity $f_{\rm rot}$ are plotted with respect to the hydrogen mass fraction in the convective core $(X_{\rm c})$ (upper panel) and $f_{\rm rot}$ obtained from $P-\Delta P$ patterns by \citet{LiG19,LiG20}. Filled (black) and open (blue) circles and open squares (red) correspond the values obtained by fitting to models without overshoot and with small overshoots; i.e., $h_{\rm os}=0.00$, $0.01$, and $0.02$ respectively. The same stars are connected with dashed lines. These results are based on models with an initial composition of $(X,Z)=(0.72,0.014)$, while black open circles denote results based on the models with $(X,Z)=(0.724,0.010)$ without overshooting. The names of two outliers, KIC~5985441 and KIC~4390625, are indicated in the upper panel.}
\label{fig:ratio}
\end{figure}

Fig.~\ref{fig:ratio} shows the ratio of the best fit rotation frequency in the convective core, $f_{\rm rot}{\rm (cc)}$, to the rotation frequency in the g-mode cavity $f_{\rm rot}$ with respect to $f_{\rm rot}$ (lower panel) and to the hydrogen mass fraction at the center, $X_{\rm c}$ (upper panel).
Filled (black) and open (blue) circles, and squares (red) are for models without overshooting, with overshooting of $h_{\rm os}=0.01$, and with $h_{\rm os}=0.02$, respectively.
Points belonging to the same stars are connected by dashed lines to show the effects of the core overshooting assumptions. 
In addition, results from metal-poor models with $Z=0.010$ (OS00) are shown by black open circles, while no connecting lines are drawn to avoid too much busyness.

Models with $h_{\rm os}=0.01$ tend to yield slightly slower convective-core rotation rates compared to the models without overshooting, although the effects are not so large to disturb the general trend.
The metal poor models tend to give slightly larger $f_{\rm rot}$(cc). Again, the tendency hardly affects the general trends seen in this figure.

The majority of $\gamma$ Dor stars we studied in this paper rotate nearly uniformly, while convective cores tend to rotate slightly faster than the surrounding g-mode cavity. However, there is a notable exception, KIC~5985441, at $f_{\rm rot}({\rm cc})/f_{\rm rot}\approx 1.2$. The $P-\Delta P$ sequence of KIC~5985441 is fitted reasonably well as shown in Fig.\,\ref{fig:k5985} with the moderate differential rotation irrespective of the assumed extent of overshooting or a metal-poor ($Z=0.010$) composition. 
Since the star is relatively faint ($K_p=15.8$~mag), no spectroscopic information  is available.

The upper panel of Fig.\,\ref{fig:ratio} shows a subtle tendency of less evolved stars (i.e., with larger $X_{\rm c}$) having larger differential rotation $f_{\rm rot}({\rm cc})/f_{\rm rot}$, 
while the lower panel seems to indicate that the differential rotations tend to be larger in stars having smaller $f_{\rm rot}$.
These tendencies, if real, may be understood as that the convective core of a star rotates slightly faster than the surrounding g-mode cavity at the beginning of the main-sequence evolution, while a part of the g-mode cavity surrounding the convective core boundary has spun up to synchronize with the convective core as evolution proceeds. 
Although it is an interesting tendency, further studies are definitely needed.

\section{$P$\,-\,$\Delta P$ relations of r modes}\label{sec:rmodes}
\begin{figure}
\includegraphics[width=\columnwidth]{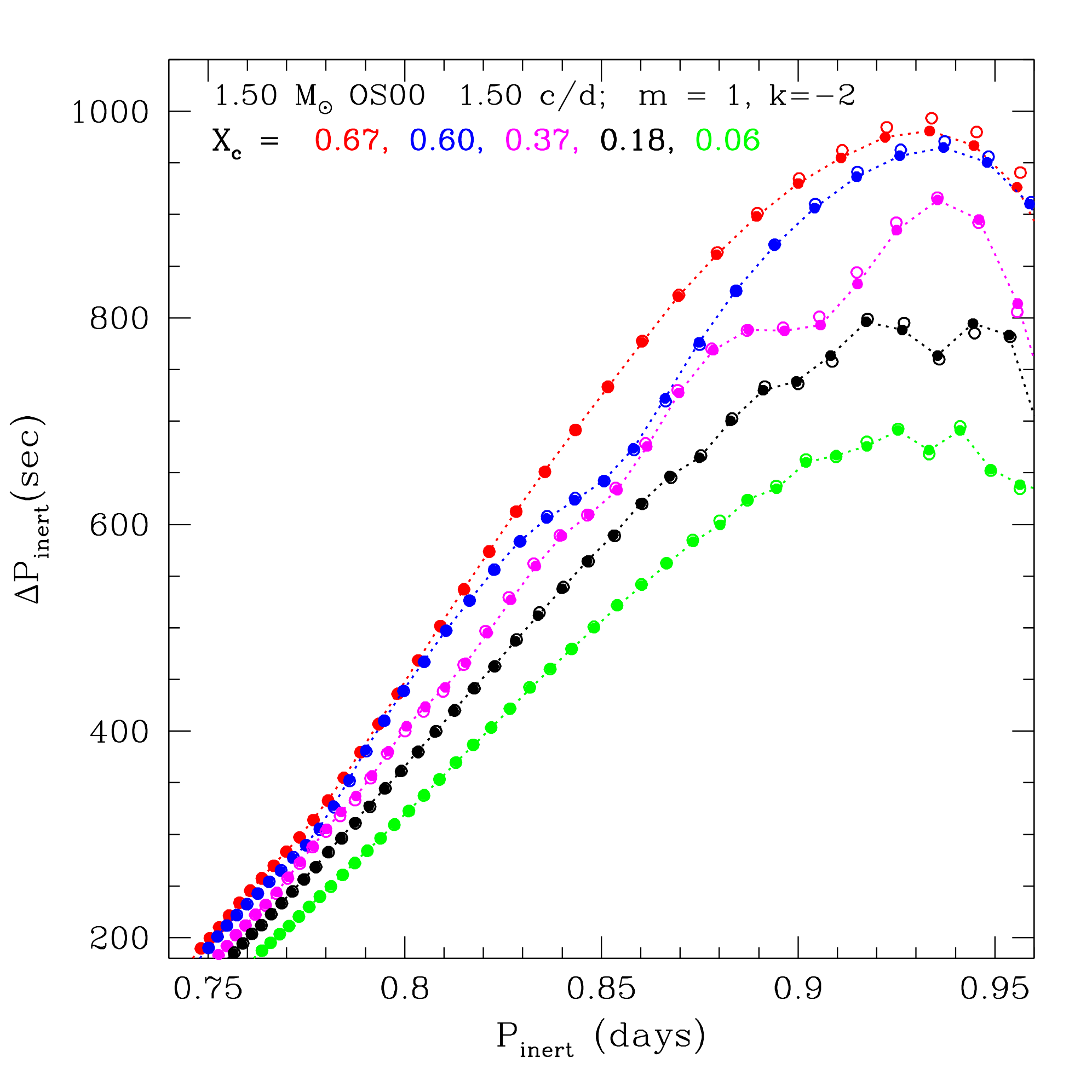}
\caption{Period spacing versus period for r modes of $(m,k)=(1,-2)$ in the inertial frame (whose temperature variations are symmetric with respect to the equator) in the 1.5-$M_\odot$ models at selected evolutionary stages (i.e., at selected central hydrogen abundance $X_{\rm c}$). A rotation frequency of 1.5~d$^{-1}$ is adopted as a typical rate, while no overshooting is included in these models. Open and filled circles are results obtained using the TAR and the expansion method, respectively. }
\label{fig:rmodes}
\end{figure}
R mode oscillations are  normal modes of global Rossby waves influenced by buoyancy. Pure Rossby waves generated by the Coriolis force which propagate only horizontally, while r modes, because of the buoyancy effect, propagate also radially. 
For this reason, the periods of high radial-order r modes in the co-rotating frame are approximately proportional to the radial order $|n|$, as for g modes. 
R modes are retrograde in the co-rotating frame, while they are observed as prograde modes in the inertial frame because the frequencies of r modes in the co-rotating frame are less than the rotation frequency. The observational frequencies are located between $(|m|-1)f_{\rm rot}$ and $|m|f_{\rm rot}$ \citep[see][for details]{sai18}.  
The period spacing of r modes in the inertial frame increases with period, which is opposite to the pattern of prograde g modes.
From this property \citet{vanr16} discovered r modes in $\gamma$ Dor stars.
\citet{LiG19,LiG20} found r modes (mostly $m=1$ even $k=-2$ modes) in many $\gamma$ Dor stars and, in some cases \citep[e.g., KIC~11775251][]{LiG20}, r-mode $P-\Delta P$ relations to show modulations similar to dipole g modes.

In order to see whether or not a resonance coupling occurs between r modes and inertial modes in the convective core, we have calculated r mode periods with the TAR and with the expansion method for 1.5-$M_\odot$ models at some main-sequence evolution stages assuming a uniform rotation of 1.50~d$^{-1}$.  
Obtained $P_{\rm inert}$\,-\,$\Delta P_{\rm inert}$ relations are shown in Fig.\,\ref{fig:rmodes}, where results with the TAR and with the expansion method are shown by open and filled circles, respectively. 
Although there are some modulations in the $P$\,-\,$\Delta P$ relations, filled and open circles always stay close to each other, indicating that the TAR works well for r modes as well as g modes, and that those modulations should be due to chemical composition gradients rather than the resonance with inertial modes.
Although the models shown in this figure have no core overshooting, we have obtained similar results for models with core overshooting.

In the co-rotating frame, frequencies of r modes of $(m,k)=(1,-2)$ are limited as $\nu_{\rm co-rot} < f_{\rm rot}/3$ \citep{sai18} so that the spin parameter should be $s > 6$, while \citep{oua20} has shown that the retrograde inertial mode of $m=1$ in the convective core has a spin parameter of $1.34-1.37$, much smaller than the spin parameters of the r modes. This explains why no resonance couplings occur between inertial modes in the convective core and r modes. 
Therefore, we should regard modulations in period spacing patterns of r modes to be caused by chemical composition gradients.

\section{Concluding remarks}
We discussed the dip in g mode period spacings which is produced by the resonance with an inertial mode in the convective core as \citet{oua20} found for the first time.
The resonance between prograde dipole g modes and an (fundamental or first-overtone) inertial mode in the core occurs at certain spin parameters of $2f_{\rm rot}/\nu_{\rm co-rot} \sim8-11$ (fundamental) and $22-25$ (first-overtone).
These values depend on evolutionary stages, on the assumptions of overshooting, and, to a lesser extent, on stellar masses. 
These properties are consistent with the result of \citet{oua20}.
We have found no resonance couplings between r modes (propagative in radiative layers) and retrograde inertial modes in the convective core, while \citet{oua20} found resonance couplings between a retrograde inertial mode and retrograde ($m=1$) g modes (as well as couplings between axisymmetric $m=0$ g and inertial modes).

From a resonance dip in the g-mode period-spacing pattern of a star, we can measure the rotation frequency of the convective core by fitting a model prediction obtained by the expansion method \citep{lee95,lee88}.
We can distinguish the true resonance dip from the modulation due to a steep change of hydrogen abundance, by comparing period spacings calculated by the expansion method and those based on the TAR, because the resonance coupling does not occur in the latter.
Thus, the resonance dip and $P$\,-\,$\Delta P$ patterns of moderately or rapidly rotating stars, are very useful to probe the rotation frequencies in the central convective core and the surrounding near-core region. 

We have selected 16 $\gamma$ Dor stars having relatively clear resonance dips, and fitted them with dipole prograde g modes of models having parameters consistent with those given in \citet{mur19}. Adopting the rotation frequency obtained by \citet{vanr16,LiG19} from the global pattern of $P$\,-\,$\Delta P$ for each star, we found a model which nicely fit the overall feature of the $P$\,-\,$\Delta P$ pattern, while the period at the resonance dip is, in most cases, slightly different from observation.
Then, we calculated g-mode periods in differentially rotating models, in which the radiative layers rotates at the same rate as before, while convective core rotates at a different rate, $f_{\rm rot}$(cc). By changing the value of $f_{\rm rot}$(cc), we have found a model which reproduces the position of the dip as well as the global pattern of the observational $P$\,-\,$\Delta P$ relation.

Seismically inferred differential rotations between stellar envelope and core of main-sequence stars have been discussed many times in the literature \citep[see e.g.][]{aer17}. In most cases the core meant so far the radiative g-mode cavity surrounding the convective core rather than the convective core itself. Because g modes do not propagate in the convective core, g modes themselves do not provide us with information on the rotation in the convective core.
Thanks to the resonance coupling with an inertial mode which is propagative in the convective core \citep{oua20}, we could obtain, in this paper, rotation frequencies of convective cores from the resonance dips in $P$\,-\,$\Delta P$ relations of prograde dipole g modes.
We found rotation frequencies in the convective core to be very close to (in many cases slightly faster than) those in the surrounding g-mode cavities. 
This is a unique new knowledge we could extract from nonradial pulsations of $\gamma$ Dor variables.

\section*{Acknowledgements}
We are grateful Rhita-Maria Ouazzani for helpful conversations.
We also thank Professor Conny Aerts for her persistent interest in the dip of period spacings, for useful comments on a draft of this paper, and for her encouragements. We thank the anonymous referee for helpful comments.
TVR gratefully acknowledges receiving support from the Research Foundation Flanders (FWO) under grant agreement N$^\circ$ 12ZB620N.

\section*{Data availability}
The data underlying this article will be shared on reasonable request to the corresponding author.




\bibliographystyle{mnras}
\bibliography{ref} 

\begin{thebibliography}{}
\makeatletter
\relax
\def\mn@urlcharsother{\let\do\@makeother \do\$\do\&\do\#\do\^\do\_\do\%\do\~}
\def\mn@doi{\begingroup\mn@urlcharsother \@ifnextchar [ {\mn@doi@}
  {\mn@doi@[]}}
\def\mn@doi@[#1]#2{\def\@tempa{#1}\ifx\@tempa\@empty \href
  {http://dx.doi.org/#2} {doi:#2}\else \href {http://dx.doi.org/#2} {#1}\fi
  \endgroup}
\def\mn@eprint#1#2{\mn@eprint@#1:#2::\@nil}
\def\mn@eprint@arXiv#1{\href {http://arxiv.org/abs/#1} {{\tt arXiv:#1}}}
\def\mn@eprint@dblp#1{\href {http://dblp.uni-trier.de/rec/bibtex/#1.xml}
  {dblp:#1}}
\def\mn@eprint@#1:#2:#3:#4\@nil{\def\@tempa {#1}\def\@tempb {#2}\def\@tempc
  {#3}\ifx \@tempc \@empty \let \@tempc \@tempb \let \@tempb \@tempa \fi \ifx
  \@tempb \@empty \def\@tempb {arXiv}\fi \@ifundefined
  {mn@eprint@\@tempb}{\@tempb:\@tempc}{\expandafter \expandafter \csname
  mn@eprint@\@tempb\endcsname \expandafter{\@tempc}}}

\bibitem[\protect\citeauthoryear{{Aerts}, {Van Reeth}  \& {Tkachenko}}{{Aerts}
  et~al.}{2017}]{aer17}
{Aerts} C.,  {Van Reeth} T.,   {Tkachenko} A.,  2017, \mn@doi [\apjl]
  {10.3847/2041-8213/aa8a62}, \href
  {https://ui.adsabs.harvard.edu/abs/2017ApJ...847L...7A} {847, L7}

\bibitem[\protect\citeauthoryear{{Borucki} et~al.,}{{Borucki}
  et~al.}{2010}]{kepler}
{Borucki} W.~J.,  et~al., 2010, \mn@doi [Science] {10.1126/science.1185402},
  \href {https://ui.adsabs.harvard.edu/abs/2010Sci...327..977B} {327, 977}

\bibitem[\protect\citeauthoryear{{Bouabid}, {Dupret}, {Salmon},
  {Montalb{\'a}n}, {Miglio}  \& {Noels}}{{Bouabid} et~al.}{2013}]{bou13}
{Bouabid} M.-P.,  {Dupret} M.-A.,  {Salmon} S.,  {Montalb{\'a}n} J.,  {Miglio}
  A.,   {Noels} A.,  2013, \mn@doi [\mnras] {10.1093/mnras/sts517}, \href
  {http://ads.nao.ac.jp/abs/2013MNRAS.429.2500B} {429, 2500}

\bibitem[\protect\citeauthoryear{{Christensen-Dalsgaard}}{{Christensen-Dalsgaard}}{2012}]{chr12}
{Christensen-Dalsgaard} J.,  2012, in {Shibahashi} H.,  {Takata} M.,
  {Lynas-Gray} A.~E.,  eds,  Astronomical Society of the Pacific Conference
  Series Vol. 462, Progress in Solar/Stellar Physics with Helio- and
  Asteroseismology. p.~503 (\mn@eprint {arXiv} {1110.5012})

\bibitem[\protect\citeauthoryear{{Christophe}, {Ballot}, {Ouazzani}, {Antoci}
  \& {Salmon}}{{Christophe} et~al.}{2018}]{chr18}
{Christophe} S.,  {Ballot} J.,  {Ouazzani} R.-M.,  {Antoci} V.,   {Salmon}
  S.~J.~A.~J.,  2018, \mn@doi [\aap] {10.1051/0004-6361/201832782}, \href
  {http://adsabs.harvard.edu/abs/2018A%26A...618A..47C} {618, A47}

\bibitem[\protect\citeauthoryear{{Dziembowski}, {Moskalik}  \&
  {Pamyatnykh}}{{Dziembowski} et~al.}{1993}]{Dzi93}
{Dziembowski} W.~A.,  {Moskalik} P.,   {Pamyatnykh} A.~A.,  1993, \mn@doi
  [\mnras] {10.1093/mnras/265.3.588}, \href
  {https://ui.adsabs.harvard.edu/abs/1993MNRAS.265..588D} {265, 588}

\bibitem[\protect\citeauthoryear{{Ekstr{\"o}m} et~al.,}{{Ekstr{\"o}m}
  et~al.}{2012}]{Geneva2012}
{Ekstr{\"o}m} S.,  et~al., 2012, \mn@doi [\aap] {10.1051/0004-6361/201117751},
  \href {https://ui.adsabs.harvard.edu/abs/2012A&A...537A.146E} {537, A146}

\bibitem[\protect\citeauthoryear{{Gabriel}, {Noels}, {Montalb{\'a}n}  \&
  {Miglio}}{{Gabriel} et~al.}{2014}]{Gabriel14}
{Gabriel} M.,  {Noels} A.,  {Montalb{\'a}n} J.,   {Miglio} A.,  2014, \mn@doi
  [\aap] {10.1051/0004-6361/201423442}, \href
  {https://ui.adsabs.harvard.edu/abs/2014A&A...569A..63G} {569, A63}

\bibitem[\protect\citeauthoryear{{Gautschy} \& {Saio}}{{Gautschy} \&
  {Saio}}{1993}]{gau93}
{Gautschy} A.,  {Saio} H.,  1993, \mn@doi [\mnras] {10.1093/mnras/262.1.213},
  \href {https://ui.adsabs.harvard.edu/abs/1993MNRAS.262..213G} {262, 213}

\bibitem[\protect\citeauthoryear{{Herwig}}{{Herwig}}{2000}]{her00}
{Herwig} F.,  2000, \aap, \href
  {https://ui.adsabs.harvard.edu/abs/2000A&A...360..952H} {360, 952}

\bibitem[\protect\citeauthoryear{{Kahraman Ali{\c{c}}avu{\textcommabelow s}},
  {Poretti}, {Catanzaro}, {Smalley}, {Niemczura}, {Rainer}  \&
  {Handler}}{{Kahraman Ali{\c{c}}avu{\textcommabelow s}} et~al.}{2020}]{kah20}
{Kahraman Ali{\c{c}}avu{\textcommabelow s}} F.,  {Poretti} E.,  {Catanzaro} G.,
   {Smalley} B.,  {Niemczura} E.,  {Rainer} M.,   {Handler} G.,  2020, \mn@doi
  [\mnras] {10.1093/mnras/staa399}, \href
  {https://ui.adsabs.harvard.edu/abs/2020MNRAS.493.4518K} {493, 4518}

\bibitem[\protect\citeauthoryear{{Keen}, {Bedding}, {Murphy}, {Schmid},
  {Aerts}, {Tkachenko}, {Ouazzani}  \& {Kurtz}}{{Keen} et~al.}{2015}]{kee15}
{Keen} M.~A.,  {Bedding} T.~R.,  {Murphy} S.~J.,  {Schmid} V.~S.,  {Aerts} C.,
  {Tkachenko} A.,  {Ouazzani} R.~M.,   {Kurtz} D.~W.,  2015, \mn@doi [\mnras]
  {10.1093/mnras/stv2107}, \href
  {https://ui.adsabs.harvard.edu/abs/2015MNRAS.454.1792K} {454, 1792}

\bibitem[\protect\citeauthoryear{{Kurtz}, {Saio}, {Takata}, {Shibahashi},
  {Murphy}  \& {Sekii}}{{Kurtz} et~al.}{2014}]{kur14}
{Kurtz} D.~W.,  {Saio} H.,  {Takata} M.,  {Shibahashi} H.,  {Murphy} S.~J.,
  {Sekii} T.,  2014, \mn@doi [\mnras] {10.1093/mnras/stu1329}, \href
  {http://adsabs.harvard.edu/abs/2014MNRAS.444..102K} {444, 102}

\bibitem[\protect\citeauthoryear{{Lee}}{{Lee}}{1988}]{lee88}
{Lee} U.,  1988, \mn@doi [\mnras] {10.1093/mnras/232.4.711}, \href
  {https://ui.adsabs.harvard.edu/abs/1988MNRAS.232..711L} {232, 711}

\bibitem[\protect\citeauthoryear{{Lee}}{{Lee}}{2006}]{lee06}
{Lee} U.,  2006, \mn@doi [\mnras] {10.1111/j.1365-2966.2005.09751.x}, \href
  {http://adsabs.harvard.edu/abs/2006MNRAS.365..677L} {365, 677}

\bibitem[\protect\citeauthoryear{{Lee} \& {Baraffe}}{{Lee} \&
  {Baraffe}}{1995}]{lee95}
{Lee} U.,  {Baraffe} I.,  1995, \aap, \href
  {http://adsabs.harvard.edu/abs/1995A%26A...301..419L} {301, 419}

\bibitem[\protect\citeauthoryear{{Lee} \& {Saio}}{{Lee} \&
  {Saio}}{1997}]{lee97}
{Lee} U.,  {Saio} H.,  1997, \apj, \href
  {http://adsabs.harvard.edu/abs/1997ApJ...491..839L} {491, 839}

\bibitem[\protect\citeauthoryear{{Lee} \& {Saio}}{{Lee} \&
  {Saio}}{2020}]{lee20}
{Lee} U.,  {Saio} H.,  2020, \mn@doi [\mnras] {10.1093/mnras/staa2250}, \href
  {https://ui.adsabs.harvard.edu/abs/2020MNRAS.tmp.2369L} {}

\bibitem[\protect\citeauthoryear{{Li}, {Bedding}, {Murphy}, {Van Reeth},
  {Antoci}  \& {Ouazzani}}{{Li} et~al.}{2019a}]{LiG19a}
{Li} G.,  {Bedding} T.~R.,  {Murphy} S.~J.,  {Van Reeth} T.,  {Antoci} V.,
  {Ouazzani} R.-M.,  2019a, \mn@doi [\mnras] {10.1093/mnras/sty2743}, \href
  {https://ui.adsabs.harvard.edu/abs/2019MNRAS.482.1757L} {482, 1757}

\bibitem[\protect\citeauthoryear{{Li}, {Van Reeth}, {Bedding}, {Murphy}  \&
  {Antoci}}{{Li} et~al.}{2019b}]{LiG19}
{Li} G.,  {Van Reeth} T.,  {Bedding} T.~R.,  {Murphy} S.~J.,   {Antoci} V.,
  2019b, \mn@doi [\mnras] {10.1093/mnras/stz1171}, \href
  {https://ui.adsabs.harvard.edu/abs/2019MNRAS.487..782L} {487, 782}

\bibitem[\protect\citeauthoryear{{Li}, {Van Reeth}, {Bedding}, {Murphy},
  {Antoci}, {Ouazzani}  \& {Barbara}}{{Li} et~al.}{2020a}]{LiG20}
{Li} G.,  {Van Reeth} T.,  {Bedding} T.~R.,  {Murphy} S.~J.,  {Antoci} V.,
  {Ouazzani} R.-M.,   {Barbara} N.~H.,  2020a, \mn@doi [\mnras]
  {10.1093/mnras/stz2906}, \href
  {https://ui.adsabs.harvard.edu/abs/2020MNRAS.491.3586L} {491, 3586}

\bibitem[\protect\citeauthoryear{{Li}, {Guo}, {Fuller}, {Bedding}, {Murphy},
  {Colman}  \& {Hey}}{{Li} et~al.}{2020b}]{LiG20EB}
{Li} G.,  {Guo} Z.,  {Fuller} J.,  {Bedding} T.~R.,  {Murphy} S.~J.,  {Colman}
  I.~L.,   {Hey} D.~R.,  2020b, \mn@doi [\mnras] {10.1093/mnras/staa2266},
  \href {https://ui.adsabs.harvard.edu/abs/2020MNRAS.497.4363L} {497, 4363}

\bibitem[\protect\citeauthoryear{{Miglio}, {Montalb{\'a}n}, {Noels}  \&
  {Eggenberger}}{{Miglio} et~al.}{2008}]{mig08}
{Miglio} A.,  {Montalb{\'a}n} J.,  {Noels} A.,   {Eggenberger} P.,  2008,
  \mn@doi [\mnras] {10.1111/j.1365-2966.2008.13112.x}, \href
  {http://adsabs.harvard.edu/abs/2008MNRAS.386.1487M} {386, 1487}

\bibitem[\protect\citeauthoryear{{Mombarg}, {Van Reeth}, {Pedersen},
  {Molenberghs}, {Bowman}, {Johnston}, {Tkachenko}  \& {Aerts}}{{Mombarg}
  et~al.}{2019}]{mom19}
{Mombarg} J.~S.~G.,  {Van Reeth} T.,  {Pedersen} M.~G.,  {Molenberghs} G.,
  {Bowman} D.~M.,  {Johnston} C.,  {Tkachenko} A.,   {Aerts} C.,  2019, \mn@doi
  [\mnras] {10.1093/mnras/stz501}, \href
  {https://ui.adsabs.harvard.edu/abs/2019MNRAS.485.3248M} {485, 3248}

\bibitem[\protect\citeauthoryear{{Mosser} et~al.,}{{Mosser}
  et~al.}{2012}]{mos12}
{Mosser} B.,  et~al., 2012, \mn@doi [\aap] {10.1051/0004-6361/201118519}, \href
  {https://ui.adsabs.harvard.edu/abs/2012A&A...540A.143M} {540, A143}

\bibitem[\protect\citeauthoryear{{Murphy}, {Fossati}, {Bedding}, {Saio},
  {Kurtz}, {Grassitelli}  \& {Wang}}{{Murphy} et~al.}{2016}]{mur16}
{Murphy} S.~J.,  {Fossati} L.,  {Bedding} T.~R.,  {Saio} H.,  {Kurtz} D.~W.,
  {Grassitelli} L.,   {Wang} E.~S.,  2016, \mn@doi [\mnras]
  {10.1093/mnras/stw705}, \href
  {http://adsabs.harvard.edu/abs/2016MNRAS.459.1201M} {459, 1201}

\bibitem[\protect\citeauthoryear{{Murphy}, {Hey}, {Van Reeth}  \&
  {Bedding}}{{Murphy} et~al.}{2019}]{mur19}
{Murphy} S.~J.,  {Hey} D.,  {Van Reeth} T.,   {Bedding} T.~R.,  2019, \mn@doi
  [\mnras] {10.1093/mnras/stz590}, \href
  {https://ui.adsabs.harvard.edu/abs/2019MNRAS.485.2380M} {485, 2380}

\bibitem[\protect\citeauthoryear{{Ouazzani}, {Salmon}, {Antoci}, {Bedding},
  {Murphy}  \& {Roxburgh}}{{Ouazzani} et~al.}{2017}]{oua17}
{Ouazzani} R.-M.,  {Salmon} S.~J.~A.~J.,  {Antoci} V.,  {Bedding} T.~R.,
  {Murphy} S.~J.,   {Roxburgh} I.~W.,  2017, \mn@doi [\mnras]
  {10.1093/mnras/stw2717}, \href
  {http://adsabs.harvard.edu/abs/2017MNRAS.465.2294O} {465, 2294}

\bibitem[\protect\citeauthoryear{{Ouazzani}, {Ligni{\`e}res}, {Dupret},
  {Salmon}, {Ballot}, {Christophe}  \& {Takata}}{{Ouazzani}
  et~al.}{2020}]{oua20}
{Ouazzani} R.-M.,  {Ligni{\`e}res} F.,  {Dupret} M.-A.,  {Salmon} S.~J.~A.~J.,
  {Ballot} J.,  {Christophe} S.,   {Takata} M.,  2020, arXiv e-prints, \href
  {https://ui.adsabs.harvard.edu/abs/2020arXiv200609404O} {p. arXiv:2006.09404}

\bibitem[\protect\citeauthoryear{{P{\'a}pics}, {Tkachenko}, {Aerts}, {Van
  Reeth}, {De Smedt}, {Hillen}, {{\O}stensen}  \& {Moravveji}}{{P{\'a}pics}
  et~al.}{2015}]{pap15}
{P{\'a}pics} P.~I.,  {Tkachenko} A.,  {Aerts} C.,  {Van Reeth} T.,  {De Smedt}
  K.,  {Hillen} M.,  {{\O}stensen} R.,   {Moravveji} E.,  2015, \mn@doi [\apjl]
  {10.1088/2041-8205/803/2/L25}, \href
  {http://adsabs.harvard.edu/abs/2015ApJ...803L..25P} {803, L25}

\bibitem[\protect\citeauthoryear{{P{\'a}pics} et~al.,}{{P{\'a}pics}
  et~al.}{2017}]{pap17}
{P{\'a}pics} P.~I.,  et~al., 2017, \mn@doi [\aap]
  {10.1051/0004-6361/201629814}, \href
  {http://adsabs.harvard.edu/abs/2017A%26A...598A..74P} {598, A74}

\bibitem[\protect\citeauthoryear{{Paxton}, {Bildsten}, {Dotter}, {Herwig},
  {Lesaffre}  \& {Timmes}}{{Paxton} et~al.}{2011}]{pax11}
{Paxton} B.,  {Bildsten} L.,  {Dotter} A.,  {Herwig} F.,  {Lesaffre} P.,
  {Timmes} F.,  2011, \mn@doi [\apjs] {10.1088/0067-0049/192/1/3}, \href
  {https://ui.adsabs.harvard.edu/abs/2011ApJS..192....3P} {192, 3}

\bibitem[\protect\citeauthoryear{{Paxton} et~al.,}{{Paxton}
  et~al.}{2013}]{pax13}
{Paxton} B.,  et~al., 2013, \mn@doi [\apjs] {10.1088/0067-0049/208/1/4}, \href
  {http://adsabs.harvard.edu/abs/2013ApJS..208....4P} {208, 4}

\bibitem[\protect\citeauthoryear{{Paxton} et~al.,}{{Paxton}
  et~al.}{2015}]{pax15}
{Paxton} B.,  et~al., 2015, \mn@doi [\apjs] {10.1088/0067-0049/220/1/15}, \href
  {https://ui.adsabs.harvard.edu/abs/2015ApJS..220...15P} {220, 15}

\bibitem[\protect\citeauthoryear{{Saio}, {Kurtz}, {Takata}, {Shibahashi},
  {Murphy}, {Sekii}  \& {Bedding}}{{Saio} et~al.}{2015}]{sai15}
{Saio} H.,  {Kurtz} D.~W.,  {Takata} M.,  {Shibahashi} H.,  {Murphy} S.~J.,
  {Sekii} T.,   {Bedding} T.~R.,  2015, \mn@doi [\mnras]
  {10.1093/mnras/stu2696}, \href
  {http://adsabs.harvard.edu/abs/2015MNRAS.447.3264S} {447, 3264}

\bibitem[\protect\citeauthoryear{{Saio}, {Kurtz}, {Murphy}, {Antoci}  \&
  {Lee}}{{Saio} et~al.}{2018a}]{sai18}
{Saio} H.,  {Kurtz} D.~W.,  {Murphy} S.~J.,  {Antoci} V.~L.,   {Lee} U.,
  2018a, \mn@doi [\mnras] {10.1093/mnras/stx2962}, \href
  {http://adsabs.harvard.edu/abs/2018MNRAS.474.2774S} {474, 2774}

\bibitem[\protect\citeauthoryear{{Saio}, {Bedding}, {Kurtz}, {Murphy},
  {Antoci}, {Shibahashi}, {Li}  \& {Takata}}{{Saio} et~al.}{2018b}]{sai18b}
{Saio} H.,  {Bedding} T.~R.,  {Kurtz} D.~W.,  {Murphy} S.~J.,  {Antoci} V.,
  {Shibahashi} H.,  {Li} G.,   {Takata} M.,  2018b, \mn@doi [\mnras]
  {10.1093/mnras/sty784}, \href
  {http://adsabs.harvard.edu/abs/2018MNRAS.477.2183S} {477, 2183}

\bibitem[\protect\citeauthoryear{{Savonije}}{{Savonije}}{2005}]{sav05}
{Savonije} G.~J.,  2005, \mn@doi [\aap] {10.1051/0004-6361:20053328}, \href
  {http://adsabs.harvard.edu/abs/2005A%26A...443..557S} {443, 557}

\bibitem[\protect\citeauthoryear{{Schmid} \& {Aerts}}{{Schmid} \&
  {Aerts}}{2016}]{sch16}
{Schmid} V.~S.,  {Aerts} C.,  2016, \mn@doi [\aap]
  {10.1051/0004-6361/201628617}, \href
  {https://ui.adsabs.harvard.edu/abs/2016A&A...592A.116S} {592, A116}

\bibitem[\protect\citeauthoryear{{Schmid} et~al.,}{{Schmid}
  et~al.}{2015}]{sch15}
{Schmid} V.~S.,  et~al., 2015, \mn@doi [\aap] {10.1051/0004-6361/201526945},
  \href {https://ui.adsabs.harvard.edu/abs/2015A&A...584A..35S} {584, A35}

\bibitem[\protect\citeauthoryear{{Takata}, {Ouazzani}, {Saio}, {Christophe},
  {Ballot}, {Antoci}, {Salmon}  \& {Hijikawa}}{{Takata} et~al.}{2020}]{tak20}
{Takata} M.,  {Ouazzani} R.~M.,  {Saio} H.,  {Christophe} S.,  {Ballot} J.,
  {Antoci} V.,  {Salmon} S.~J.~A.~J.,   {Hijikawa} K.,  2020, \mn@doi [\aap]
  {10.1051/0004-6361/201936297}, \href
  {https://ui.adsabs.harvard.edu/abs/2020A&A...635A.106T} {635, A106}

\bibitem[\protect\citeauthoryear{{Townsend}}{{Townsend}}{2003}]{tow03a}
{Townsend} R.~H.~D.,  2003, \mn@doi [\mnras]
  {10.1046/j.1365-8711.2003.06379.x}, \href
  {http://adsabs.harvard.edu/abs/2003MNRAS.340.1020T} {340, 1020}

\bibitem[\protect\citeauthoryear{{Townsend}}{{Townsend}}{2005}]{tow05}
{Townsend} R.~H.~D.,  2005, \mn@doi [\mnras]
  {10.1111/j.1365-2966.2005.09585.x}, \href
  {http://adsabs.harvard.edu/abs/2005MNRAS.364..573T} {364, 573}

\bibitem[\protect\citeauthoryear{{Triana}, {Moravveji}, {P{\'a}pics}, {Aerts},
  {Kawaler}  \& {Christensen-Dalsgaard}}{{Triana} et~al.}{2015}]{tri15}
{Triana} S.~A.,  {Moravveji} E.,  {P{\'a}pics} P.~I.,  {Aerts} C.,  {Kawaler}
  S.~D.,   {Christensen-Dalsgaard} J.,  2015, \mn@doi [\apj]
  {10.1088/0004-637X/810/1/16}, \href
  {https://ui.adsabs.harvard.edu/abs/2015ApJ...810...16T} {810, 16}

\bibitem[\protect\citeauthoryear{{Unno}, {Osaki}, {Ando}, {Saio}  \&
  {Shibahashi}}{{Unno} et~al.}{1989}]{unno}
{Unno} W.,  {Osaki} Y.,  {Ando} H.,  {Saio} H.,   {Shibahashi} H.,  1989,
  {Nonradial oscillations of stars}

\bibitem[\protect\citeauthoryear{{Van Reeth} et~al.,}{{Van Reeth}
  et~al.}{2015a}]{vanr15apjs}
{Van Reeth} T.,  et~al., 2015a, \mn@doi [\apjs] {10.1088/0067-0049/218/2/27},
  \href {http://adsabs.harvard.edu/abs/2015ApJS..218...27V} {218, 27}

\bibitem[\protect\citeauthoryear{{Van Reeth} et~al.,}{{Van Reeth}
  et~al.}{2015b}]{vanr15}
{Van Reeth} T.,  et~al., 2015b, \mn@doi [\aap] {10.1051/0004-6361/201424585},
  \href {http://adsabs.harvard.edu/abs/2015A%26A...574A..17V} {574, A17}

\bibitem[\protect\citeauthoryear{{Van Reeth}, {Tkachenko}  \& {Aerts}}{{Van
  Reeth} et~al.}{2016}]{vanr16}
{Van Reeth} T.,  {Tkachenko} A.,   {Aerts} C.,  2016, \mn@doi [\aap]
  {10.1051/0004-6361/201628616}, \href
  {http://adsabs.harvard.edu/abs/2016A%26A...593A.120V} {593, A120}

\bibitem[\protect\citeauthoryear{{Van Reeth} et~al.,}{{Van Reeth}
  et~al.}{2018}]{vanr18}
{Van Reeth} T.,  et~al., 2018, \mn@doi [\aap] {10.1051/0004-6361/201832718},
  \href {https://ui.adsabs.harvard.edu/abs/2018A&A...618A..24V} {618, A24}

\bibitem[\protect\citeauthoryear{{Zahn}}{{Zahn}}{2002}]{zahn02}
{Zahn} J.~P.,  2002, in {Aerts} C.,  {Bedding} T.~R.,   {Christensen-Dalsgaard}
  J.,  eds,  Astronomical Society of the Pacific Conference Series Vol. 259,
  IAU Colloq. 185: Radial and Nonradial Pulsationsn as Probes of Stellar
  Physics. p.~58

\bibitem[\protect\citeauthoryear{{Zwintz} et~al.,}{{Zwintz}
  et~al.}{2017}]{zwi17}
{Zwintz} K.,  et~al., 2017, \mn@doi [\aap] {10.1051/0004-6361/201731784}, \href
  {http://adsabs.harvard.edu/abs/2017A%26A...608A.103Z} {608, A103}

\makeatother
\end{thebibliography}



\appendix
\section{KIC 1431379: A large dip caused by chemical composition gradient}
\label{sec:bigdip}
\begin{figure}
\includegraphics[width=0.49\textwidth]{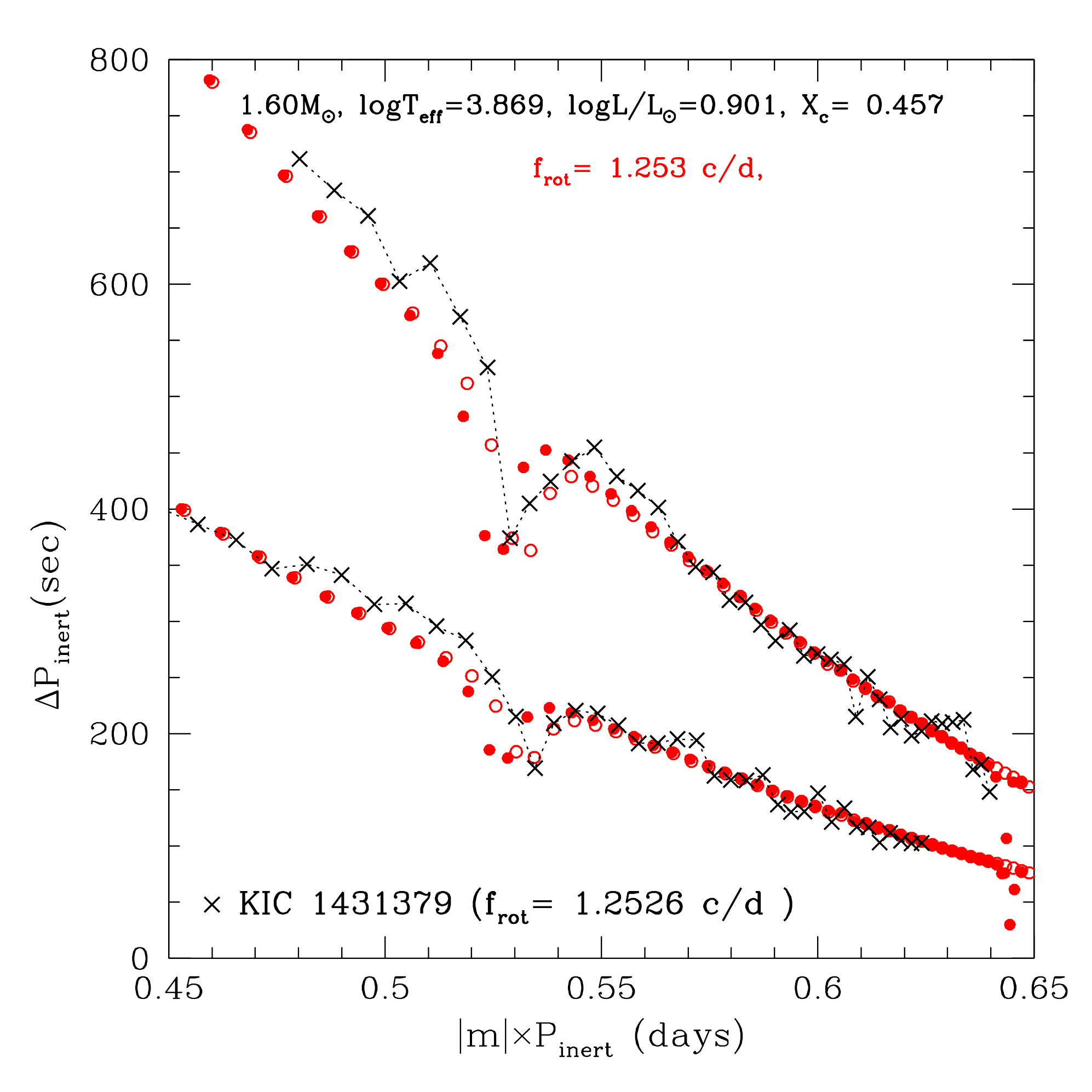}
\caption{Period-$\Delta P$ diagram for KIC 1431379 \citep[crosses][]{LiG19}. The large dip is produced by the stratification of the hydrogen abundance surrounding the convective core. Open and filled circles present $P_{\rm inert}$\,-\,$\Delta P$ computed for a $1.6\,M_\odot$ model by using TAR and expansion method, respectively. To include $m=-2$ modes, $|m|P_{\rm inert}$ is adopted horizontal axis. }
\label{fig:k1431}
\end{figure}

This section presents a case, in which observed large dips are fitted with theoretical dips caused by chemical composition gradients.  
Fig.\,\ref{fig:k1431} shows $P-\Delta P$ relations of KIC~1431379 compared with prograde sectoral $m=-1$ and $-2$ g modes of a 1.6-$M_\odot$ model uniformly rotating at 1.253~d$^{-1}$ \citep{LiG19} without overshooting. Because, the $P-\Delta P$ relations  obtained by the expansion method (filled circles) behave similarly to those obtained with the TAR (open circles), large dips for both $m=-1$ and $-2$ relations  should be caused by a steep gradient of chemical composition \citep{mig08,bou13}.
This model predicts a dip caused by the resonance coupling with a core inertial mode to appear in the $m=-1$ sequence at a period of about 0.65 days, which is just beyond the longest period detected in KIC~1431379.
Although dips caused by chemical composition gradients are sensitive to evolutionary stages, we could also find a OS02 model with overshooting $h_{\rm os}=0.02$ (not shown) which fit the dips similarly well as in Fig.\,\ref{fig:k1431}.

\section{Other $\gamma$ Dor stars}\label{sec:otherfits}
\begin{figure*}
\includegraphics[width=0.33\textwidth]{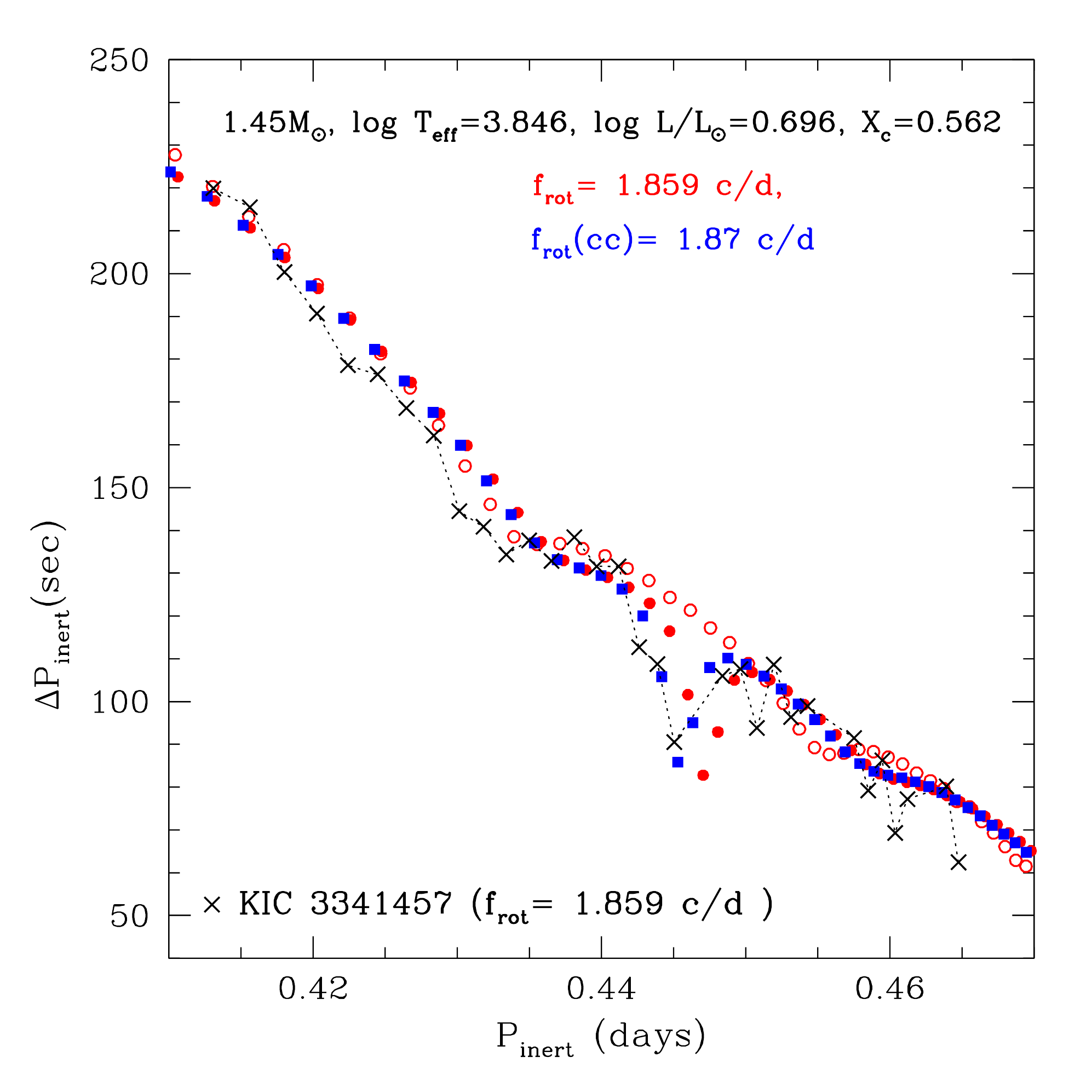}
\includegraphics[width=0.33\textwidth]{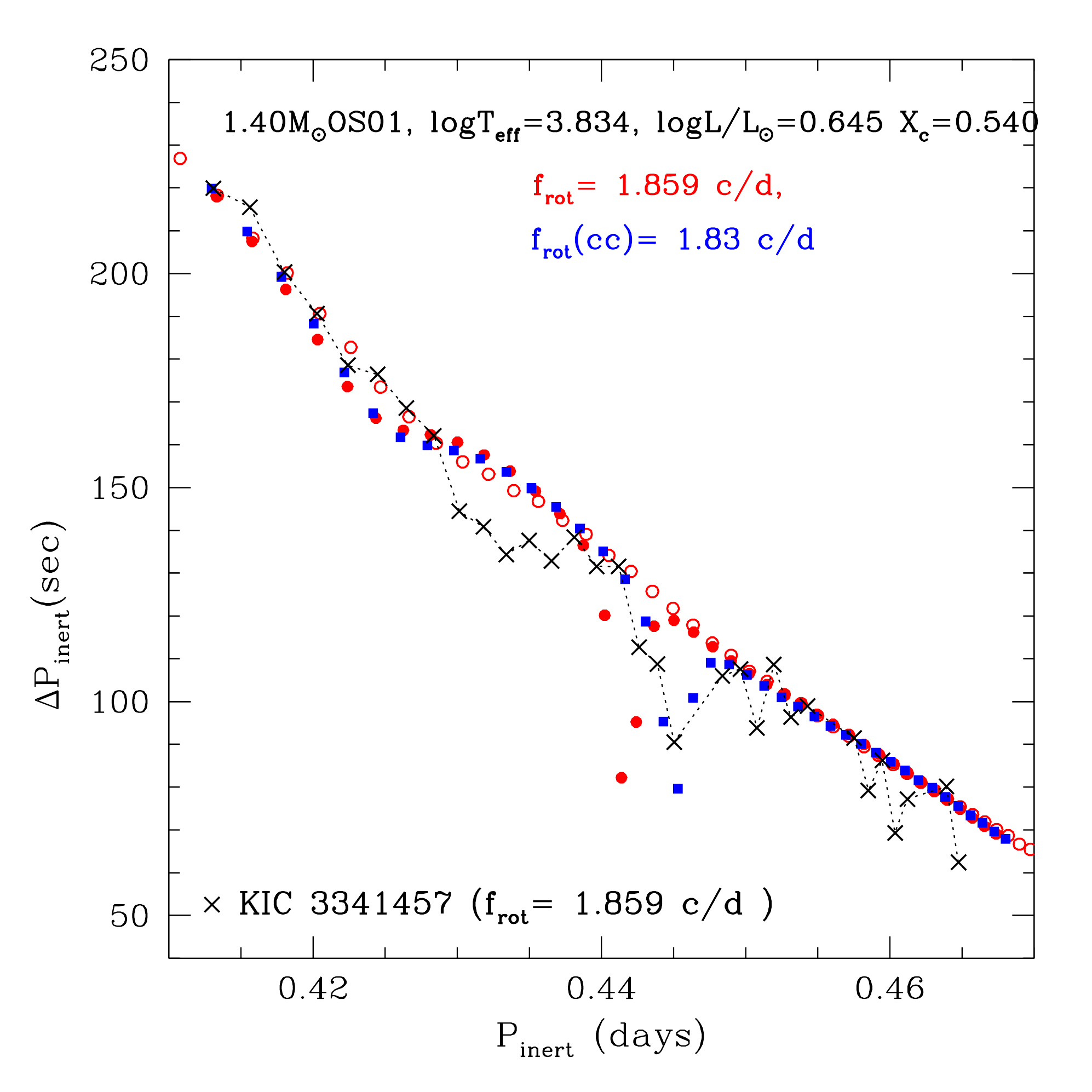}
\includegraphics[width=0.33\textwidth]{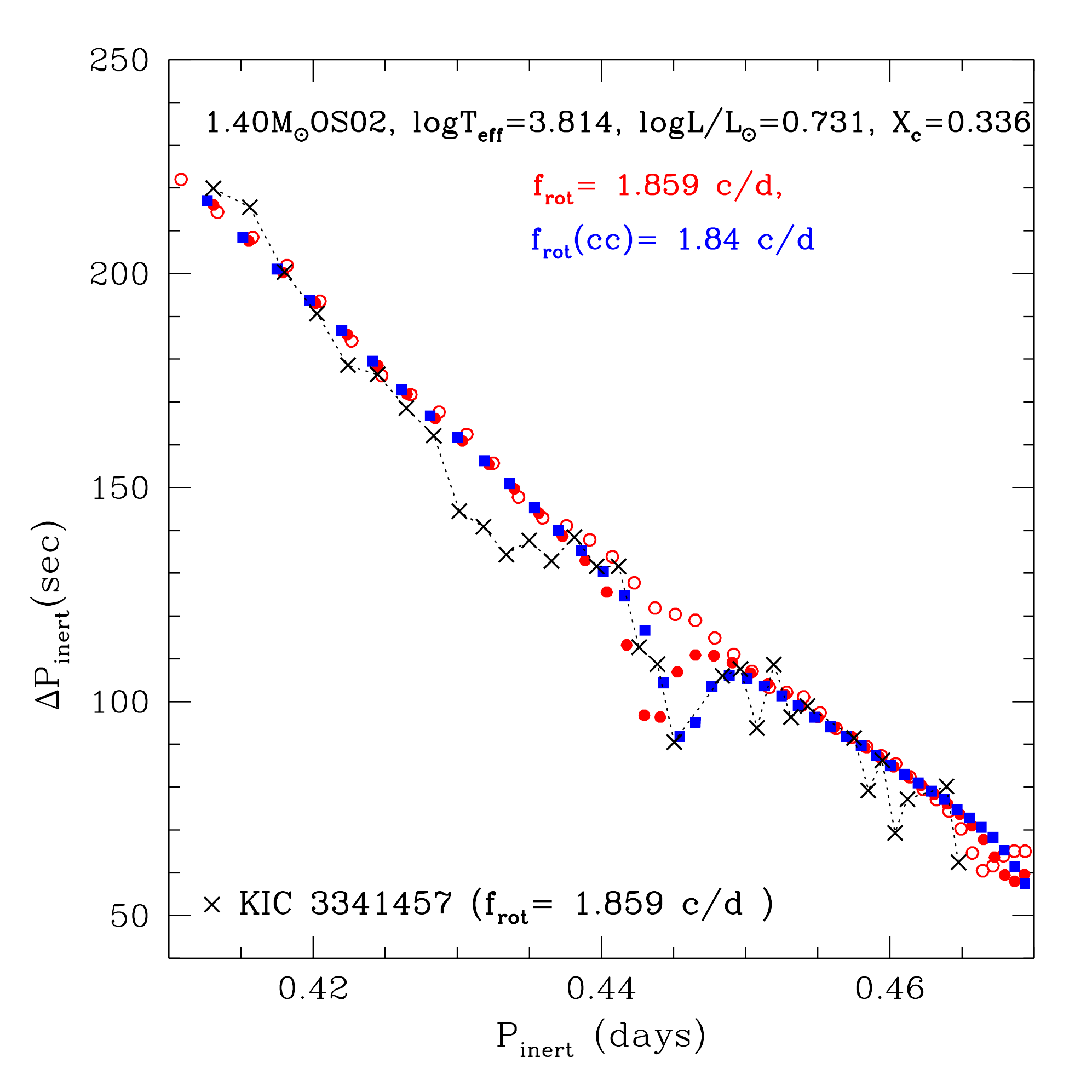}
\caption{KIC~3341457 is an eclipsing binary with an orbital frequency of $1.893$\,d$^{-1}$. \citet{LiG20EB} obtained $f_{\rm rot}=1.859 \pm 0.001$\,d$^{-1}$   from the $P$\,-\,$\Delta P$ patterns of prograde dipole g modes and of r modes. A dip in the g-mode $P$\,-\,$\Delta P$ sequence (crosses) at $0.445$\,days can be reproduced assuming $f_{\rm rot}$(cc) slightly different from the $f_{\rm rot}$ of KIC~3341457 irrespective of overshooting assumptions, indicating that rotation throughout the stellar interior is synchronized with the orbital motion in this short period eclipsing binary. 
The model without overshooting (left panel) also reproduce roughly the broad modulation in the period range $0.43 - 0.44$\,days.}
\label{fig:k334}
\end{figure*}

\begin{figure*}
\includegraphics[width=0.33\textwidth]{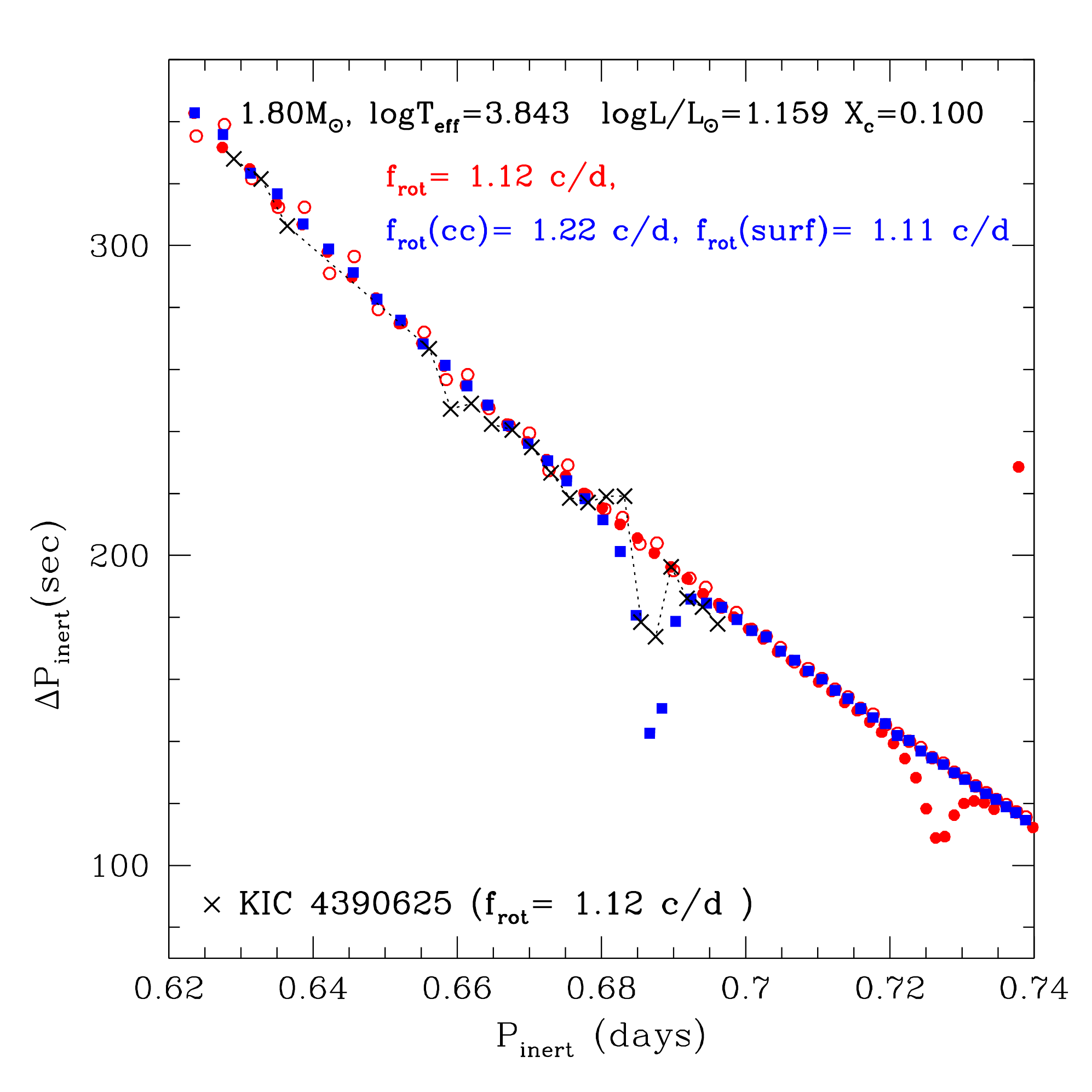}
\includegraphics[width=0.33\textwidth]{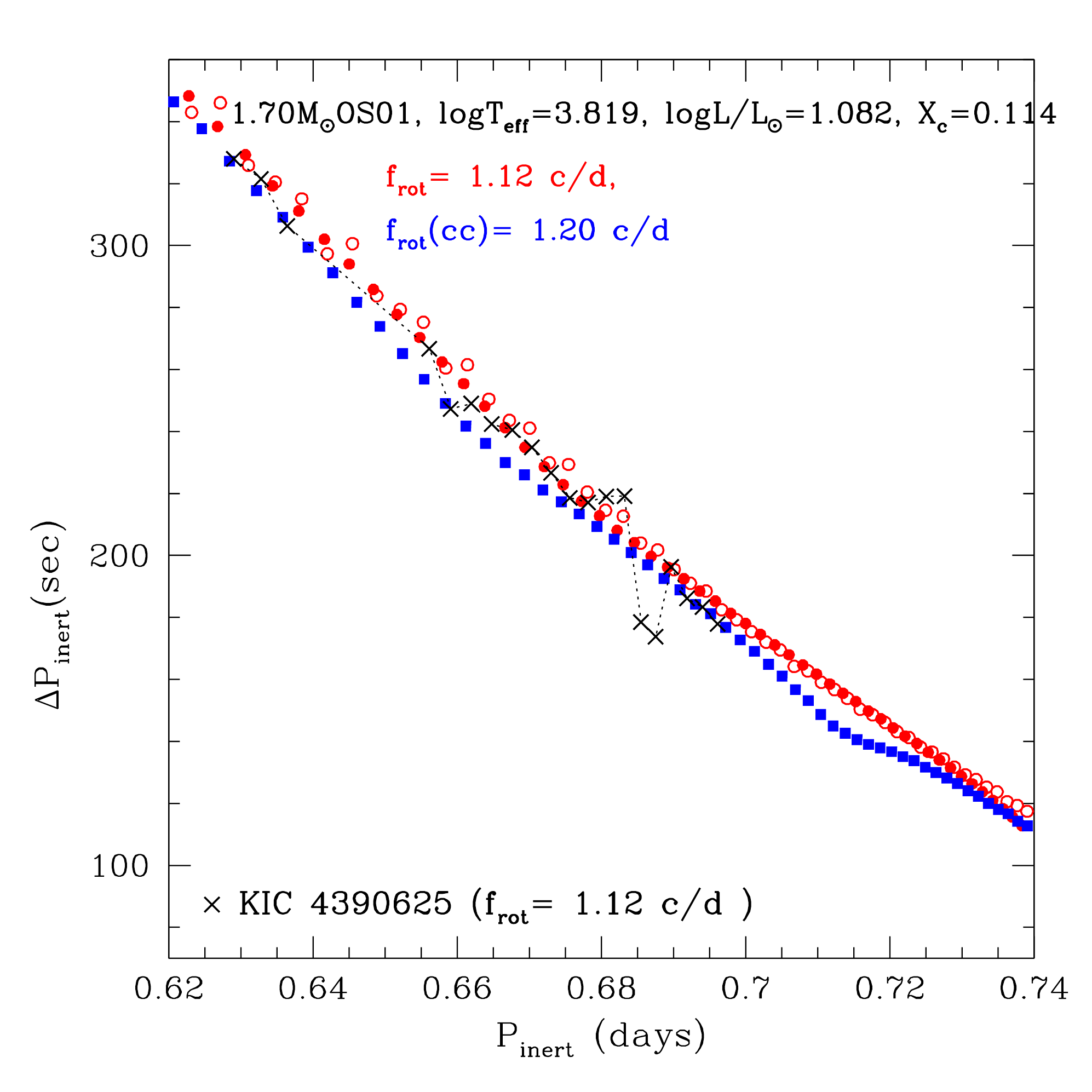}
\includegraphics[width=0.33\textwidth]{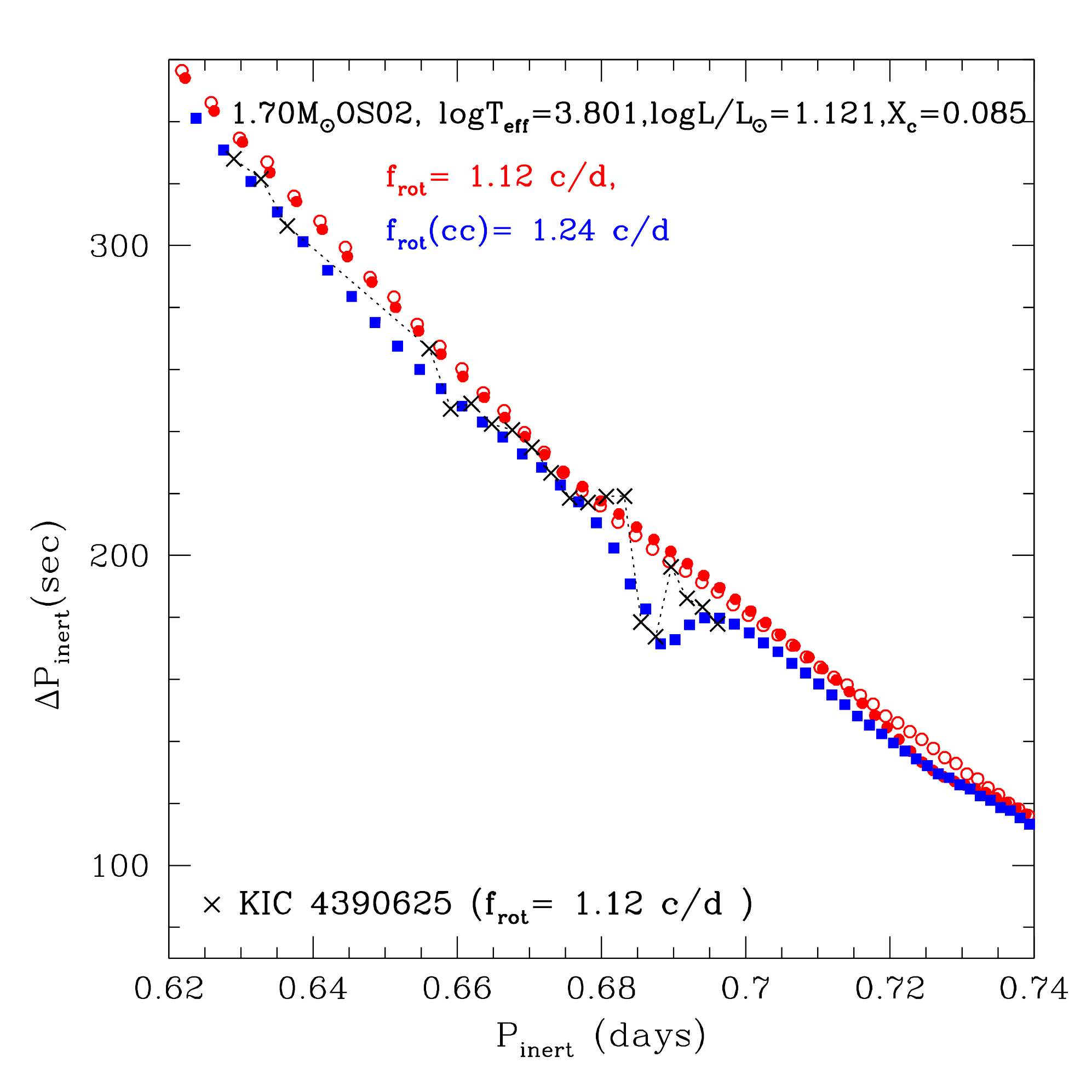}
\caption{KIC~4390625: \citet{LiG20} obtained a near-core (g-cavity) rotation $f_{\rm rot}=1.12 \pm 0.01$\,d$^{-1}$ from the $P$\,-\,$\Delta P$ patterns of prograde $m=-1,-2$ g modes. The $P$\,-\,$\Delta P$ relation of the dipole g modes of KIC~4390625 (crosses) has a dip at a period of 0.686\,days, which is attributable to couplings with the fundamental inertial mode in the convective core. The best-fit model (blue squares) without overshooting (left panel) assumes that rotation frequencies at the surface and in the convective-core are 1.11\,d$^{-1}$ and 1.22\,d$^{-1}$, respectively. To fit the $P$\,-\,$\Delta P$ pattern of KIC~4390625 with an OS01 model (middle panel), the model should have a central hydrogen mass fraction of about 0.1. However, such an evolved model has no narrow-resonance dip (see Fig.\,\ref{fig:pdp_overshoot}). Therefore, no OS01 model can fit the resonance dip of KIC~4390625, while an OS02 model (right panel) can fit the resonance dip assuming that the convective core rotates at 1.24~d$^{-1}$, although the predicted dip is slightly broader than observed.}
\label{fig:k439}
\end{figure*}

\begin{figure*}
\includegraphics[width=0.33\textwidth]{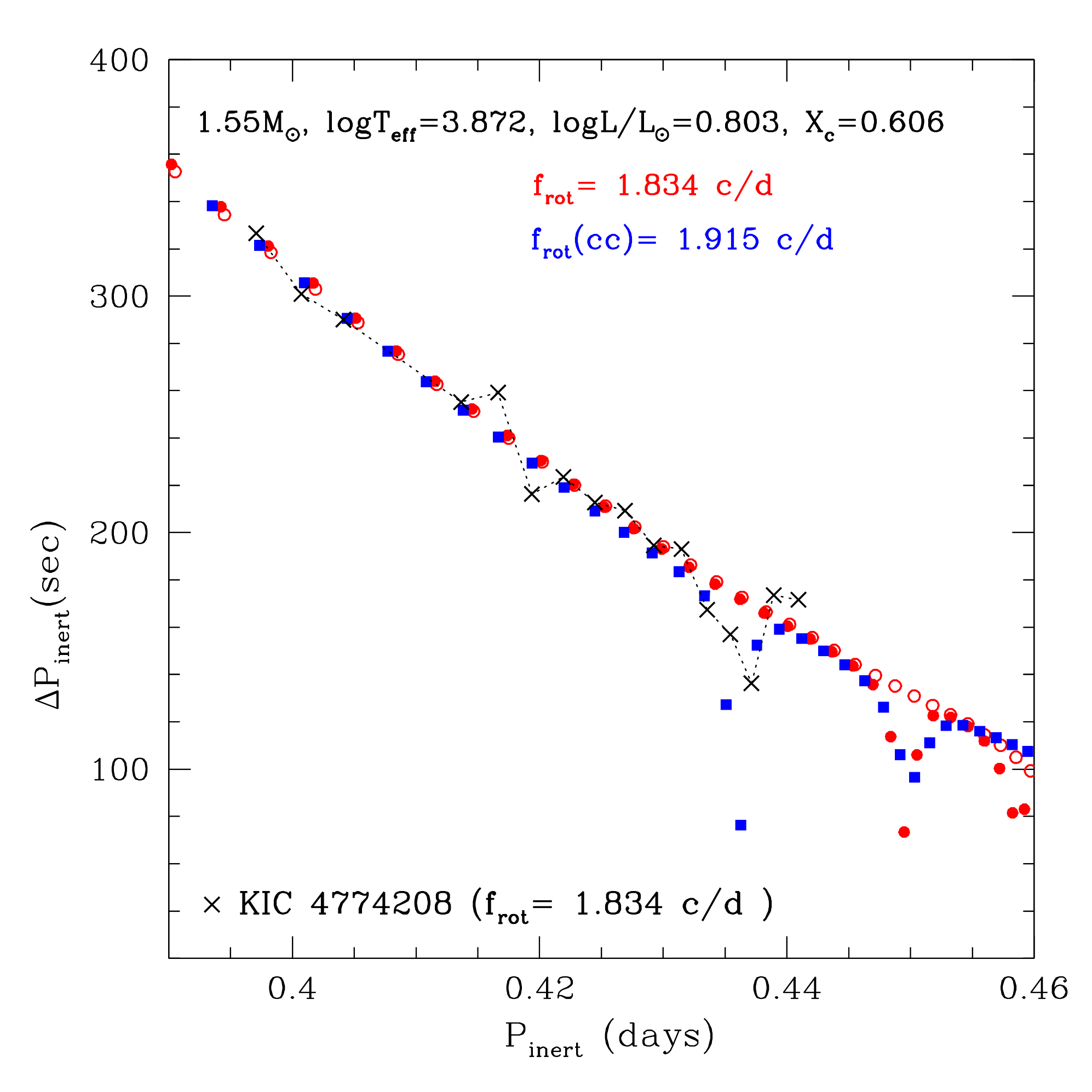}
\includegraphics[width=0.33\textwidth]{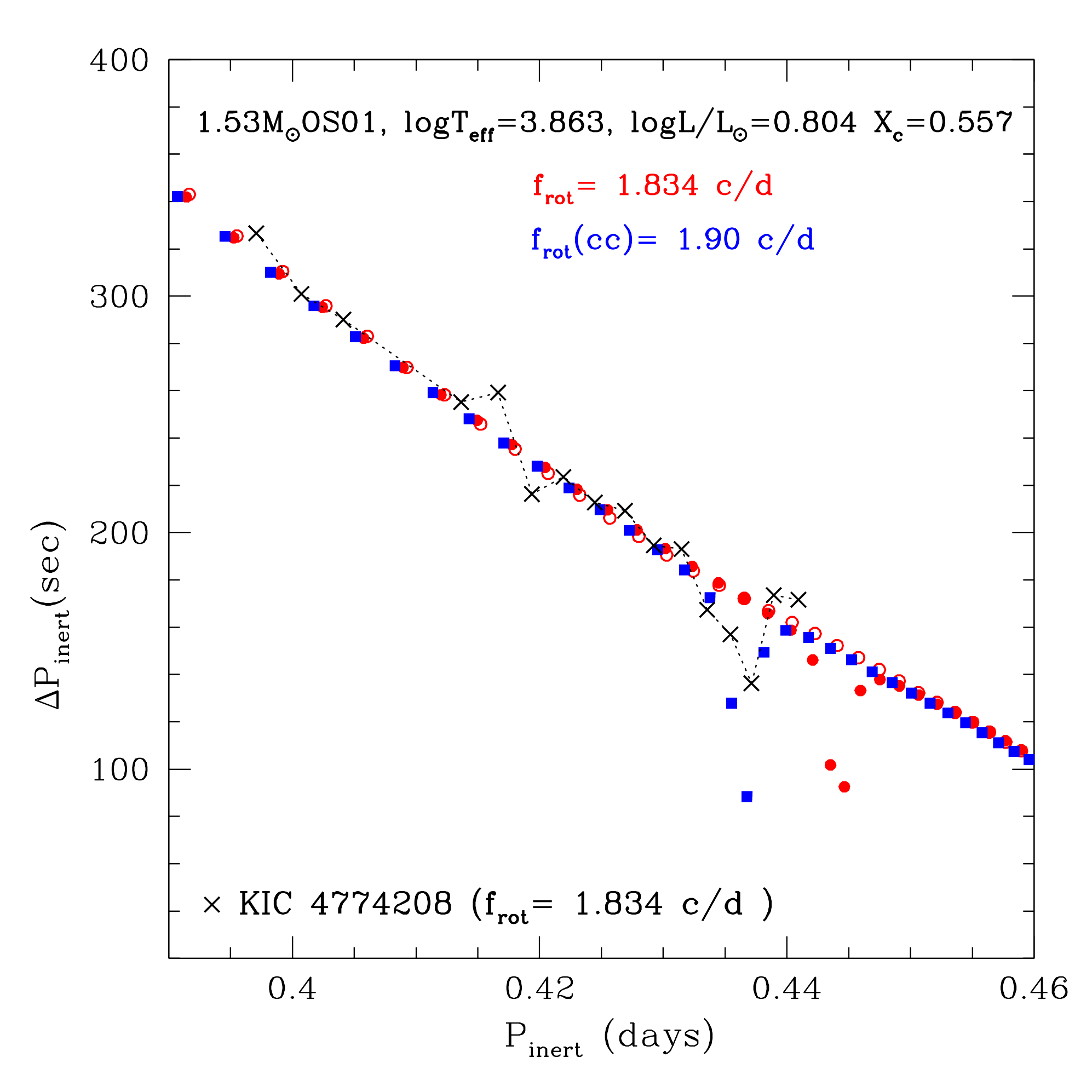}
\includegraphics[width=0.33\textwidth]{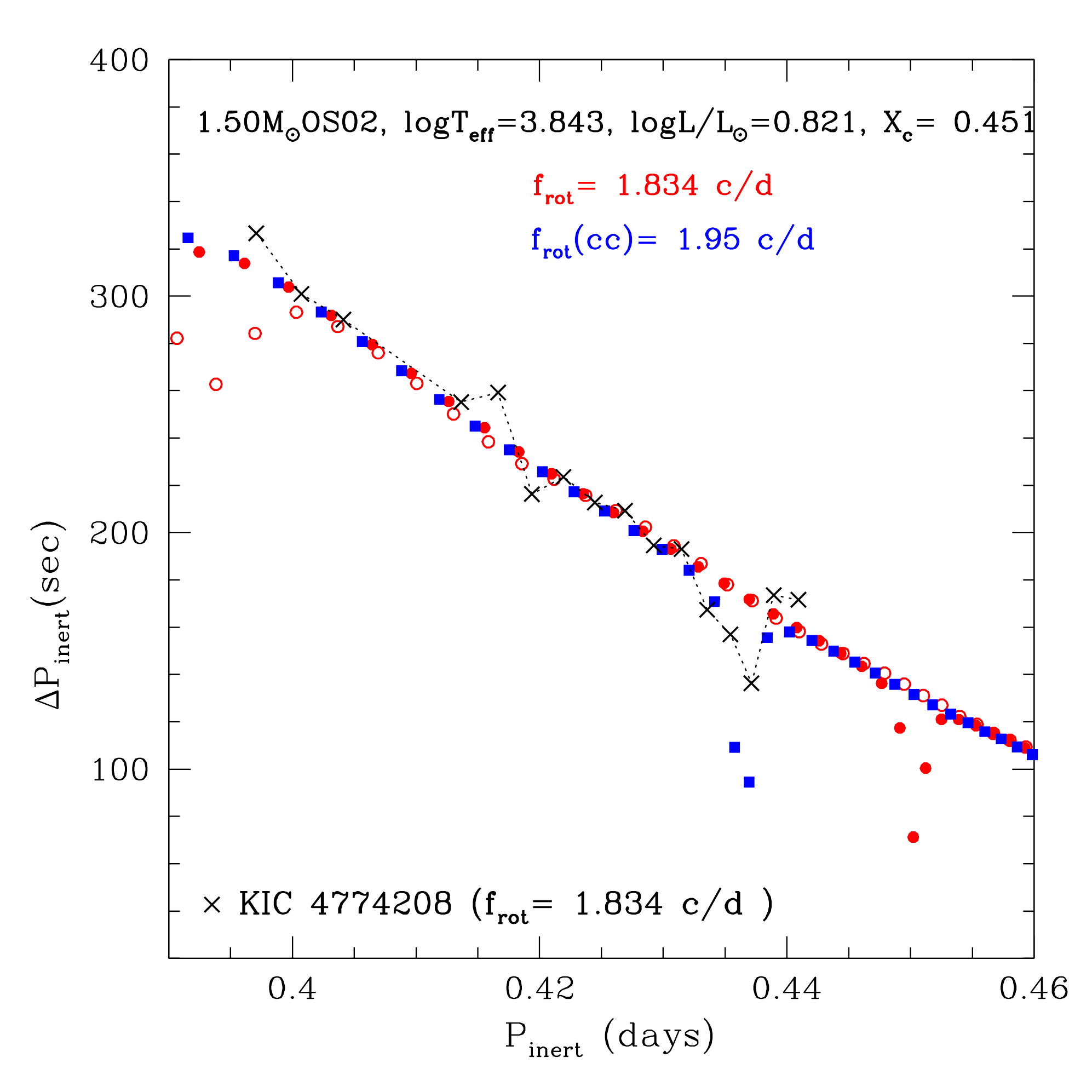}
\caption{KIC 4774208: \citet{LiG19} obtained $f_{\rm rot}= 1.834\pm0.001$\,d$^{-1}$ from the $P$\,-\,$\Delta P$ relations of prograde dipole g modes and of r modes. The g-mode $P$\,-\,$\Delta P$ relation has a dip at $\sim0.435$\,days attributable to resonance couplings with the fundamental inertial mode of the convective core. Each model with or without core overshooting can fit the period at the resonance  dip by assuming a rotation rate of the convective core slightly faster than that of the g-mode cavity, while the predicted dips are considerably deeper than observed.}
\label{fig:k477}
\end{figure*}

\begin{figure*}
\includegraphics[width=0.33\textwidth]{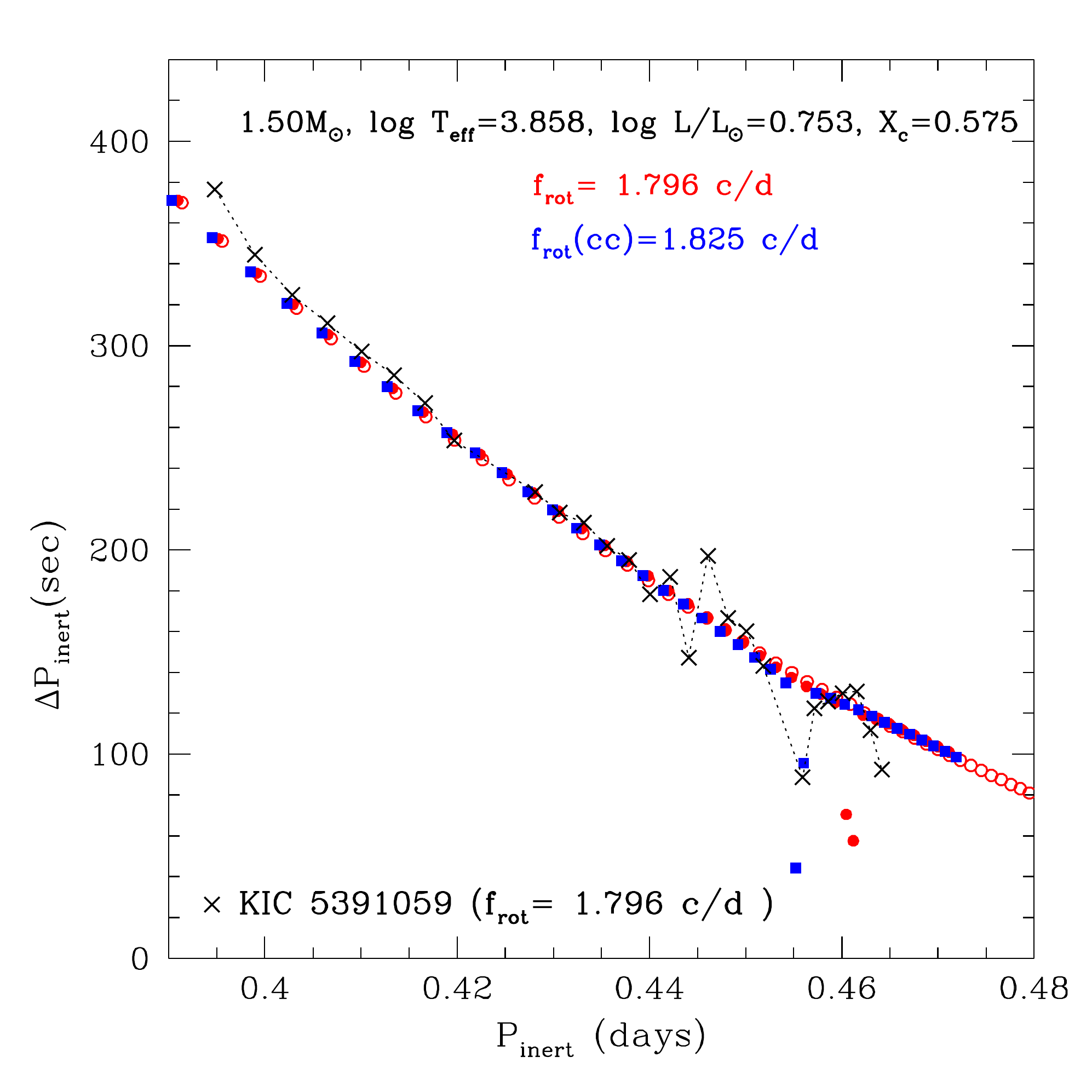}
\includegraphics[width=0.33\textwidth]{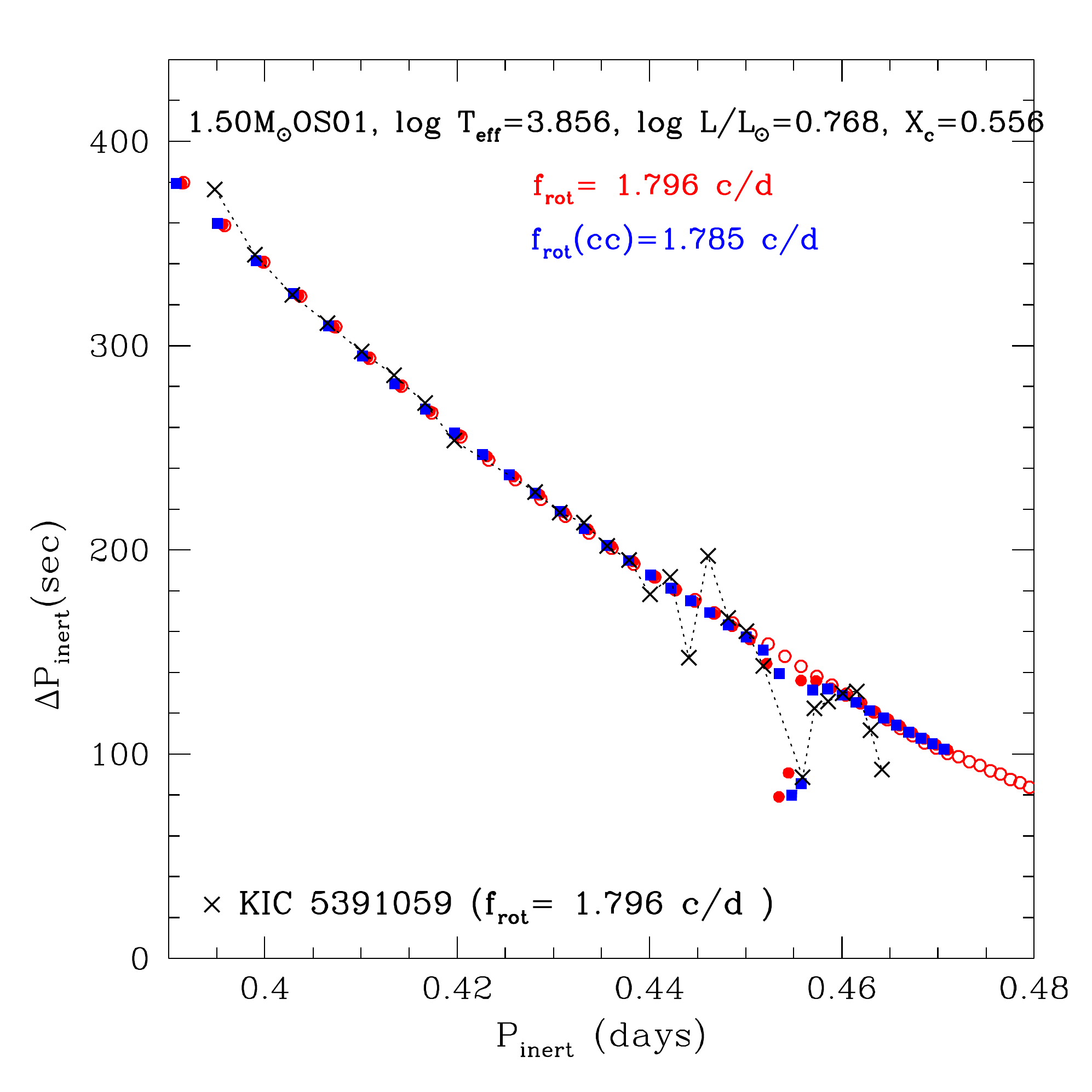}
\includegraphics[width=0.33\textwidth]{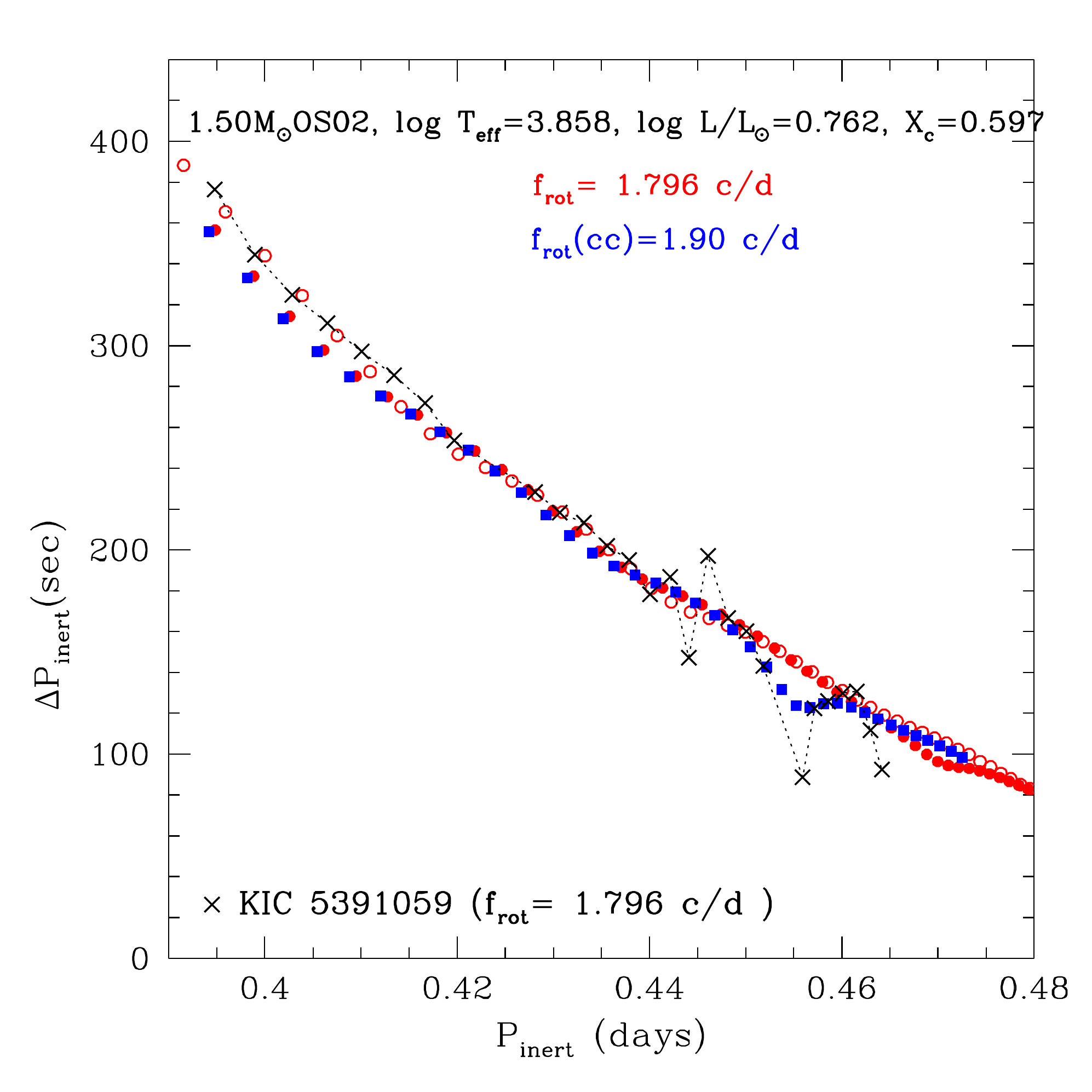}
\caption{KIC 5391059: \citet{LiG19} obtained $f_{\rm rot}= 1.796\pm0.001$\,d$^{-1}$ from the $P$\,-\,$\Delta P$ relations of prograde dipole g modes and of r modes. The g-mode $P$\,-\,$\Delta P$ relation has a dip at $\sim0.455$\,days attributable to resonance couplings with the fundamental inertial mode in the convective core.  The OS01 model (middle panel) fits best with the depth of the dip, while the dip predicted by the model without overshooting is too deep, and the OS02 model has a too shallow dip. }
\label{fig:k539}
\end{figure*}

\begin{figure*}
\includegraphics[width=0.33\textwidth]{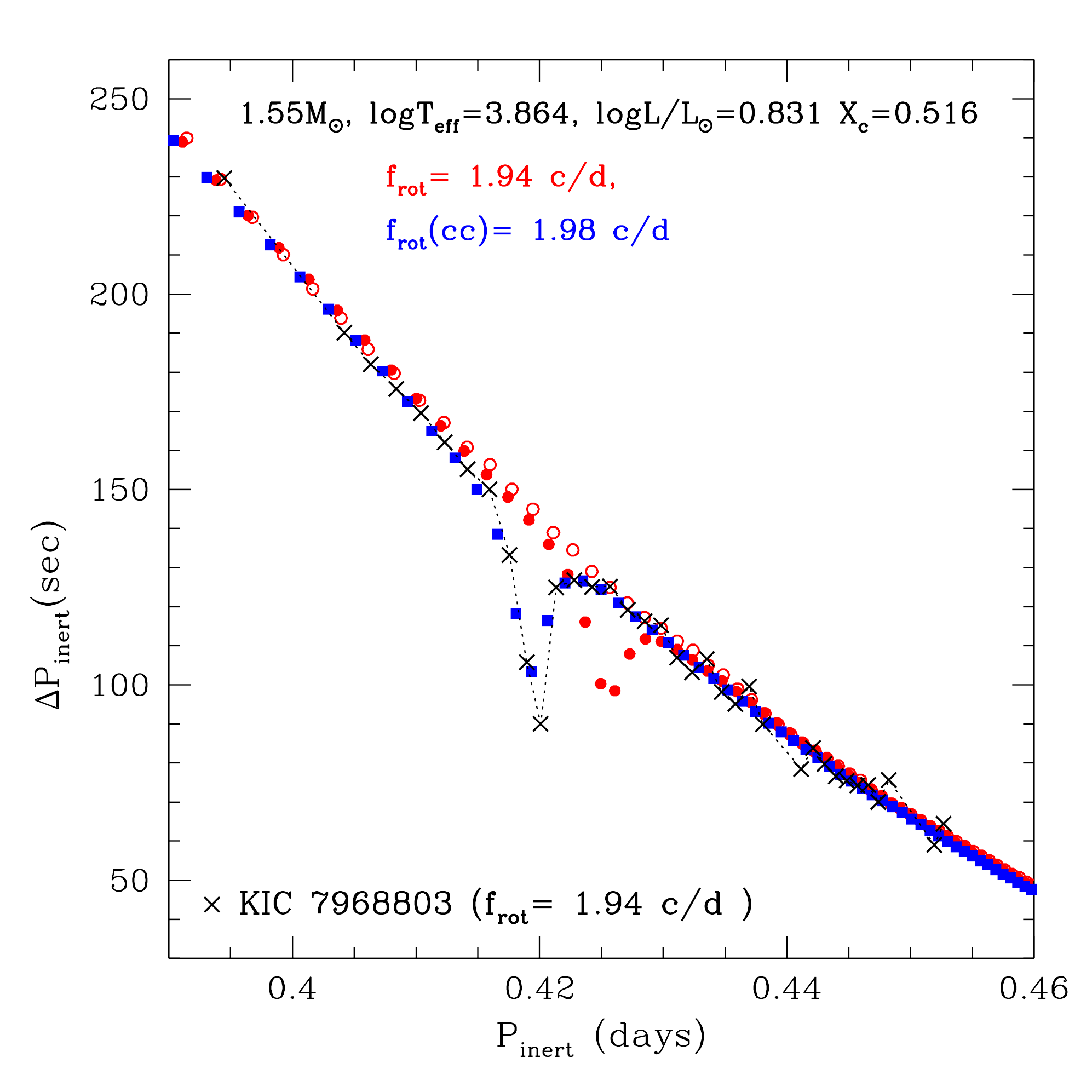}
\includegraphics[width=0.33\textwidth]{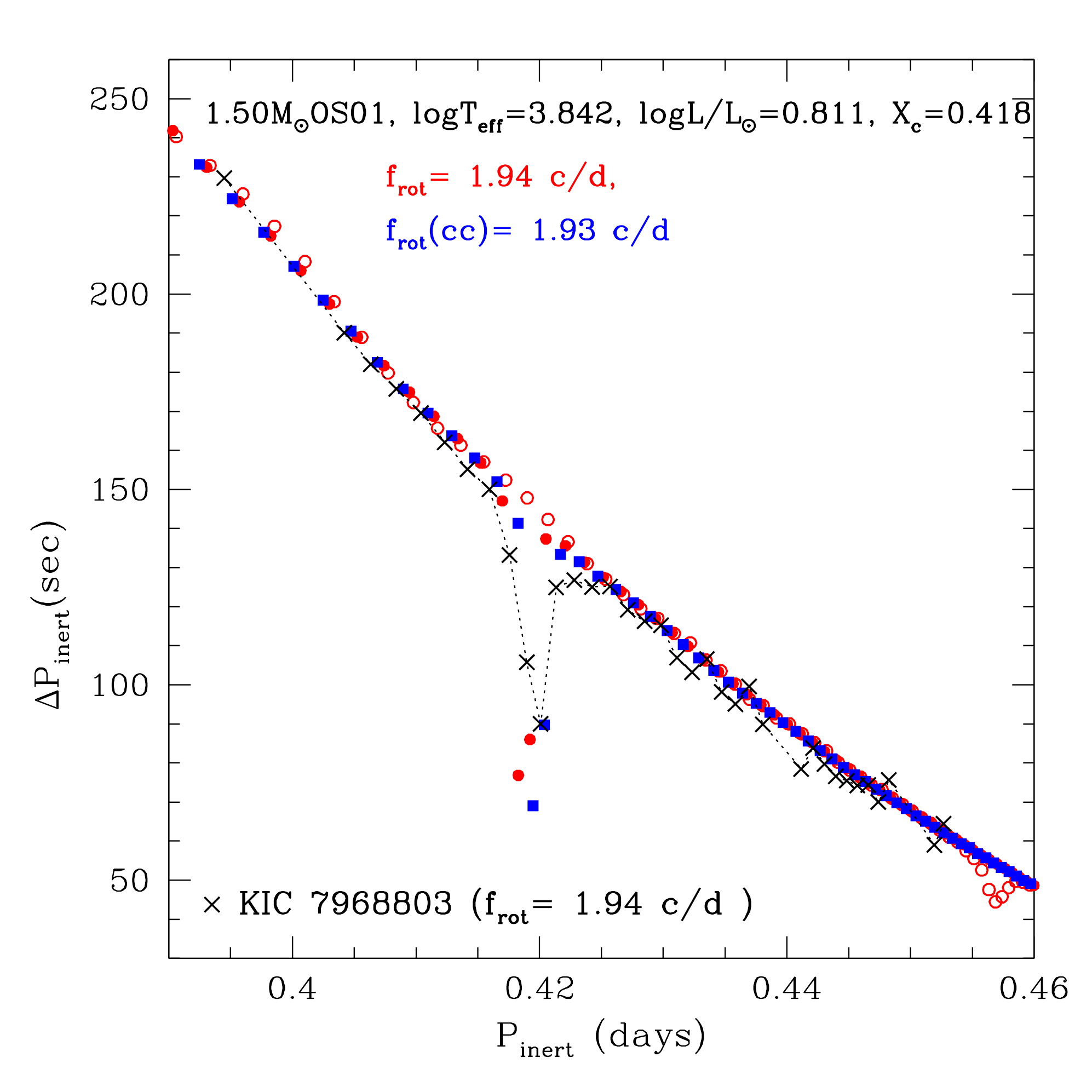}
\includegraphics[width=0.33\textwidth]{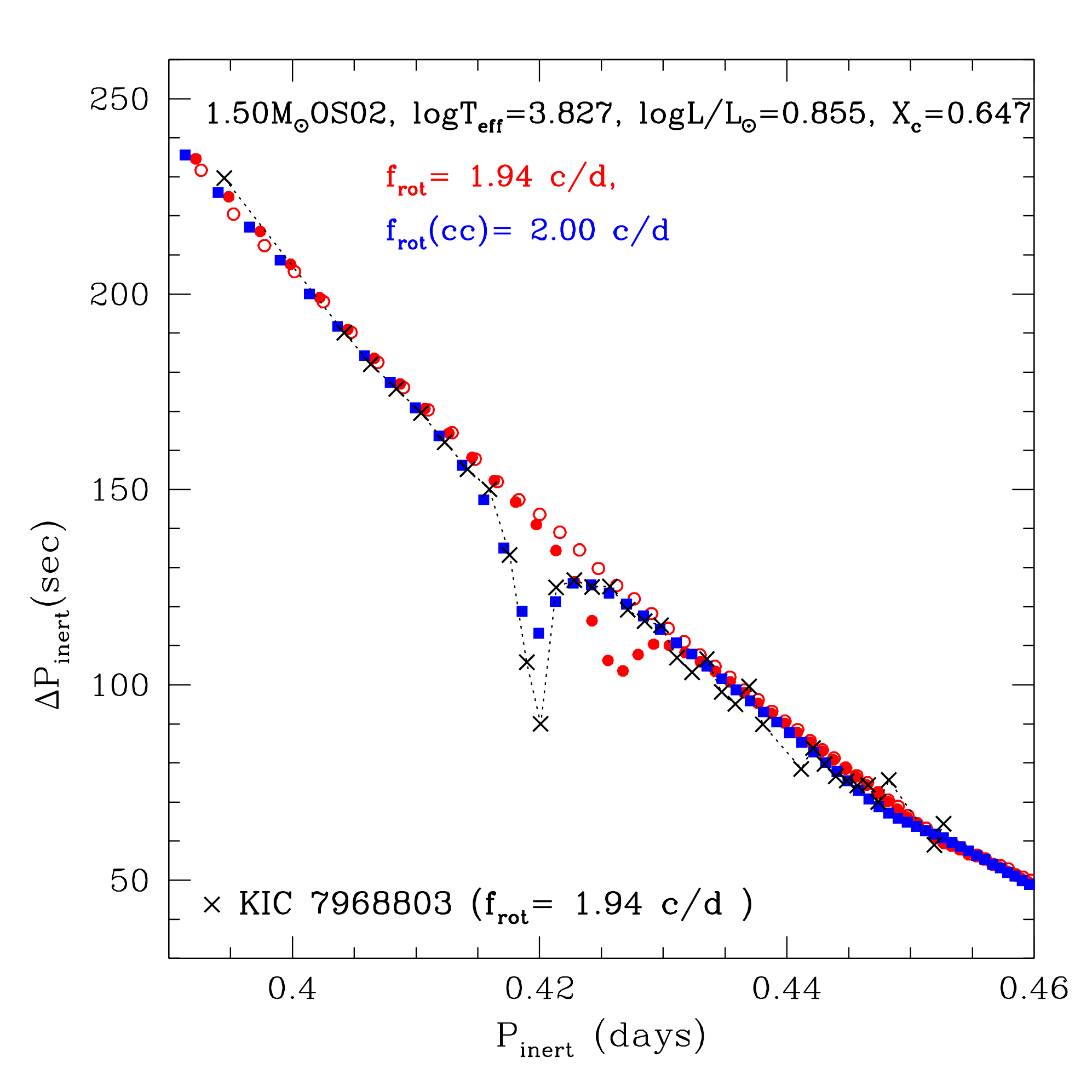}
\caption{KIC 7968803: \citet{LiG20} obtained $f_{\rm rot}= 1.94\pm0.01$~d$^{-1}$ from the $P$\,-\,$\Delta P$ sequences of prograde $m=-1$ and $-2$ g modes. The deep dip in the $P$\,-\,$\Delta P$ relation (crosses) of the dipole g modes at a period of $0.42$~days is explained by resonance couplings with an inertial mode in the convective core rotating nearly synchronously with the surround g-mode cavity (filled blue squares).}
\label{fig:k796}
\end{figure*}

\begin{figure*}
\includegraphics[width=0.33\textwidth]{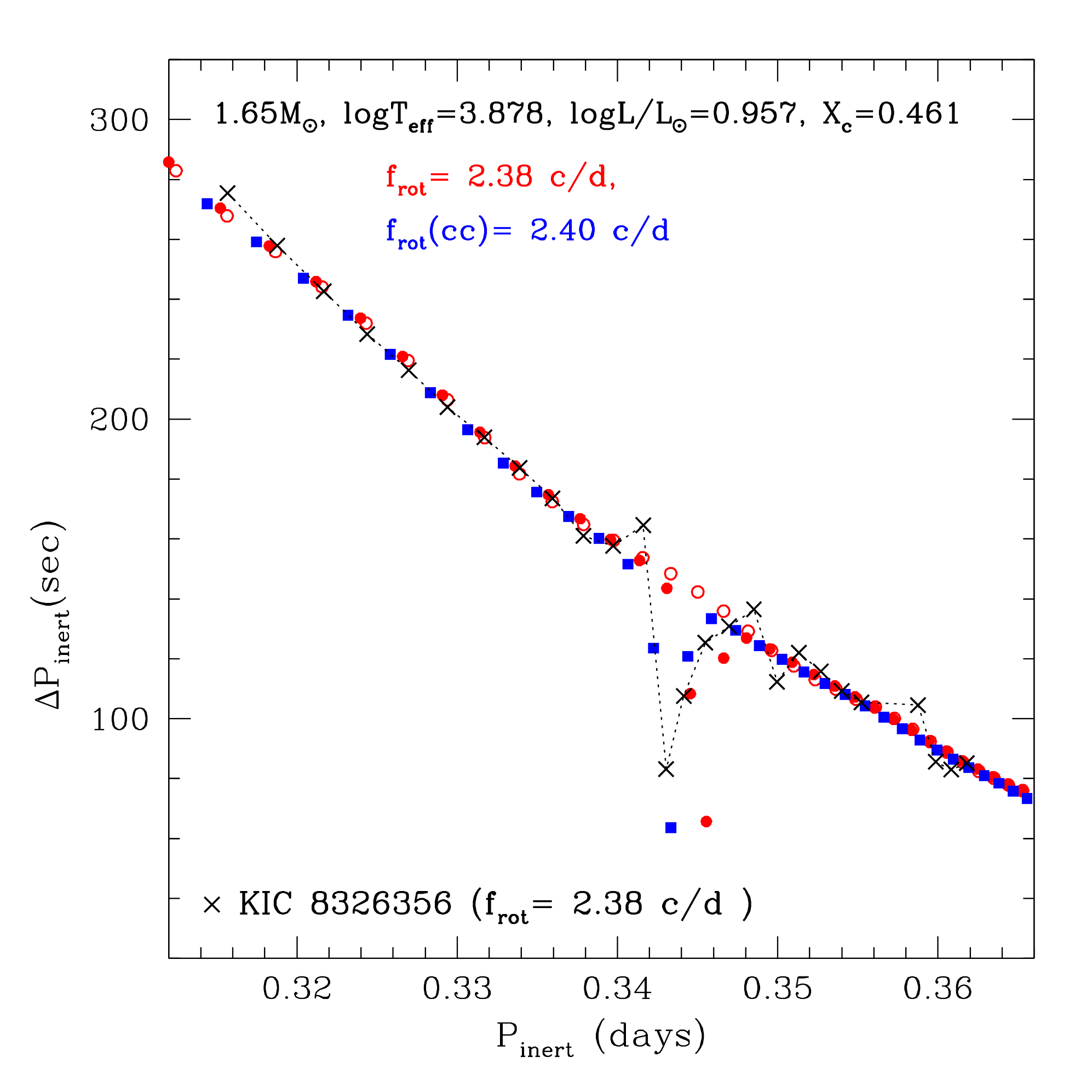}
\includegraphics[width=0.33\textwidth]{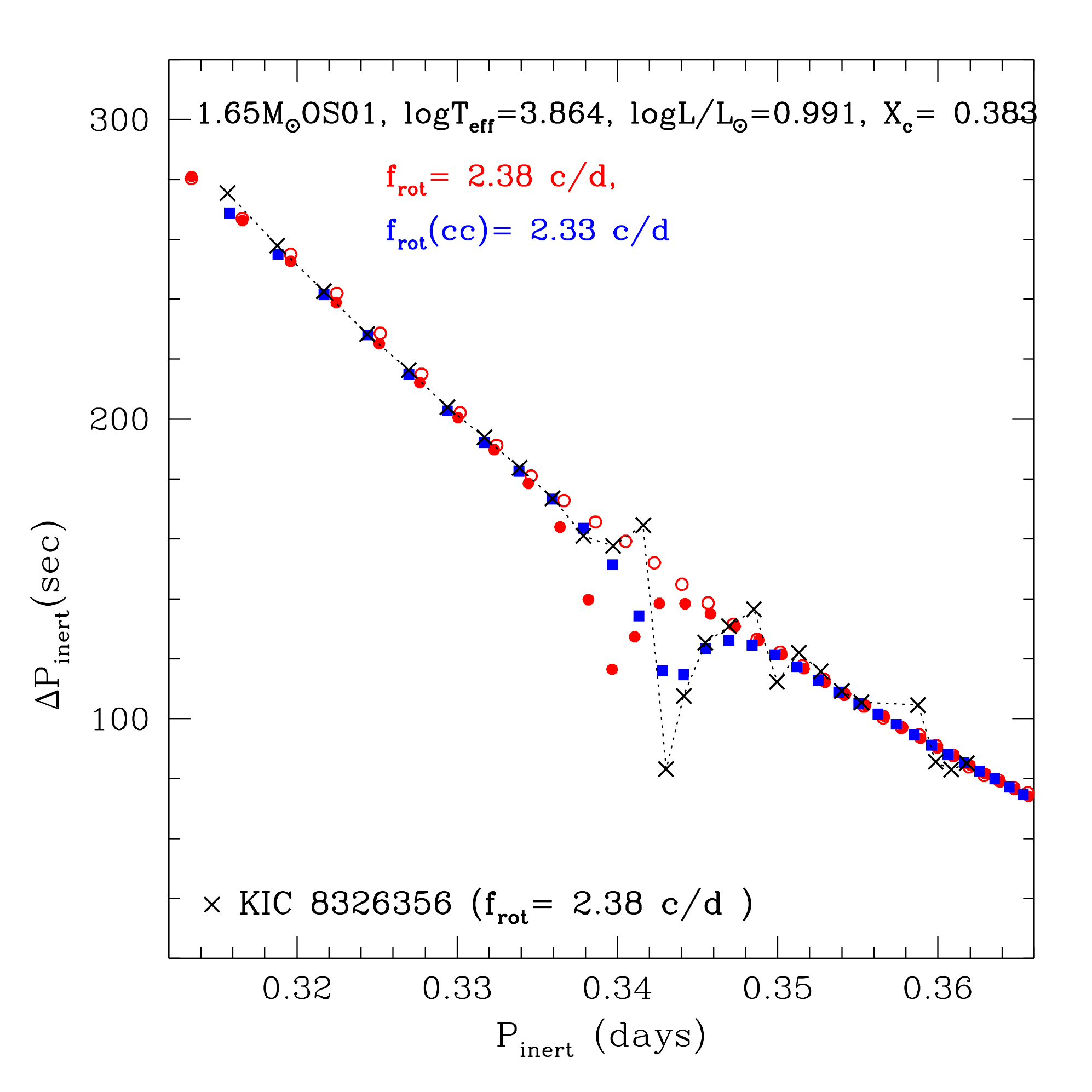}
\includegraphics[width=0.33\textwidth]{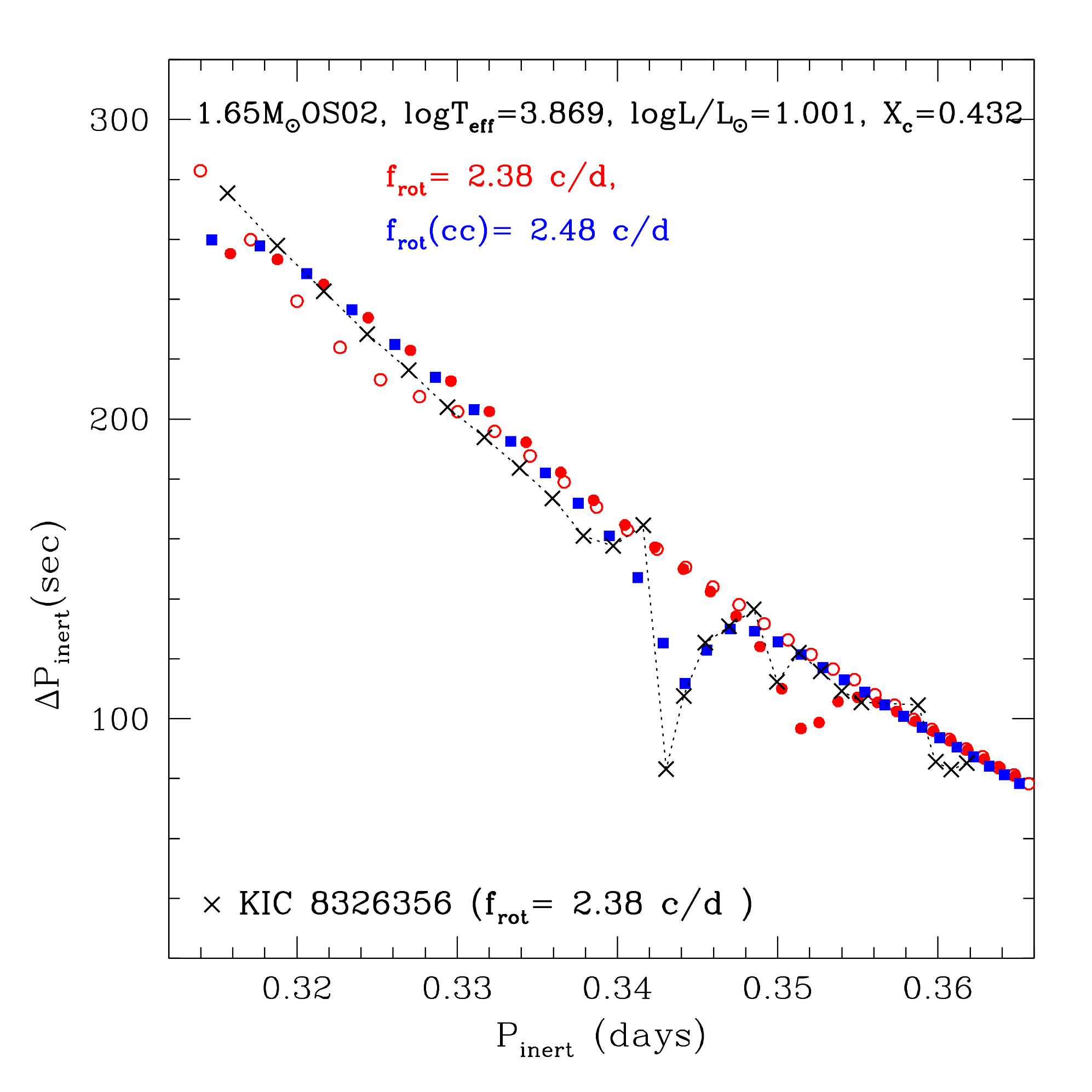}
\caption{KIC 8326356: \citet{LiG20} obtained $f_{\rm rot}=2.38\pm0.02$~d$^{-1}$ from the $P$\,-\,$\Delta P$ sequence of prograde dipole g modes. The pronounced dip at $P_{\rm inert}\approx 0.343$~days can be explained as the resonance dip with the fundamental inertial mode in the convective core rotating nearly synchronously with the surrounding g mode cavity (filled blue squares). The effects of different overshooting assumptions are small.}
\label{fig:k832}
\end{figure*}

\begin{figure*}
\includegraphics[width=0.33\textwidth]{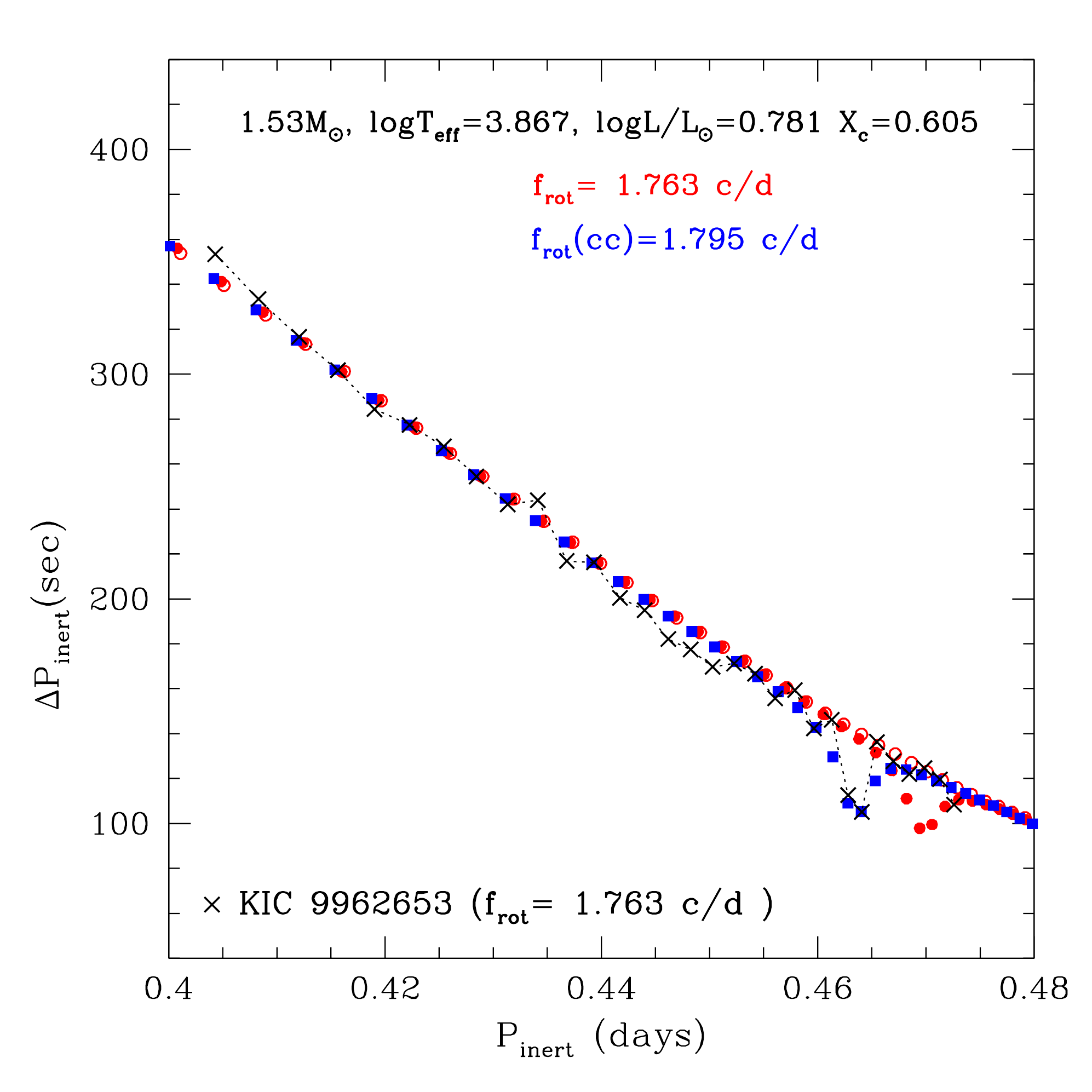}
\includegraphics[width=0.33\textwidth]{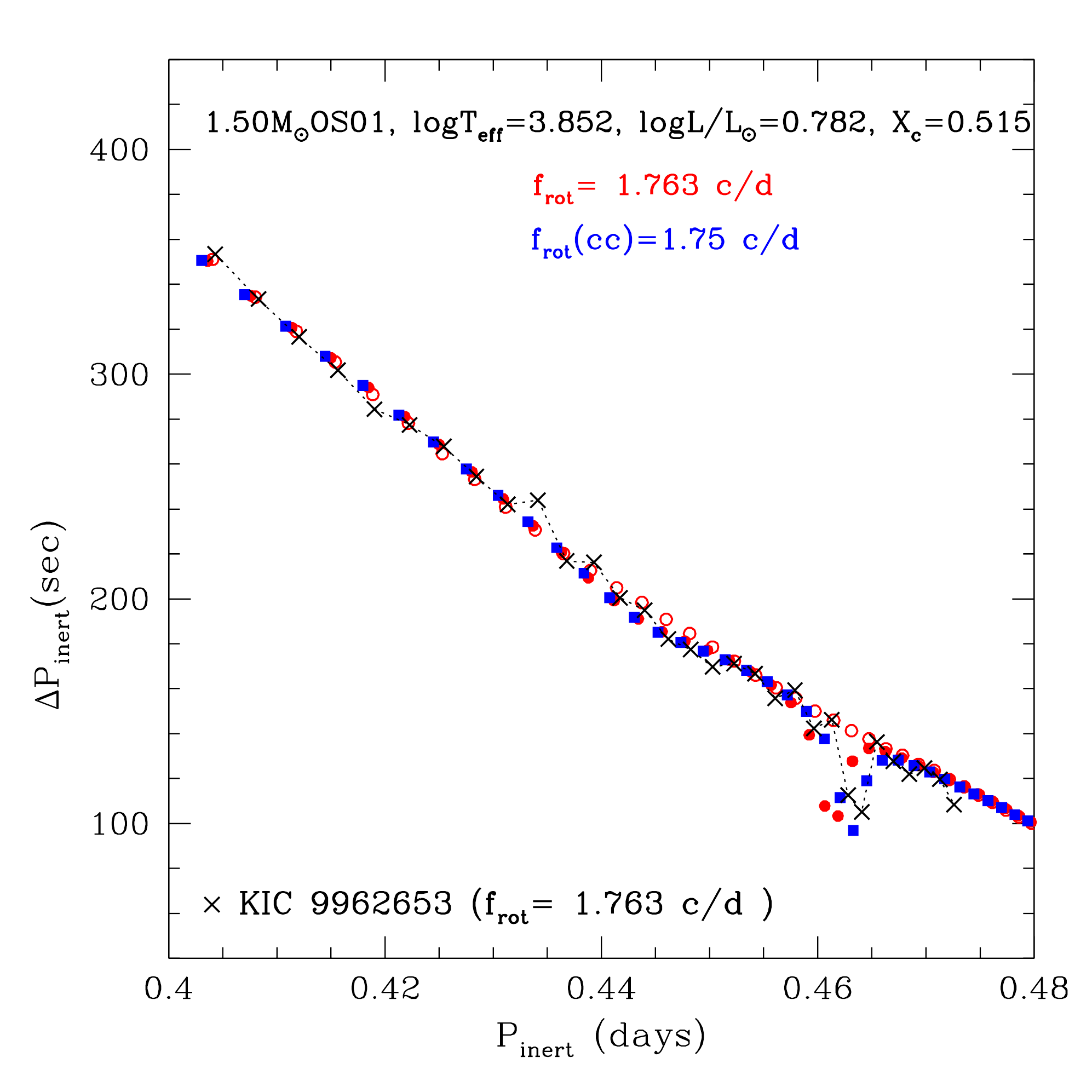}
\includegraphics[width=0.33\textwidth]{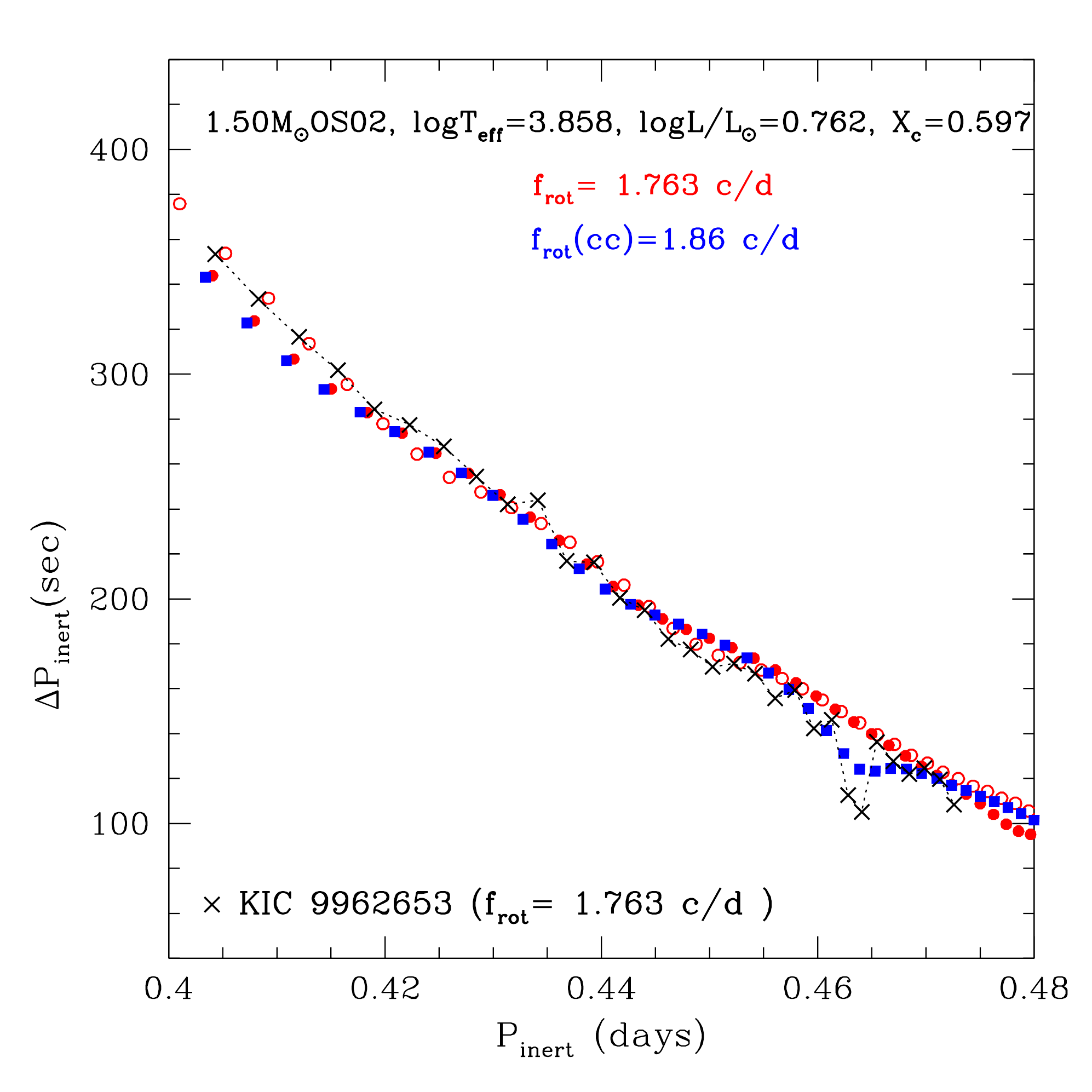}
\caption{KIC~9962653: \citet{LiG19} obtained a rotation rate of $1.763\pm 0.001$\,d$^{-1}$ from the $P$\,-\,$\Delta P$ relations of prograde dipole g modes and of r modes. A resonance dip appears at 0.464~days in the g-mode $P$\,-\,$\Delta P$ relation (crosses). The 1.53-$M_\odot$ model without overshooting (left panel) with a convective core rotating at $f_{\rm rot}{\rm (cc)}=1.795~{\rm d}^{-1}$ slightly faster than $f_{\rm rot}$ reproduces well the resonance $\Delta P$ dip of KIC~9962653 as well as the overall shape of the $P$\,-\,$\Delta P$ relation.
A similarly good fit is obtained by a 1.50-$M_\odot$ model with OS01 ($h_{\rm os}=0.01$) overshooting (middle panel) with $f_{\rm rot}{\rm(cc)}=1.75~{\rm d}^{-1}$ very slightly slower than $f_{\rm rot}$ in the g mode cavity.
However, a $\Delta P$ dip predicted by a 1.50-$M_\odot$ models with larger overshooting (OS02, $h_{\rm os}=0.02$; right panel) is too shallow and too broad.     
}
\label{fig:k996}
\end{figure*}

\begin{figure*}
\includegraphics[width=0.33\textwidth]{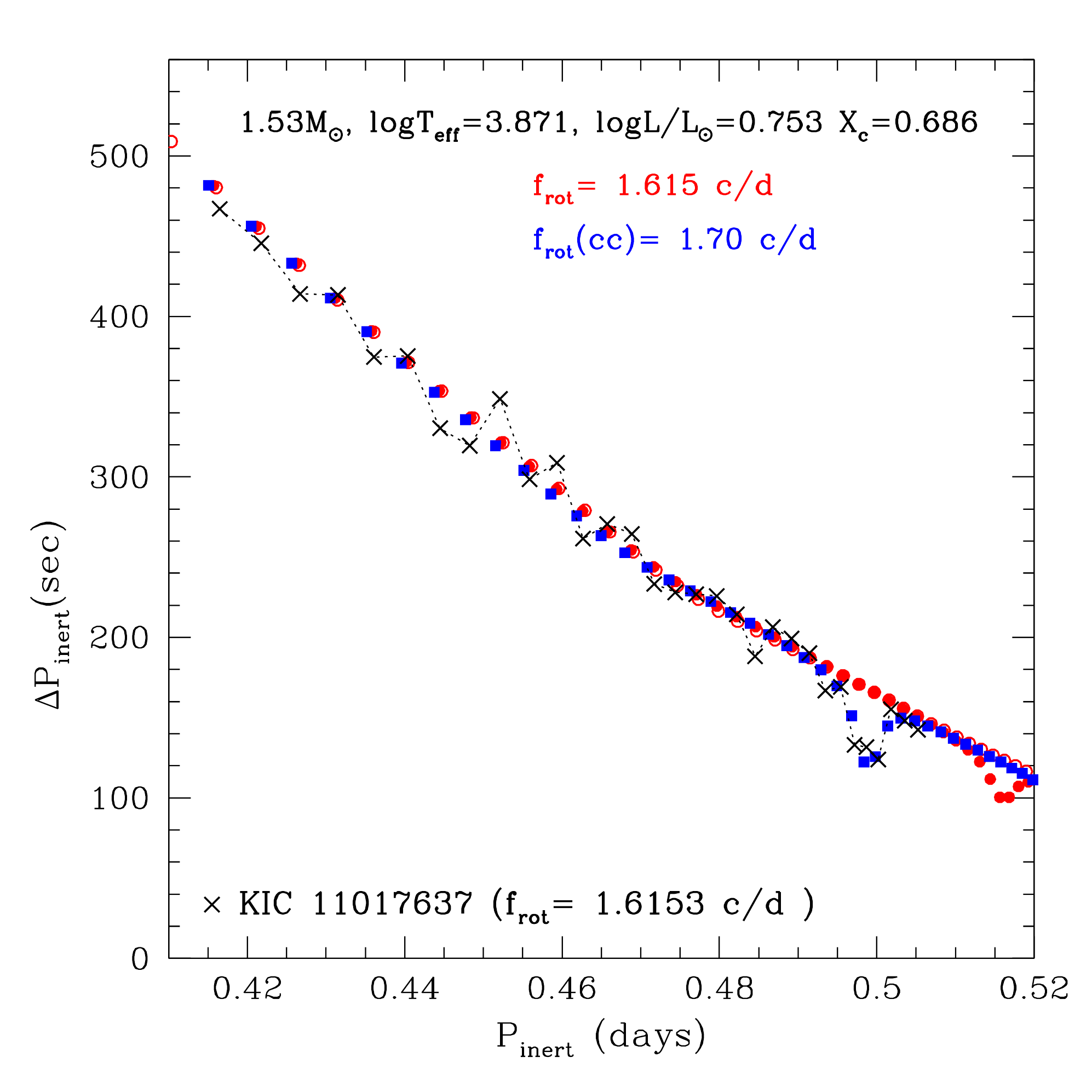}
\includegraphics[width=0.33\textwidth]{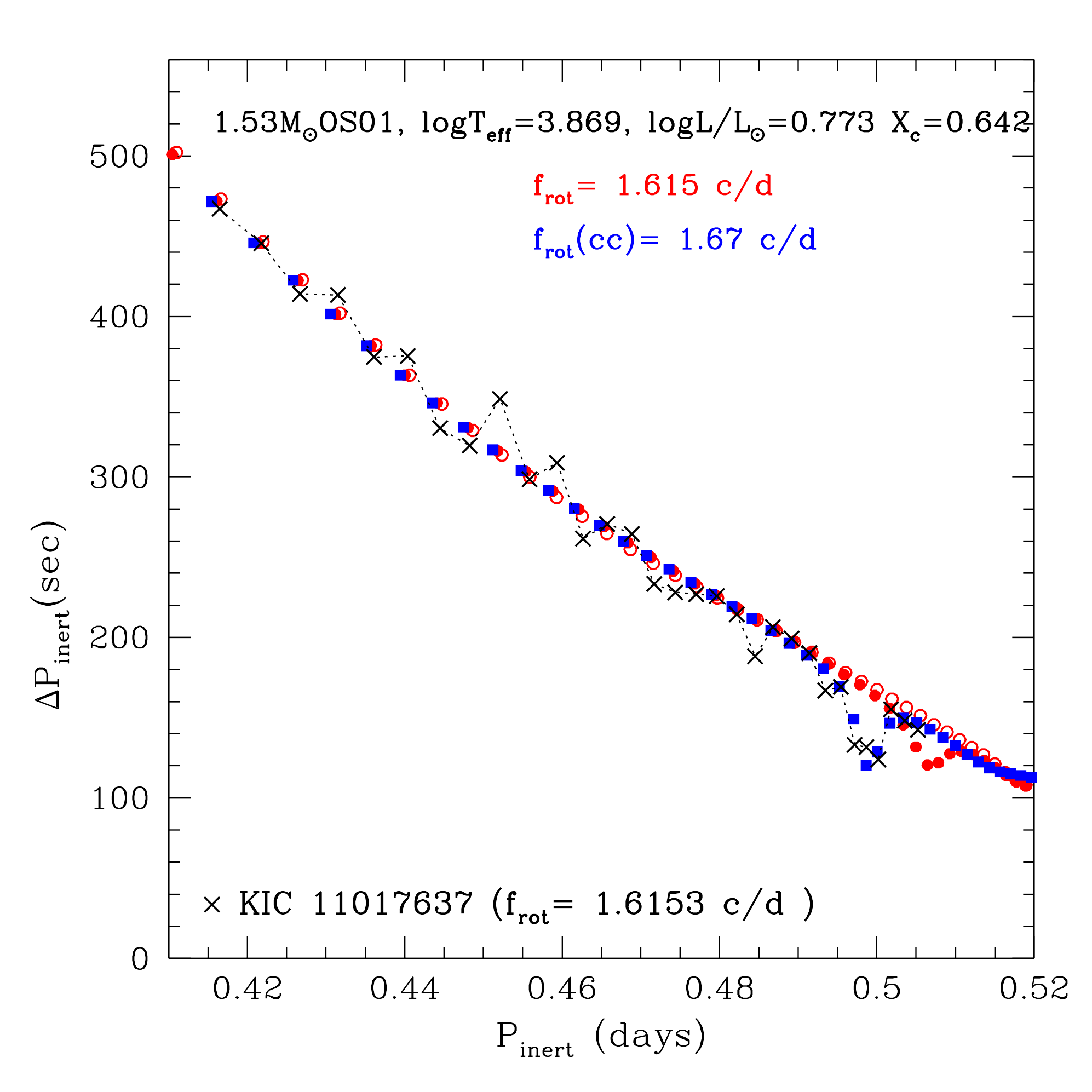}
\includegraphics[width=0.33\textwidth]{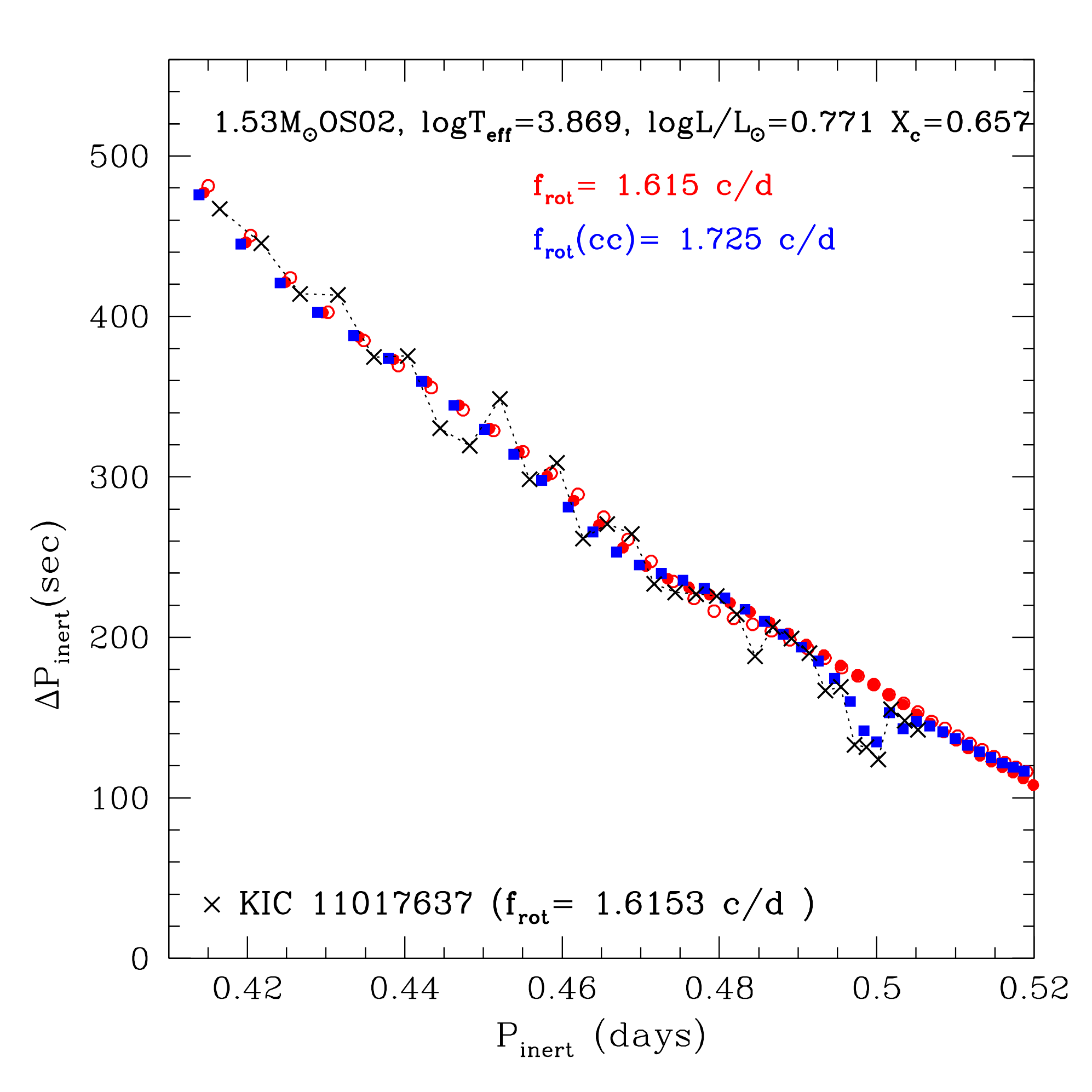}
\caption{KIC~11017637: \citet{LiG19} obtained $f_{\rm rot}=1.6153\pm 0.0008$\,d$^{-1}$ from the $P$\,-\,$\Delta P$ relations of prograde dipole g modes and r modes.
The g-mode $P$\,-\,$\Delta P$ relation (crosses connected with dotted lines) has a dip at a period of $\sim0.5$\,days, which 
can be fitted well, irrespective of the assumptions on core overshooting, if the convective core rotates slightly faster than the radiative g-mode cavity.}
\label{fig:k110}
\end{figure*}

\begin{figure*}
\includegraphics[width=0.33\textwidth]{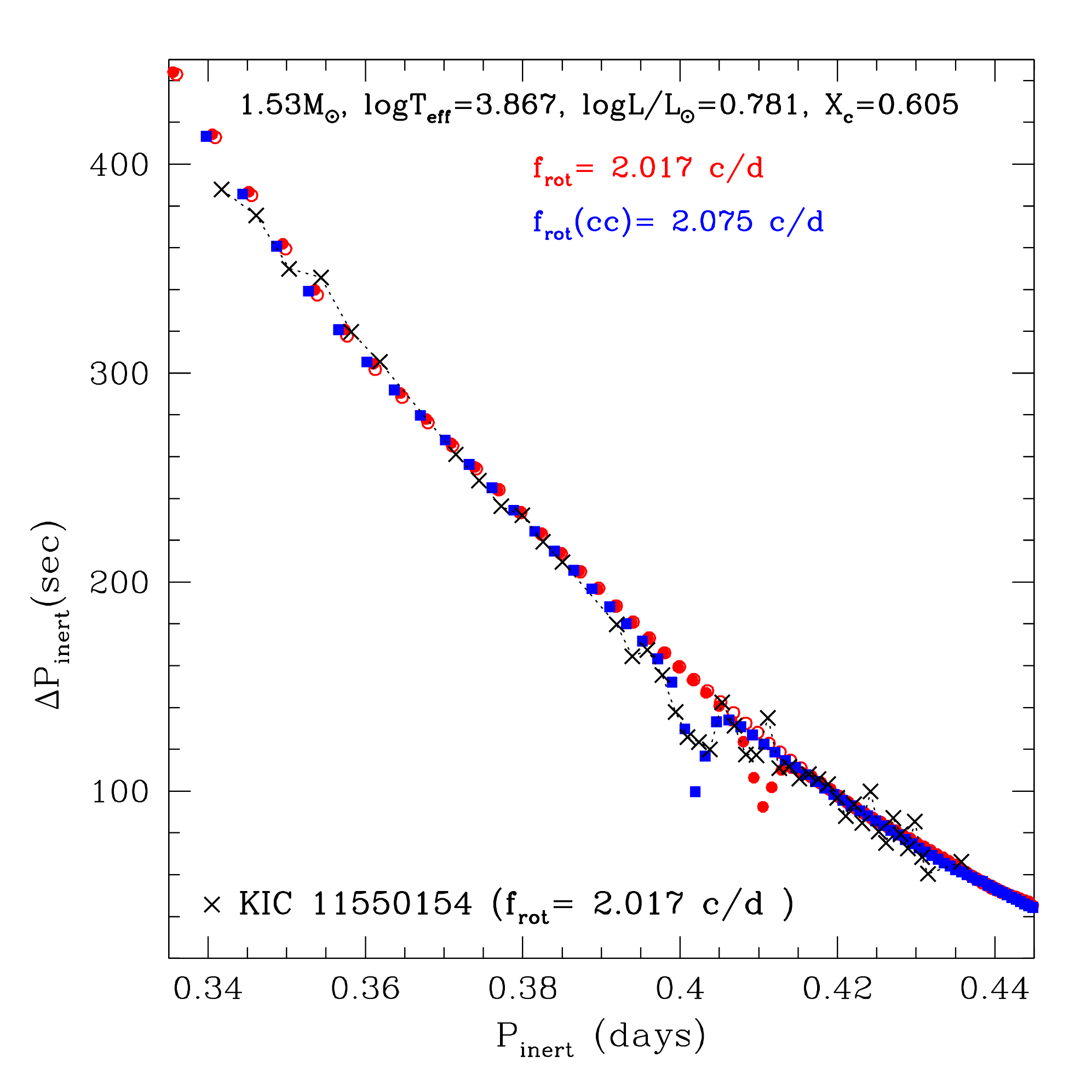}
\includegraphics[width=0.33\textwidth]{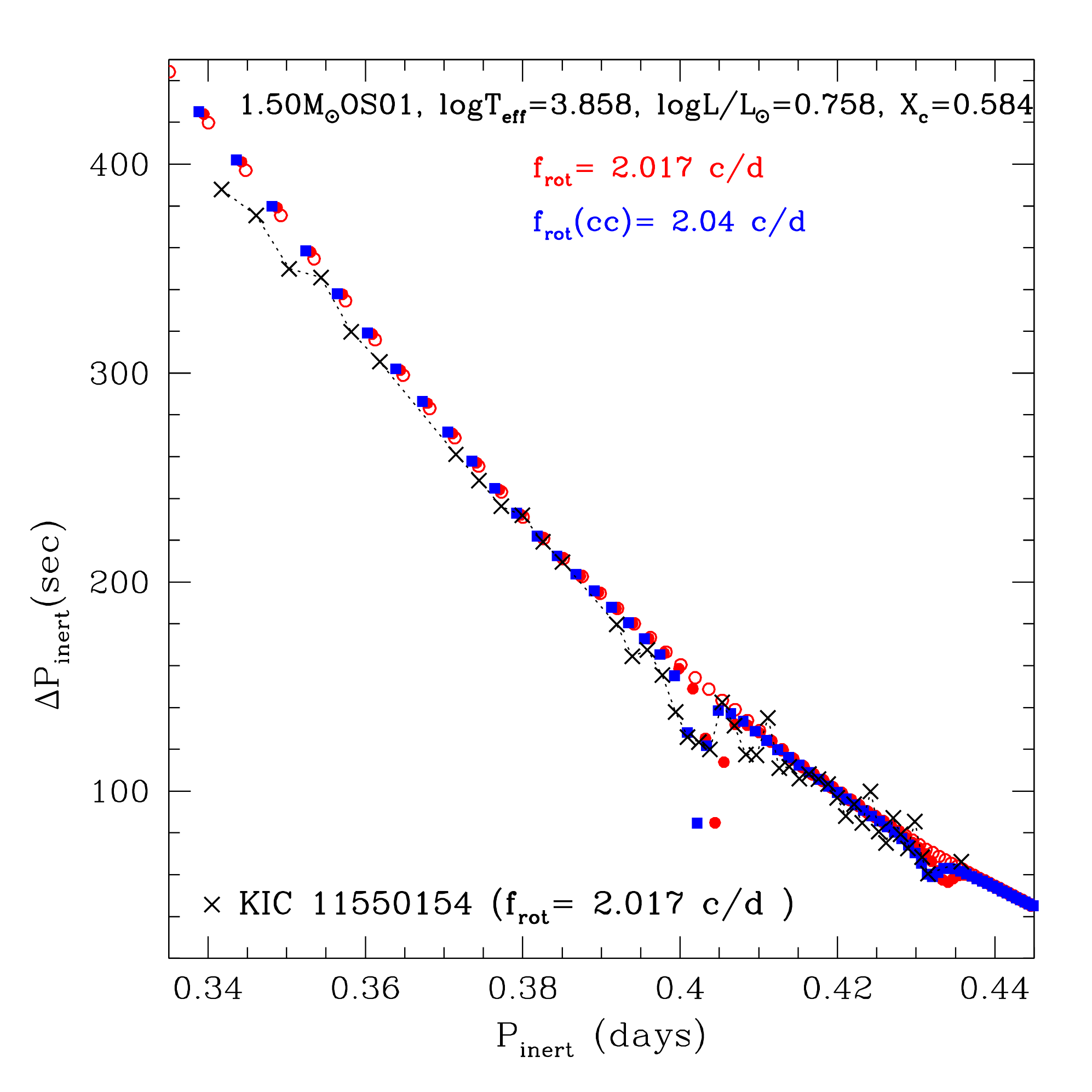}
\includegraphics[width=0.33\textwidth]{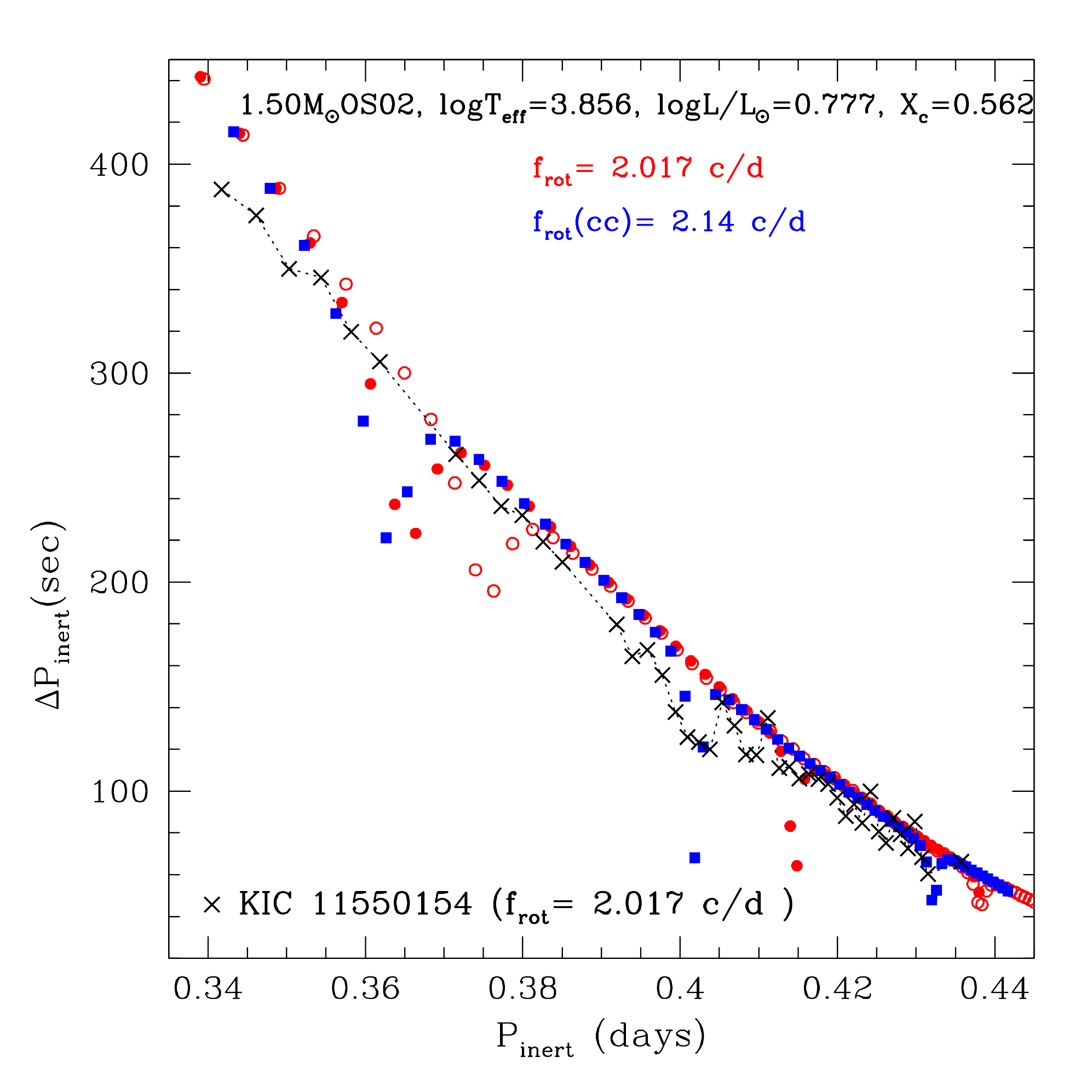}
\caption{KIC 11550154: \citet{LiG19} obtained $f_{\rm rot}= 2.017 \pm 0.001$\,d$^{-1}$ from the $P$\,-\,$\Delta P$ relations of prograde $m=-1, -2$ g modes and $m=1$ r modes. The resonance dip at $\sim0.40$\,days is reasonably fitted with the predictions (filled blue squares) of OS00 (left panel) and OS01 (middle panel) models assuming that the convective core rotates slightly faster than the surrounding g-mode cavity. 
The OS02 model (right panel) shows a large dip deviating from the observational $P$\,-\,$\Delta P$ relation (even for the case of TAR; open circles) in the period range $0.36 - 0.38$~days, which are related with a radiative zone with small Brunt B\"ais\"al\"a frequency produced by the overshooting (see Fig.\,\ref{fig:pdp_overshoot}).}
\label{fig:k1155}
\end{figure*}

\begin{figure*}
\includegraphics[width=0.33\textwidth]{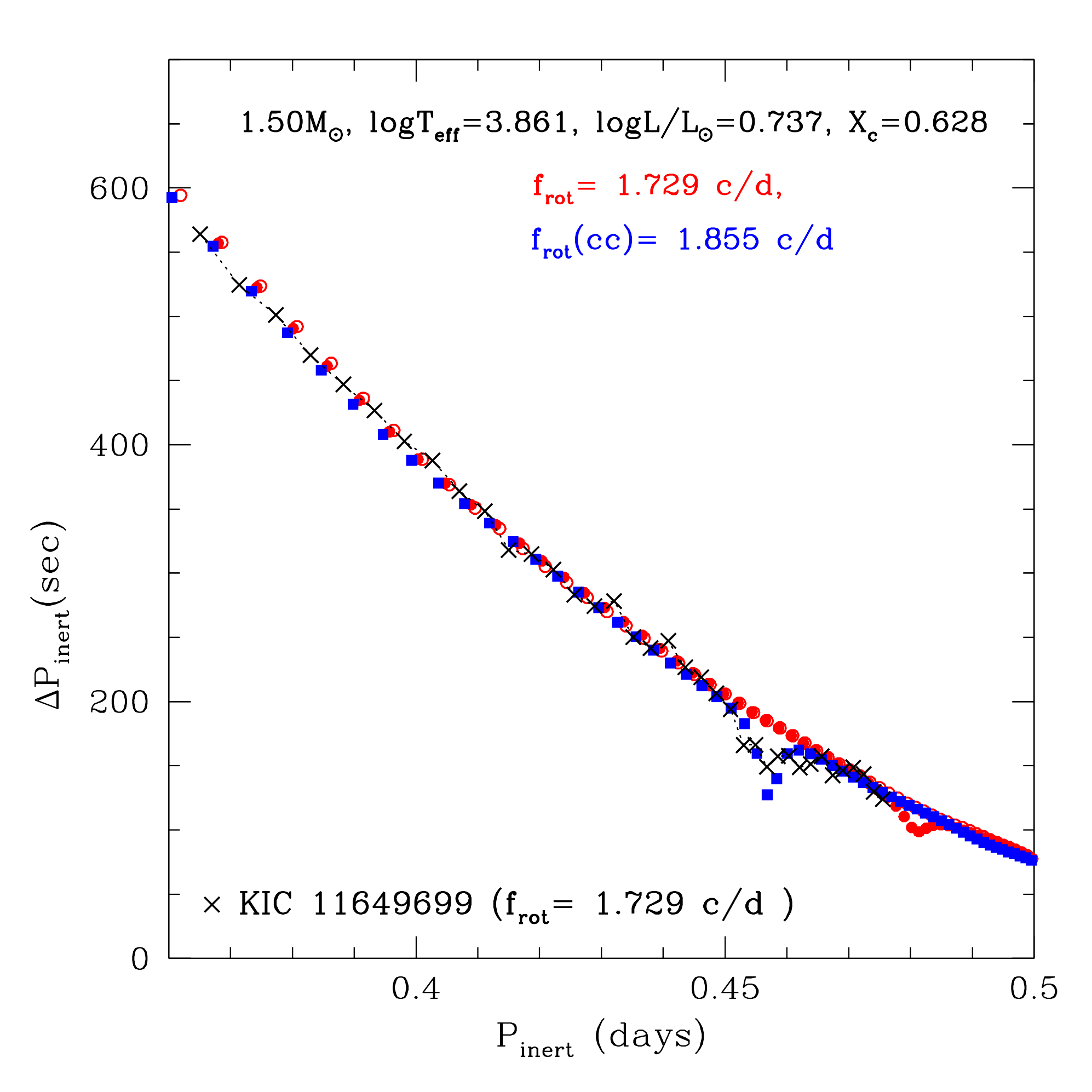}
\includegraphics[width=0.33\textwidth]{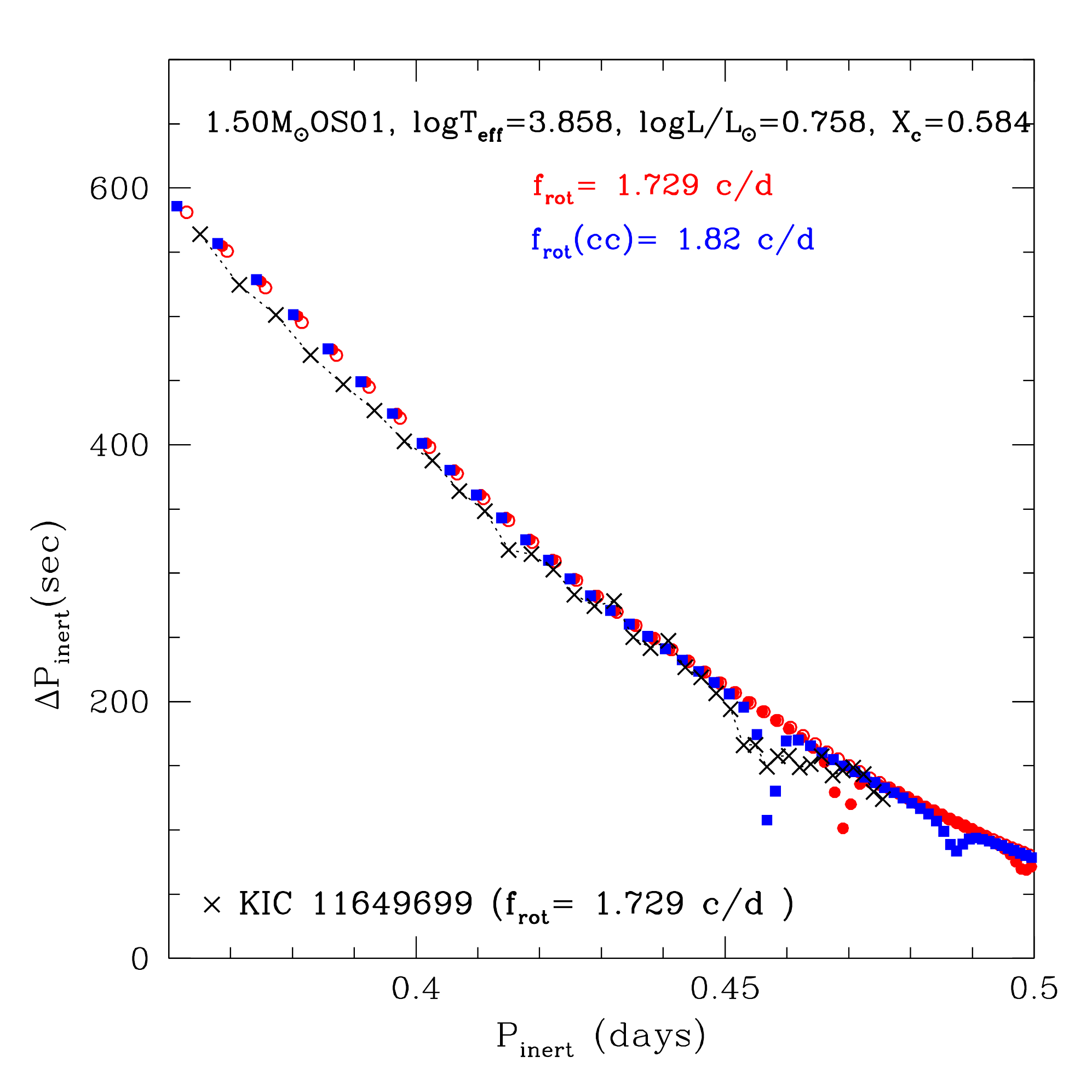}
\includegraphics[width=0.33\textwidth]{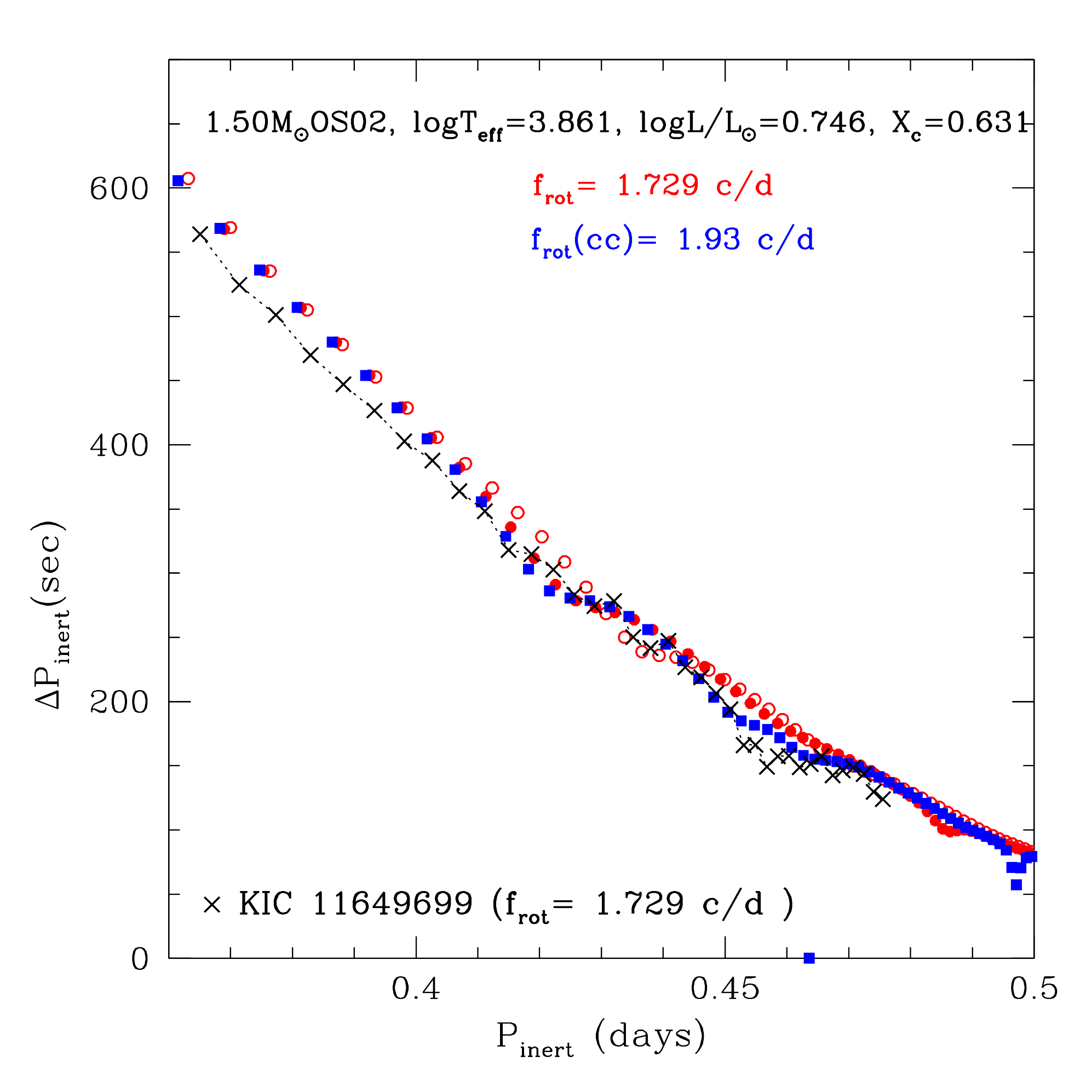}
\caption{KIC 11649699: \citet{LiG19} obtained $f_{\rm rot}=1.729\pm0.002$\,d$^{-1}$ from the $P$\,-\,$\Delta P$ relations of prograde $m=-1,-2$ g modes and $m=1$ r modes. The $P$\,-\,$\Delta P$ relation of the prograde dipole g modes of KIC~1164966 is shown by crosses connected with dotted lines.  The small dip around a period of 0.455~days is fitted well with a model without overshooting (left panel) with a convective core rotating slightly faster than the radiative g-mode cavity.}
\label{fig:k116}
\end{figure*}

\begin{figure*}
\includegraphics[width=0.33\textwidth]{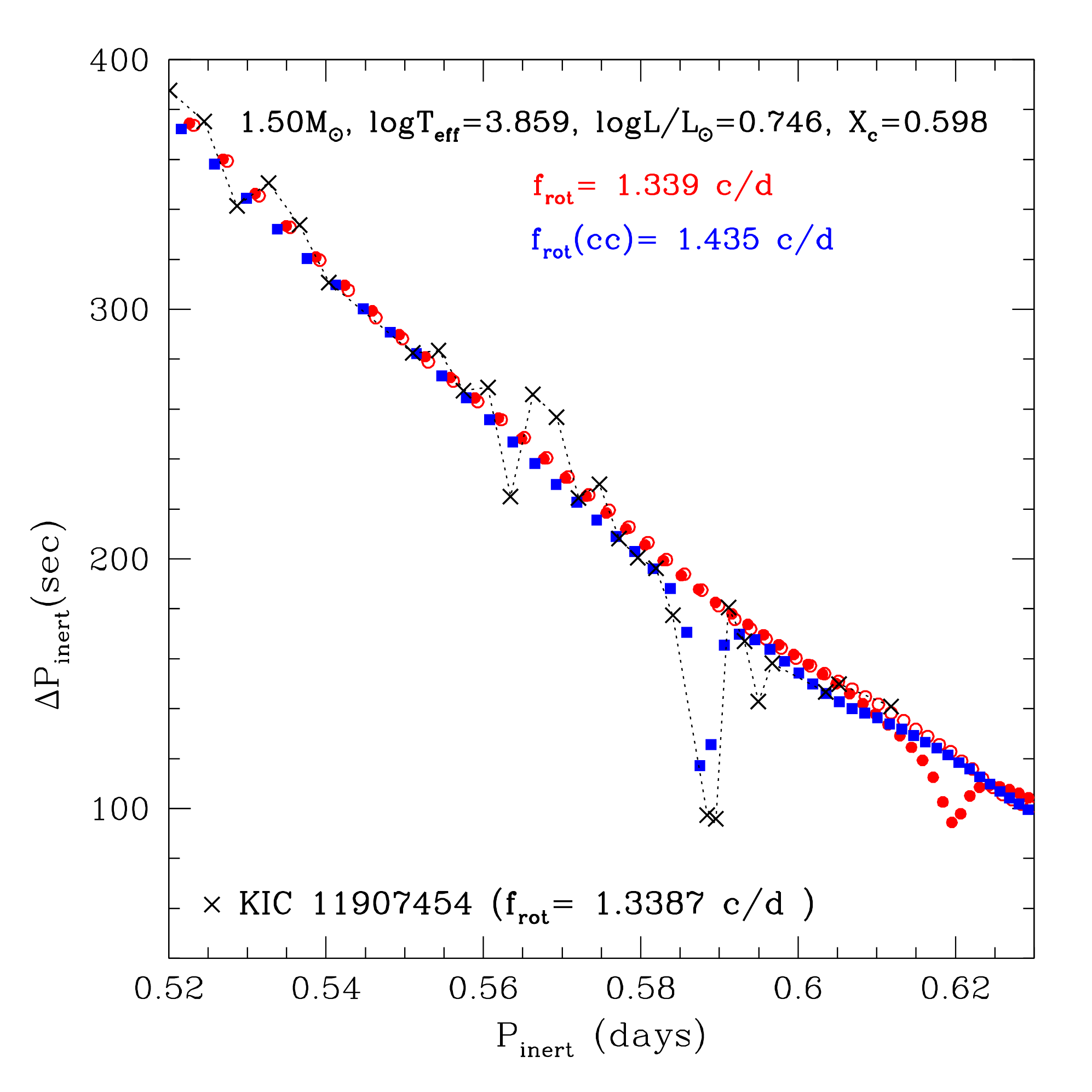}
\includegraphics[width=0.33\textwidth]{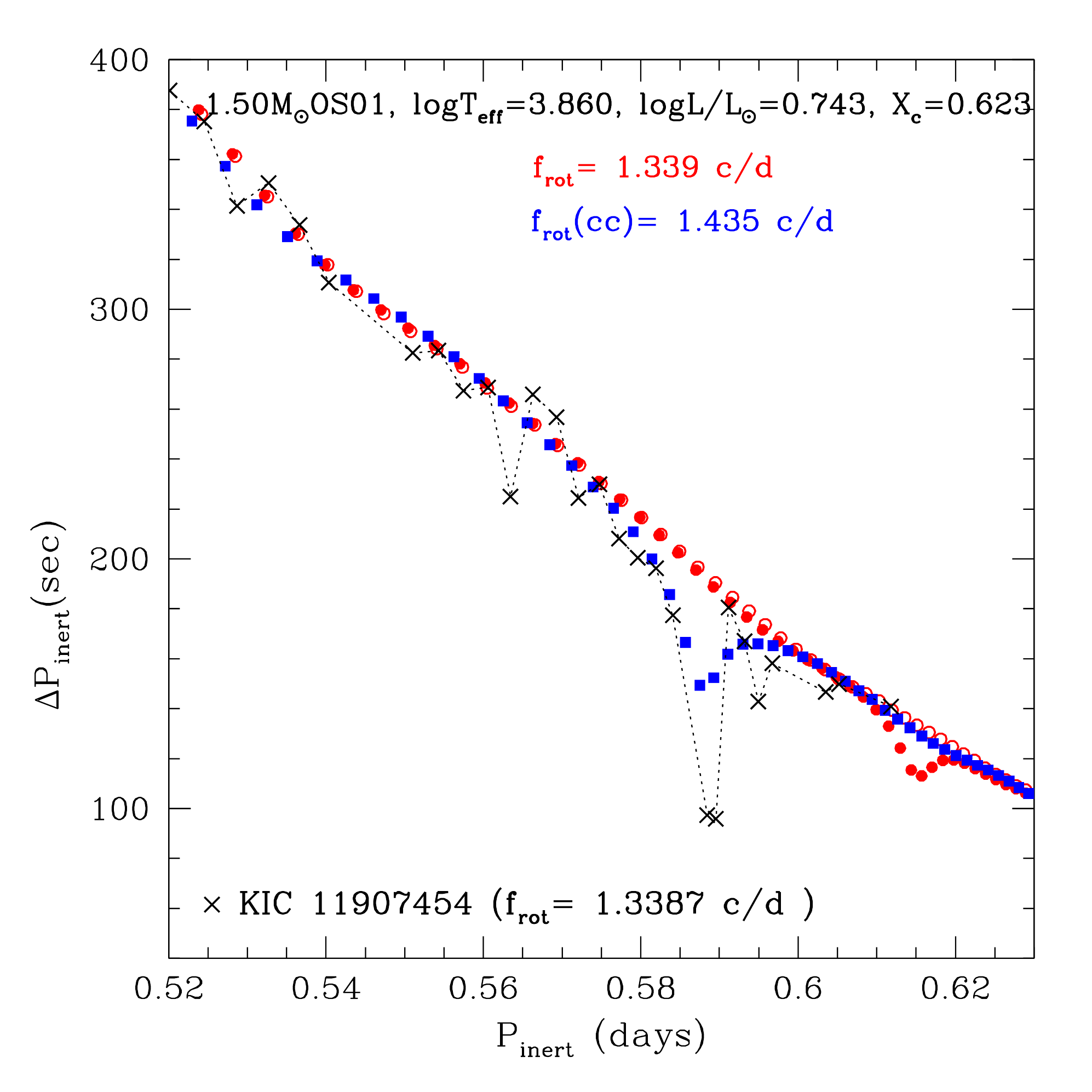}
\includegraphics[width=0.33\textwidth]{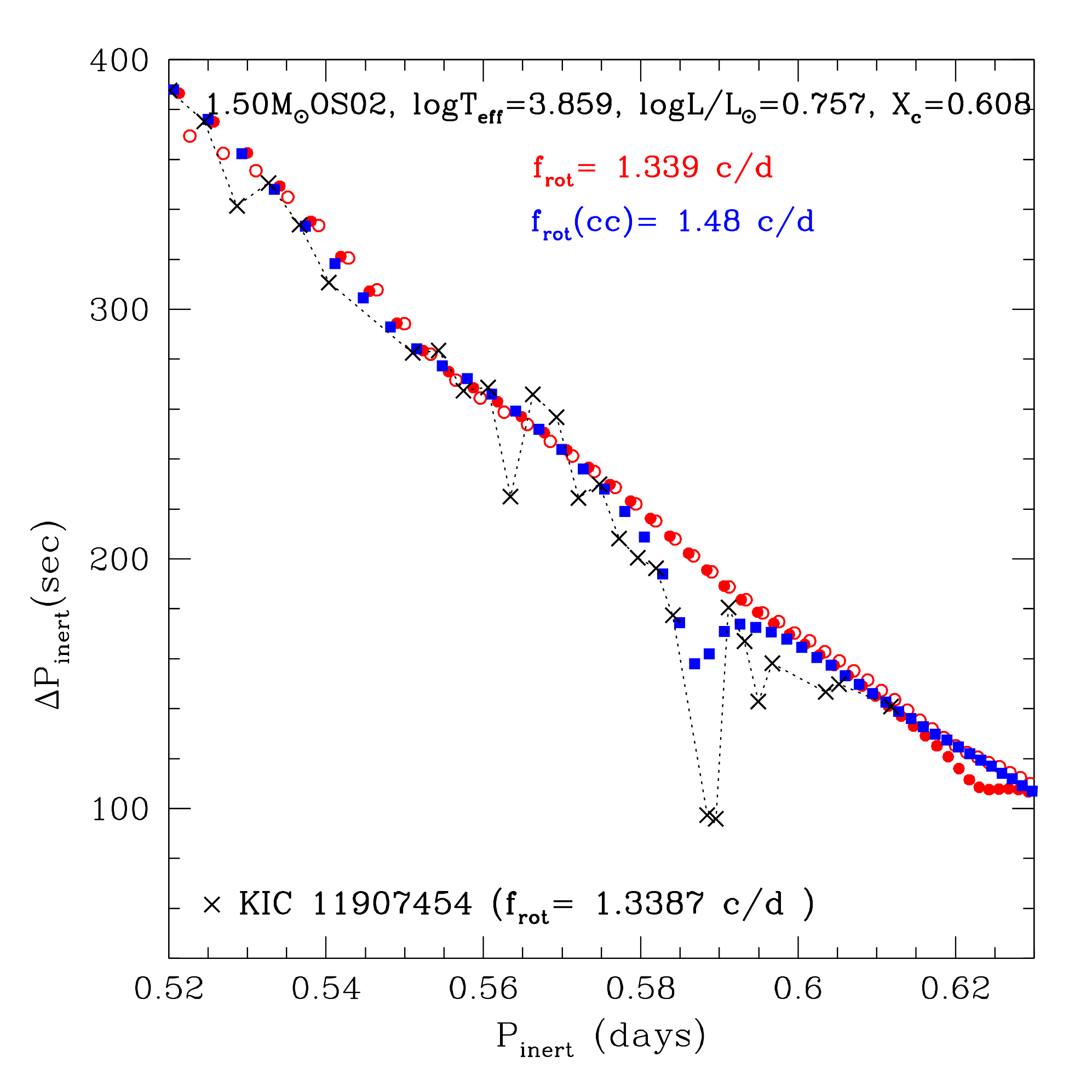}
\caption{KIC~11907454: \citet{LiG19} obtained $f_{\rm rot}=1.3387\pm 0.0006$\,d$^{-1}$ from the $P$\,-\,$\Delta P$ relations of prograde dipole g modes and r modes. The g-mode $P$\,-\,$\Delta P$ relation (crosses) has a deep dip at a period of about 0.59~days, which is reproduced best by a model without overshooting (left panel) with a convective core rotating slightly faster than the surrounding g-mode cavity. Although the period of the dip can be fitted with models with overshooting, the predicted depth is shallower. }
\label{fig:k119}
\end{figure*}

\begin{figure*}
\includegraphics[width=0.33\textwidth]{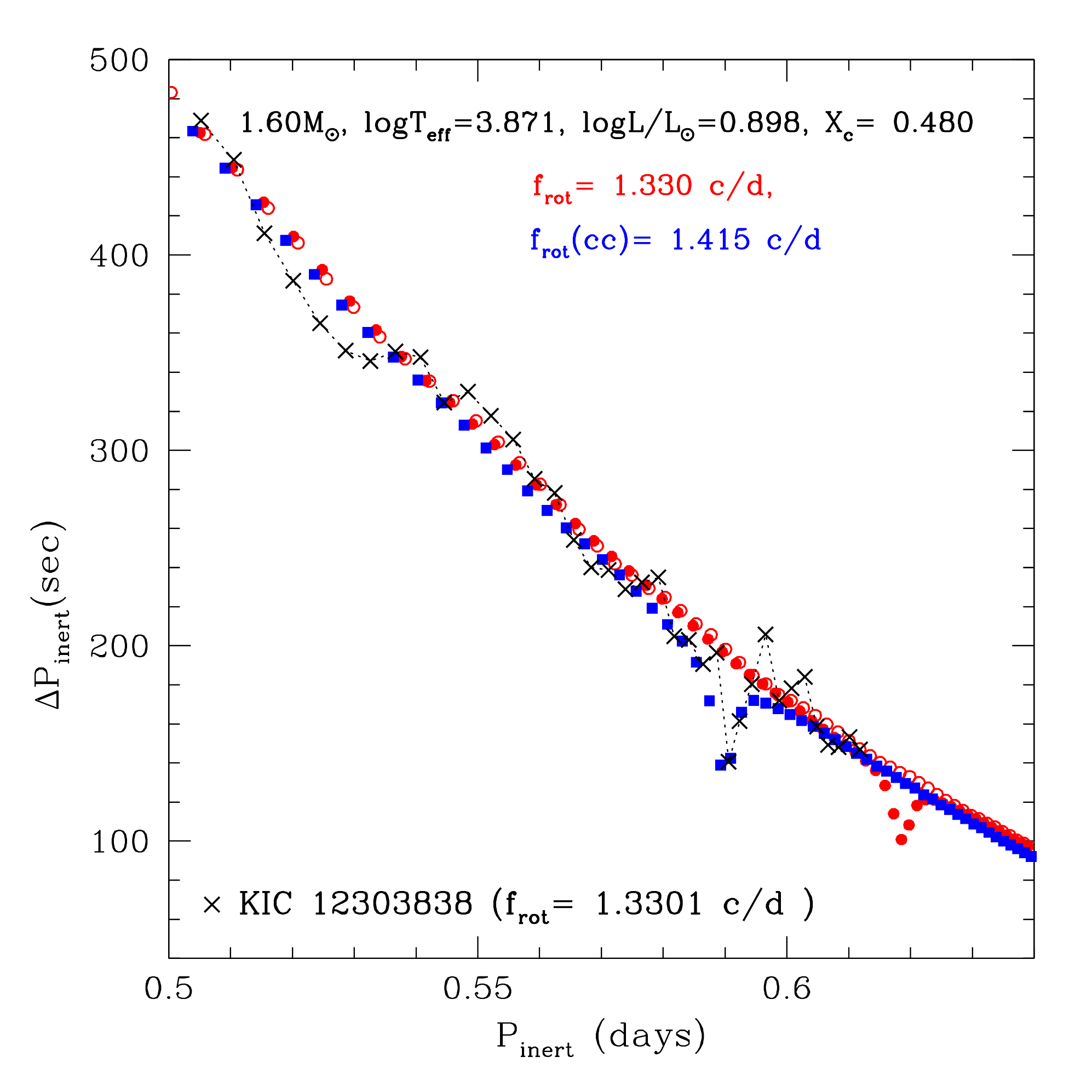}
\includegraphics[width=0.33\textwidth]{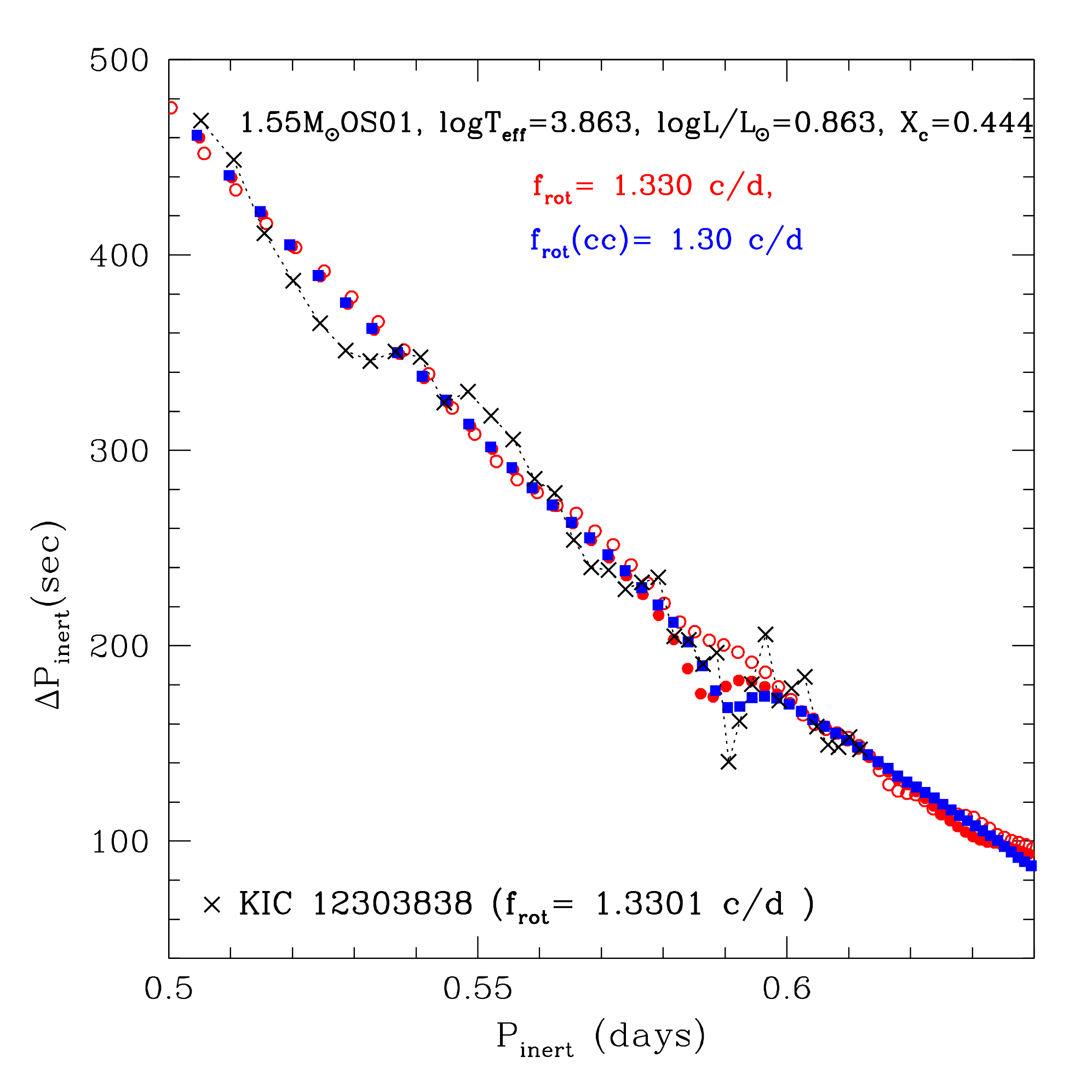}
\includegraphics[width=0.33\textwidth]{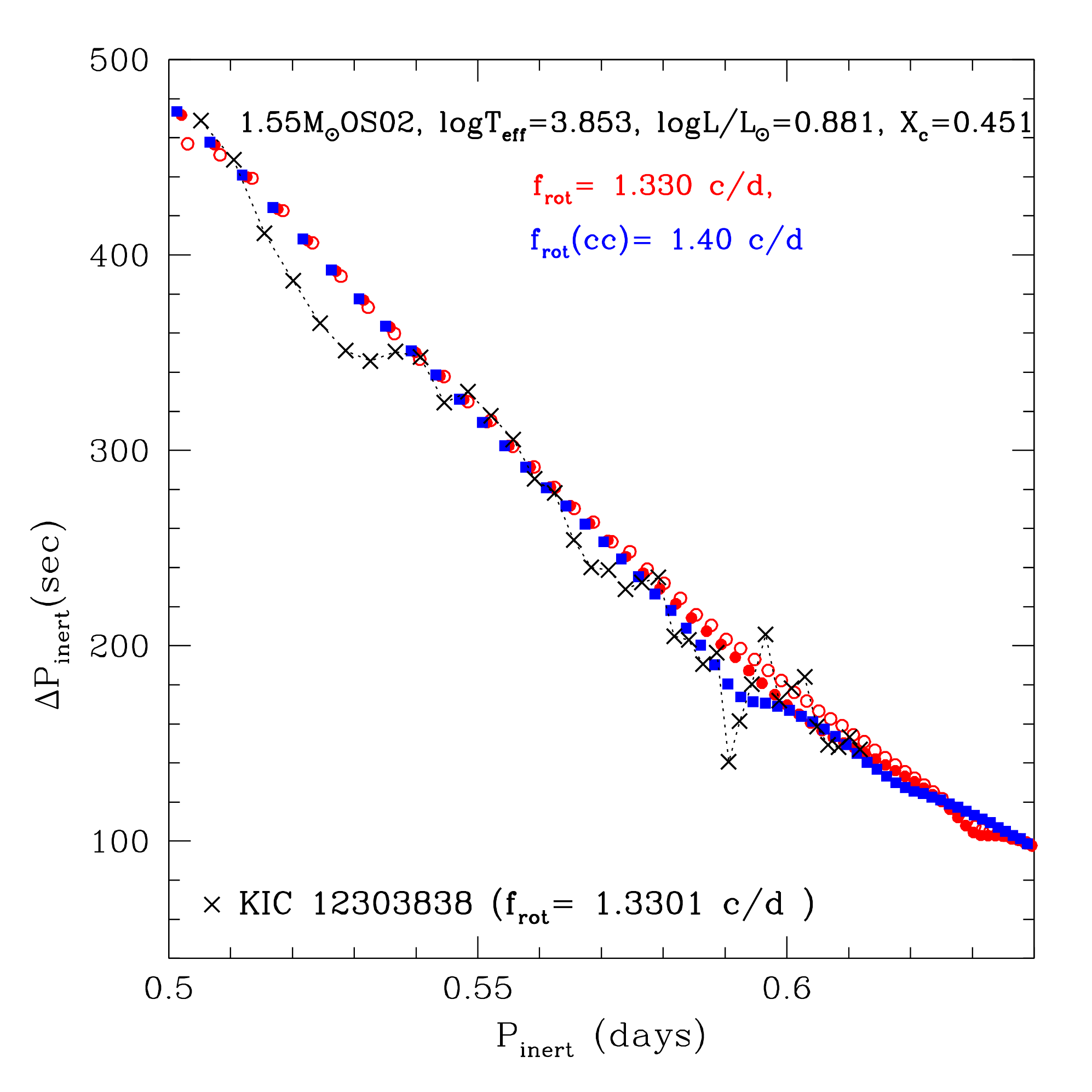}
\caption{KIC~12303838: \citet{LiG19} obtained $f_{\rm rot}=1.3301\pm0.0007$\,d$^{-1}$ from the $P$\,-\,$\Delta P$ relations of prograde dipole g modes and r modes. The g-mode $P$\,-\,$\Delta P$ relation has a resonance dip at a period of 0.59~days, which is fitted well with the model without overshooting (left panel) with a convective core rotating slightly faster ($1.415~{\rm d}^{-1}$) than the surrounding g-mode cavity ($1.330~{\rm d}^{-1}$). Models with overshooting can also reproduce the period at the dip, while they (OS2 model in particular) predict the depth shallower than the observed one. 
}
\label{fig:k123}
\end{figure*}

This Appendix section shows fittings of $P$\,-\,$\Delta P$ relations (in the same format as Fig.\,\ref{fig:k529}) for $\gamma$ Dor stars in Table~\ref{tab:sum} but not shown in \S\ref{sec:fit};
KIC~3341457 (Fig.\,\ref{fig:k334}),
KIC~4390625 (Fig.\,\ref{fig:k439}),  
KIC~4774208 (Fig.\,\ref{fig:k477}), 
KIC~5391059 (Fig.\,\ref{fig:k539}),
KIC~7968803 (Fig.\,\ref{fig:k796}),
KIC~8326356 (Fig.\,\ref{fig:k832}), 
KIC~9962653 (Fig.\,\ref{fig:k996}),
KIC~11017637 (Fig.\,\ref{fig:k110}), 
KIC~11550154 (Fig.\,\ref{fig:k1155}), 
KIC~11649699 (Fig.\,\ref{fig:k116}), 
KIC~11907454 (Fig.\,\ref{fig:k119}), 
and KIC~12303838 (Fig.\,\ref{fig:k123}). 
The periods and period spacings of these stars are adopted from \citet{LiG19,LiG20}.
They are shown by crosses connected with dotted lines in each figure.

\bsp	
\label{lastpage}
\end{document}